\documentclass[11pt,a4paper]{article}

\usepackage{jheppub}
\usepackage{chessfss}
\usepackage{dsfont}

\newcommand{\beqa}{\begin{eqnarray}}
\newcommand{\eeqa}{\end{eqnarray}}

\def\W{{\cal W}} 
\def\Tr{\rm Tr}
\newcommand{\tr}{\text{tr}\,}

\newcommand{\beq}{\begin{equation}}
\newcommand{\eeq}{\end{equation}}
\newcommand{\bea}{\begin{eqnarray}}
\newcommand{\eea}{\end{eqnarray}}

\newcommand{\CC}{{\mathcal C}}

\newcommand{\CN}{{\mathcal N}}

\newcommand{\CS}{{\mathcal S}}

\newcommand\SU{{\rm SU}}

\newcommand{\be}{\begin{equation}}
\newcommand{\ee}{\end{equation}}

\newcommand{\complex}{{\mathds C}} 
\newcommand{\real}{{\mathds R}} 

\def\+{{+\!\!\!+}}

\def\0{\nonumber}

\def\W{{\cal W}}

\def\Tr{\rm Tr}

\def\1{{\bf 1}}

\setboardfontsize{8}

\title{$\CN=1$ Geometries via M-theory}

\preprint{SISSA 36/2013/MATE-FISI, CALT-68-2850}

\author[1, 2]{Giulio Bonelli,}
\author[3]{Simone Giacomelli,}
\author[4]{Kazunobu Maruyoshi,}
\author[1]{and Alessandro Tanzini}

\affiliation[1]{International School of Advanced Studies (SISSA) 
\\ via Bonomea 265, 34136 Trieste, Italy \\and INFN, Sezione di Trieste}
\affiliation[2]{ICTP \\ Strada Costiera 11, 34014 Trieste, Italy}
\affiliation[3]{Scuola Normale Superiore \\ Piazza dei Cavalieri 7, Pisa, Italy}
\affiliation[4]{California Institute of Technology \\ Pasadena, CA 91125, USA}

\emailAdd{bonelli@sissa.it}
\emailAdd{si.giacomelli@sns.it}
\emailAdd{maruyosh@caltech.edu}
\emailAdd{tanzini@sissa.it}

\abstract{
  We provide an M-theory geometric set-up to describe four-dimensional ${\cal N}=1$ gauge theories. 
  This is realized by a generalization of Hitchin's equation. 
  This framework encompasses a rich class of theories including superconformal and confining ones.  
  We show how the spectral data of the generalized Hitchin's system encode 
  the infrared properties of the gauge theory in terms of ${\cal N}=1$ curves. 
  For $\CN=1$ deformations of ${\cal N}=2$ theories in class ${\cal S}$, we show how the superpotential
  is encoded in an appropriate choice of boundary conditions at the marked points in different S-duality frames.
  We elucidate our approach in a number of cases -- including Argyres-Douglas points, confining
  phases and gaugings of $T_N$ theories -- and display
  new results for linear and generalized quivers.
}


\begin{document}
\setcounter{tocdepth}{2}
\maketitle
\section{Introduction}
\label{sec:intro}
  M-theory geometric description of supersymmetric gauge theories is a very powerful access key 
  to the non-perturbative aspects of the formers and complements their perturbative Lagrangian description. 
  In this context, a very powerful approach to four-dimensional ${\cal N}=2$ gauge theories
  has been introduced in \cite{Witten:1997sc,Gaiotto:2009we}. 
  In this paper we propose a general geometric set-up for four-dimensional ${\cal N}=1$ gauge theories 
  which naturally encompasses previous well known examples studied 
  from the point of view of rotated M5-brane systems in \cite{Hori:1997ab,Witten:1997ep,Brandhuber:1997iy,deBoer:1997ap}.
  
  The approach we follow consists in considering the six-dimensional $(2,0)$ theory compactified 
  on a Riemann surface ${\cal C}$ with marked points with a suitable topological twist.
  This is realized by wrapping an M5-brane bound system around ${\cal C}\times \real^{1,3}$,
  where $\CC$ is the base of a local Calabi-Yau (CY) threefold $X$ obtained as the total space of 
  a rank 2 holomorphic vector bundle with normalized determinant over ${\cal C}$.
  For theories with $\CN=1$ superconformal invariance in the infrared,
  this set-up has been considered in \cite{Benini:2009mz,Bah:2012dg}.
  See also \cite{Maruyoshi:2009uk,Bah:2011je,Beem:2012yn,Gadde:2013fma} for relevant field theoretical arguments.
  
  The dynamics of the M5-branes is determined by M2-branes suspended between them.
  These are particles in the spacetime and therefore are effectively open strings in $X$ 
  ending on the holomorphic two-cycle ${\cal C}$.
  Therefore the dynamics of the M5-branes reduces to that of B-branes in $X$ and as such it is described 
  by the appropriate dimensional reduction of the holomorphic Chern-Simons theory on ${\cal C}$.
  Its equations of motion give rise to a generalization of the Hitchin system 
  with suitable singular boundary conditions at the marked points.
  
  The spectral curve of this generalized Hitchin system describes the geometry of the $N$ M5-brane bound state 
  as an $N$-branched cover of ${\cal C}$. 
  This is encoded in a specific overdetermined algebraic system that we study in detail. 
  This system describes a rich class of ${\cal N}=1$ theories including superconformal and confining ones, 
  and contains the relevant infrared data of the four-dimensional ${\cal N}=1$ gauge theory.
  We show how to obtain the generalized Konishi anomaly equation \cite{Cachazo:2002ry}, 
  the Dijkgraaf-Vafa curve \cite{Dijkgraaf:2002dh}, the $\CN=1$ curves presented in \cite{Hori:1997ab,Witten:1997ep},
  and the curve describing the ${\cal N}=1$ gauge theory coupled to $T_N$ theories,
  related with \cite{Tachikawa:2011ea, Maruyoshi:2013hja}.
  In particular the factorization of both the Seiberg-Witten curve and the Dijkgraaf-Vafa curve 
  is elucidated and worked out in new examples.
  
  The above geometrical objects compute the vev of chiral ring observables.
  The moduli space of vacua of the ${\cal N}=1$ theory is described by the moduli space of solutions 
  of the generalized Hitchin system associated to it.
  A particularly important class of theories is the one obtained via turning on an ${\cal N}=1$ breaking superpotential
  of adjoint chiral fields to ${\cal N}=2$ theories.
  This sets the boundary conditions on the generalized Hitchin field and provides an obstruction in the 
  ${\cal N}=2$ Hitchin moduli space. 
  Therefore the generalized system has a reduced moduli space corresponding to the lifting of the flat directions 
  in the Coulomb branch of the ${\cal N}=2$ theory, induced by the superpotential.
  Indeed, we see that in various examples the Coulomb moduli is completely fixed, leading to isolated vacua.
  The dual geometric set-up to the above case is described as arising via a complex structure rotation 
  in the target space geometry, which connects the ${\cal N}=1$ breaking potential to purely geometrical data.
  
  We show how our set-up reproduces several known examples previously studied in the literature, 
  namely ${\cal N}=1$ super Yang-Mills (SYM) theory, supersymmetric QCD (SQCD) 
  and ${\cal N}=1^*$ theories as trial cases.
  The new results which are obtained concern the extension of the factorization condition 
  of the Seiberg-Witten curve for linear and generalized quivers.
  We also discuss a subtle difference in the interpretation of the boundary conditions 
  which has to be taken into account when considering gauge theories with $\SU(N)$ group, rather than ${\rm U}(N)$, 
  via a shift of the boundary condition by the meson vevs.
  
  The use of the generalized Hitchin system requires the introduction of a vector field on ${\cal C}$
  whose choice looks crucial in order to establish the S-duality frame of the underlying theory. 
  For $\SU(2)$ generalized quivers we show that the factorization condition is an easy consequence 
  of the generalized Hitchin system and provides a simple recipe for computing the chiral condensates
  when combined with the appropriately chosen vector field.
  We exemplify the relation between the choice of the vector field and the S-duality frame 
  for the corresponding gauge theory in the case of $\SU(2)$ theory with $N_f=4$. 

  This paper is organized as follows. 
  In section \ref{sec:Hitchin} we consider the general geometric set-up generating 
  four-dimensional ${\cal N}=1$ gauge theories from M5-branes in the local CY curve.
  We discuss the generalized Hitchin system and its spectral curve data,
  and we focus on the theories obtained from ${\cal N}=2$ ones via superpotential breaking.
  In section \ref{sec:A1} we see how the framework provided in section \ref{sec:Hitchin} is applied to 
  $\CN=1$ $\SU(2)$ gauge theories.
  We show that the factorization condition which follows from the generalized Hitchin system is 
  strong enough to determine the $\CN=1$ curves of various $\SU(2)$ gauge theories.
  We also give some results for the $\CN=1^*$ gauge theory with $\SU(2)$ gauge group.
  In section \ref{sec:AN}, we study the higher rank theories 
  including $\SU(N)$ SYM theory and the $T_N$ theories coupled to $\CN=1$ $\SU(N)$ gauge groups.
  In section \ref{sec:Konishi} the $\CN=1$ curve of linear quiver theory is obtained from the Konishi anomaly equations.
  We discuss the factorization condition of this theory in detail.
  Section \ref{sec:discussion} contains our final remarks and discusses some open issues.
  In appendix \ref{sec:Hori} we present some of our results by an alternative method 
  making use of the type IIA branes picture.

\section{Generalized Hitchin system}
\label{sec:Hitchin}

\subsection{${\cal N}=1$ M-theory engineering}
  ${\cal N}=1$ supersymmetric gauge theories in four dimensions can be engineered 
  by considering the reduction of $N$ M5-branes on the CY geometry
    $$\real^{1,3}\times X \times \real$$
  where the $N$ M5-branes wrap $\real^{1,3}\times{\cal C}_{g,n}$, ${\cal C}_{g,n}$ 
  being an holomorphic two-cycle with genus $g$ and $n$ marked points in the CY threefold $X$.
  
  Let us concentrate on the geometry in the vicinity of the brane system and specify the CY threefold 
  to be a local curve over ${\cal C}$ \footnote{Here and in the following we drop for simplicity the subscripts: 
  ${\cal C}\equiv{\cal C}_{g,n}$.}, namely $X={\rm tot}V_{\cal C}$,
  where $V_{\cal C}$ is a rank two holomorphic vector bundle with 
  ${\rm det}V_{\cal C}=K_{\cal C}$ the canonical line bundle on ${\cal C}$,
  to ensure supersymmetry by the usual CY condition.
  
  The geometry of the $N$ M5-branes encodes the definition of the ${\cal N}=1$ gauge theory 
  and it is described by the following generalization of the Hitchin system
    \bea
    \bar D\vec\Phi
    &=&     0 \0, \\
    \vec\Phi\wedge\vec\Phi
    &=&     0,
            \label{gh}
    \eea
  which should be supplemented by suitable boundary conditions at the marked points.
  Eqs.~\eqref{gh} are the Uhlenbeck-Yau equations in ${\rm tot}V_{\cal C}$ dimensionally reduced to the base ${\cal C}$.
  $\vec\Phi$ is a section in $V_{\cal C}$ transforming in the adjoint representation of a given gauge bundle. 
  The D-term equation is traded for the complexification of the gauge group 
  and a stability assumption on the gauge bundle.
  
  Notice that the BPS M5-branes, as seen from the internal CY geometry viewpoint, are B-branes of complex dimension one.
  The M2-branes stretching between them are particles in $\real^{1,3}$ and open strings in $X$ 
  with boundary on the B-branes.
  Indeed, the action functional generating the equations \eqref{gh} is the dimensional reduction 
  to ${\cal C}$ of the holomorphic Chern-Simons theory on $X$ and reads
    \bea
    S
     =     \frac{1}{2} \int_{{\cal C}} {\rm Tr} \vec{\Phi} \wedge \bar D \vec{\Phi}.
    \eea
  
  The geometry of the $N$ M5-brane bound state is described as an $N$-branched cover of ${\cal C}$ 
  in the form of the spectral curve for the commuting pair $\vec\Phi$.
  This can be easily written in components as follows.
  By specifying $\vec\Phi=(\phi_1,\phi_2)$ in components the generalized Hitchin equations \eqref{gh} read
    \bea
    \bar D\phi_i
     =     0\ \ \ \ 
           (i=1,2),~~~~~~~
    [\phi_1,\phi_2]
     =     0
           \label{ghc}
    \eea
  whose spectral curve is given by the overdetermined algebraic system
    \bea
    \det(x_1-\phi_1)
    &=&     0, \nonumber \\
    \det(x_2-\phi_2)
    &=&     0,\nonumber \\
    \det(x_1x_2-\phi_1\phi_2)
    &=&     0,
            \label{curvas}
    \eea
  spanning an $N$-sheeted cover of ${\cal C}$ described by the simultaneous eigenvalues of $\phi_1$ and $\phi_2$.
 
  The set-up we just described is seemingly quite rich and can be used to define 
  a wide class of ${\cal N}=1$ supersymmetric gauge theories in four dimensions, 
  whose moduli space of vacua is described by the moduli space of solutions of the generalized Hitchin's system. 
  In particular maximally confining vacua correspond to isolated solutions.
  An important subset of these theories is ${\cal N}=1$ deformation of ${\cal N}=2$ gauge theories 
  in class ${\cal S}$ by superpotential terms depending on the adjoint chiral fields.
  This subset includes the theories considered extensively in
  \cite{Douglas:1995nw, Elitzur:1996gk, Cachazo:2001jy,Dijkgraaf:2002dh}.
  We will mainly focus on this subset in this paper. Let us remark however that more general genuine
  ${\cal N}=1$ theories can be described within our approach. 
  
  In order to see the ${\cal N}=1$ theory as a deformation of an ${\cal N}=2$ theory in class ${\cal S}$, 
  one should rotate the target space complex structure as 
    \be
    {\rm tot}V_{\cal C}
    \quad\to\quad
    \complex\times T^*{\cal C}.
    \label{rotationz}
    \ee
  In the following, for the sake of simplicity, we will restrict to the case 
  in which $V_{\cal C}=K_{\cal C} {\cal L}^{-1}\oplus {\cal L}$ is split in the sum of two line bundles, 
  although more general cases can be considered.
  The rotation of the complex structure \eqref{rotationz} can be encoded 
  in a $t$-dependent coordinate redefinition $(x_1,x_2,z)\to(x,w,t)$ which has to be a canonical transformation 
  to map (almost everywhere) the CY holomorphic top forms as
    \bea
    dx_1\wedge dx_2\wedge d z\quad \sim \quad dx\wedge dw\wedge dt,
    \nonumber 
    \eea
  where $t$ and $z$ are generically different coordinates on ${\cal C}$, 
  $x$ is a coordinate in the canonical bundle $K_{\cal C}$ and $w$ in ${\mathbb C}$.
  The change of coordinates is then obtained from a generating potential $\Psi$ valued in $K_{\cal C}$.
  For example one can have 
    \bea
    w
     =     \partial_x\Psi(x,x_2,t)\,\,~~ {\rm and}\quad 
    x_1\partial_t z
     =     \partial_{x_2}\Psi(x,x_2,t).
    \eea
  The canonical transformation can be extended to the commuting pairs $(\phi_1,\phi_2)\to(\Phi,\varphi)$, 
  where $\Phi$ and $\varphi$ are the components of the $\vec\Phi$ doublet in the rotated target space geometry.
  
  As we will describe in the following subsection the pair $(\Phi,\varphi)$ subject 
  to some precise boundary conditions describes ${\cal N}=1$ breaking of ${\cal N}=2$ theories in class ${\cal S}$.
  We remark that a non trivial constraint on the canonical transformation is that it produces 
  the correct boundary conditions. 
  This in general cannot be done and therefore in this setting this is the condition 
  such that the ${\cal N}=1$ initial theory can be seen as a deformation ${\cal N}=2$ one.
  For example, the simplest canonical transformation is the coordinate rotation induced 
  by meromorphic sections $\sigma_1$ of ${\cal L}$, $\sigma_2$ of $K_{\cal C}{\cal L}^{-1}$ and $\tau$ of ${\cal L}^{-1}$
    \bea
    \Phi
     =     \sigma_1\phi_1+\sigma_2\phi_2,~~~~
    \varphi
     =     \tau\phi_2,
           \label{mianonna}
    \eea
  where $\Psi=\tau x_2\left(x-\frac{\sigma_2x_2}{2}\right)$ and $\sigma_1\tau=\partial_t z$.
  This rotation has a non trivial effect on the behavior of the fields at the divisor dual to ${\cal L}$ 
  and can induce spurious singularities in the Hitchin field.
  In particular, if ${\cal L}$ is positive we expect to have a genuine  ${\cal N}=1$ theory.
  For example for ${\cal C}={\mathbb P}^1$, if ${\cal L}={\cal O}(p)$ and $p>0$, 
  then one gets $\CN=1$ superconformal theories studied in \cite{Bah:2012dg}.
  
  In what follows we focus on the study of ${\cal N}=1$ deformation of theories in class ${\cal S}$.

\subsection{${\cal N}=1$ breaking of theories in class ${\cal S}$}
  ${\cal N}=2$ gauge theories are obtained as particular cases of ${\cal N}=1$ theories 
  which are constrained in their spectrum and couplings.
  These constraints can be implemented in the M-theory engineering 
  by imposing the existence of higher supersymmetry in the geometry, namely by requiring the $\SU(2)$ $R$-symmetry 
  to get realized as a spacetime symmetry.
  In concrete we need the local CY geometry threefold to have a trivial fiber 
  and therefore to be a local K3 manifold.
  This produces the familiar M-theory geometric background
    $$\real^{1,3}\times K3 \times \real^3$$
  where the local K3 geometry is uniquely determined by the holomorphic two-cycle ${\cal C}$
  to be the total space of its canonical bundle $K_{\cal C}$.
  More explicitly the holomorphic vector bundle is $V_{\cal C}=K_{\cal C}\oplus \complex$ 
  and $\real\times\complex=\real^3$.
  
  The constraints on the M5-brane system should still be consistently enforced to guarantee 
  the overall ${\cal N}=2$ supersymmetry.
  The generalized Hitchin equations \eqref{ghc} now read
    \bea
    \bar D\Phi=0,
    \\
    \bar D\varphi =0,
    \\
    \left[\varphi,\Phi\right]=0,
    \label{eh}
    \eea
  where $\Phi$ is a $(1,0)$-form on ${\cal C}$ and $\varphi$ is a scalar on ${\cal C}$ 
  both transforming in the adjoint representation of the gauge bundle.

  The M5-brane system should occupy a point in the $\real^3$ factor of the geometry and this can be enforced 
  by setting to vanish the field $\varphi$ reducing the extended Hitchin system \eqref{eh} 
  to the well known Hitchin equation
    $$\bar D \Phi=0.$$
  This should be supplemented by appropriate boundary conditions at the punctures 
  describing the M5-brane asymptotic geometry.
  Notice the important fact that the above procedure of eliminating $\varphi$
  is independent on the boundary conditions imposed on $\Phi$.
  
  The complete geometry is obtained by the Seiberg-Witten (SW) curve in the form 
  of the spectral curve of the Hitchin system which follows directly from \eqref{curvas}. 
  By specifying $\varphi=0$ and $x_2=0$ one gets
    $$\det[x -\Phi]=0,$$ 
  where $x_1\to x$ is the fiber coordinate on the local K3 geometry around ${\cal C}$.
  Notice that in order to fix $\varphi=0$
  it is enough to set to zero its boundary values at the singularities, whenever the gauge bundle is irreducible.

  In conclusion, by these two steps, namely the holonomy reduction of the target {\it and} 
  the assignment of vanishing boundary conditions to $\varphi$ at the punctures, 
  we could enforce the $R$-symmetry required to get an enhancement to an ${\cal N}=2$ gauge theory.
  
  Corresponding to the two steps above, we have therefore two seemingly different ways 
  to break ${\cal N}=2$ supersymmetry to ${\cal N}=1$ within the family of theories 
  which we are considering in this paper.
  One is by rotating the M5-branes in a direction orthogonal to the local K3 by fixing suitable boundary conditions,
  the other one is obtained by embedding the system 
  in a more general local CY geometry and rotating the complex structure.
  Actually both these steps, in the perturbative regimes, should be equivalent to switch on 
  an $\mathcal{N}=1$ superpotential.
  Let us describe the two deformations in more detail.

\subsubsection{Breaking via M5-branes rotation}
\label{subsubsec:1st}
  {\it The first} way of ${\cal N}=1$ deformation corresponds to consider once more the extension of the Hitchin system
    \bea
    \bar D\Phi
     =     0,
           \nonumber \\
    \bar D\varphi
     =     0,
           \nonumber \\
    \left[\varphi,\Phi\right]
     =     0,
           \label{eh}
    \eea
  but now with non-vanishing boundary conditions on $\varphi$.
  As we noticed, if the field $\varphi$ had vanishing boundary conditions at the punctures, then it vanishes everywhere.
  Changing the boundary conditions for $\varphi$ at the punctures therefore corresponds to break to ${\cal N}=1$.
  The allowed boundary conditions for $\varphi$ shall be given in terms of the field $\Phi$ itself
  in order to automatically satisfy the equation $[\varphi,\Phi]=0$.
  Henceforth the boundary conditions encode the supersymmetry breaking superpotential.

  To write the boundary conditions we need to compare the $\Phi$ and $\varphi$ on the Riemann surface ${\cal C}$
  and this is obtained by introducing a (possibly singular) holomorphic vector field on ${\cal C}$. 
  Let us denote this vector field by $\zeta=\zeta^t\partial_t$ 
  and consider then the system with boundary conditions at the punctures $\{P_i\}_{i=1,\ldots,n}$
    \be
    \varphi(P)\sim W'_i(V) \quad {\rm as}\quad P\sim P_i,
    \label{bcc}
    \ee
  where $V=i_\zeta\Phi=\zeta^t\Phi_t$ and $W'_i$ are polynomials in $V$.
  Let us notice that one of the boundary conditions in \eqref{bcc} can be set to zero, say $W'_1=0$, 
  by the change of variables $\varphi \to \varphi - W'_1(V)$
  unless $V$ introduces spurious singularities by the vector field $\zeta$.

  The rotated M5-brane is then described by the spectral curve of (\ref{eh}), 
  namely by the curve in the space $\complex\times T^*{\cal C}$ spanned by the overdetermined system
    \bea
    \det[x -\Phi]
    &=&     0,
            \nonumber \\
    \det[xw -\Phi\varphi]
    &=&     0,
            \nonumber \\
    \det[w -\varphi]
    &=&     0,
    \label{gs}
    \eea
  where $w$ and $x$ are the coordinates in $\complex$ and the fiber of $T^*{\cal C}$ respectively.
  As it will become evident in the following sections, 
  the resulting system describes both the original Hitchin spectral curve at the singular point 
  on the $\CN=2$ Coulomb branch which is not lifted by the $\CN=1$ deformation, 
  and the ${\cal N}=1$ curve describing the solution of the generalized Konishi anomaly equation in a unified way.
  The elementary, but crucial, observation we will use is that, once fixed by the boundary conditions, 
  the algebraic system (\ref{gs}) is solved as an algebraic relation 
  by the fields $(\Phi,\varphi)$ (Hamilton-Cayley theorem for the commuting pair).

  In this framework the fixing of the ${\cal N}=2$ Coulomb moduli to their ${\cal N}=1$ vacuum values is 
  due to the fact that the extended Hitchin system's moduli space has a reduced dimension 
  with respect to the Hitchin system itself 
  because of the obstructions induced by the non-vanishing boundary conditions on $\varphi$.
  More precisely, the equations \eqref{gs} describe a K\"ahler submanifold in the Hitchin moduli space 
  obtained by specifying the Hitchin's Hamiltonians to the ${\cal N}=1$ condensates. 
  Indeed, as we will show in section \ref{sec:A1}, the second equation of \eqref{gs} can be identified 
  for Lagrangian theories with the generalized Konishi anomaly equations encoding the ${\cal N}=1$ chiral condensates. 

\subsubsection{Breaking via complex structure rotation}
\label{subsubsec:2nd}
  {\it The second} way to break supersymmetry down to ${\cal N}=1$ corresponds to rotate
  the target space complex structure as
    \be
    \complex\times T^*{\cal C}\quad\to\quad {\rm tot}V_{\cal C}
    \label{rotation}\ee
  This can be implemented via a suitable canonical transformation, as discussed in the previous subsection
  as a coordinate redefinition $(x,w,t)\to(x_1,x_2,z)$ preserving (almost everywhere) the CY holomorphic top form.
  The canonical transformation can be extended to the commuting pairs $(\Phi,\varphi)\to(\phi_1,\phi_2)$
  setting the equivalence between the descriptions in \eqref{eh} and \eqref{gh}
  {\it if accompanied by the induced correspondence of the boundary conditions in the appropriate way.}

  The choice of the canonical transformation is indeed crucial in order to avoid a mismatch in the boundary conditions. 
  In particular it is important to select adapted sections to match them. 
  We will consider this issue in detail in the following examples.

\subsection{M-theory curve and factorization condition}
\label{subsec:factorization}
  In the following sections we will mainly exploit the first description involving the $\mathcal{N}=2$ Hitchin field. 
  The curve is given by the generalized Hitchin system we discussed before, 
  which includes the $\mathcal{N}=2$ curve and a degree $N$ equation for the coordinate $w$ 
  (the spectral curve for the $\varphi$ field) which has the form 
    \bea
    w^N
     =     \sum_{k=2}^{N}w^{N-k}V_k(t),
           \label{N=1curve}
    \eea
  where $V_k(t)$ are meromorphic functions on ${\cal C}$ 
  with poles at the punctures, since the coordinate $w$ lives in ${\cal O}(0)$. 

  A curve exactly of the form written above, which encodes the matrix of gauge couplings, has been found recently 
  for $\mathcal{N}=1$ gauge theory coupled to $T_N$ theories \cite{Tachikawa:2011ea, Maruyoshi:2013hja}. 
  It is thus natural to expect that what we need is precisely the generalization of 
  these ``$\mathcal{N}=1$ Seiberg-Witten curves''. 
  We will indeed see that these curves emerge naturally for the class of theories considered 
  in \cite{Tachikawa:2011ea,Maruyoshi:2013hja}, once we impose the correct boundary conditions at the punctures. 
  Thus we refer to \eqref{N=1curve} as $\CN=1$ curve.
  
  These two equations must then be supplemented by a third equation, which encodes the restrictions 
  on the Coulomb branch coordinates of the underlying $\mathcal{N}=2$ theory. 
  From the perspective of the generalized system, this is encoded in the spectral equation for the product
  (the second equation in \eqref{gs}). 
  How can we understand it from the brane perspective?
  What we are trying to do is to construct softly broken $\mathcal{N}=2$ theories obtained 
  by compactifying the six-dimensional $\mathcal{N}=(2,0)$ theory of type $A_{N-1}$ on $\CC$. 
  Indeed, the Seiberg-Witten (SW) curve associated to the underlying $\mathcal{N}=2$ theory is an $N$-sheeted covering 
  of $\CC$ and the same should be required to hold in the present context if we want this picture to make sense. 
  It is easy to see that this property is not satisfied for generic values of the parameters: 
  for fixed $t$ the SW curve gives us $N$ possible solutions for $x$, 
  and the $\mathcal{N}=1$ curve provides $N$ solutions for $w$. 
  This generically leads to $N^2$ possible pairs $(x,w)$.
  We should then add to this system the condition that $w$ is fixed 
  once we have chosen $x$ and $t$ (and of course $x$ is fixed once $w$ and $t$ are chosen),
  which reduces the solutions to $N$ possible pairs $(x,w)$.
  Indeed, this is what the second equation in \eqref{gs} implies. 
  This simply translates into the requirement that $w$ is a rational function of $x$ and $t$. 
  
  Introducing the vector field $\zeta$, 
  we write the SW curve in terms of $v=\lambda(\zeta)$, where $\lambda=xdt$ is the SW differential, as 
    $$v^N=\sum_kv^{N-k}f_k(t),$$ 
  where $f_k(t)$'s are meromorphic functions.
  The factorization condition for a generic rank $N$ theory will be of the form 
    \beq
    \label{facct}
    w
     =     \frac{P_{N-1}(v)}{Q_{N-1}(v)},
    \eeq 
  (of course an analogous equation holds with $v$ and $w$ interchanged) 
  where $P_{N-1}(v)$ and $Q_{N-1}(v)$ are polynomials in $v$ of degree $N-1$, 
  whose coefficients are meromorphic functions on $\CC$ 
  (rational functions for theories on the sphere, elliptic functions for genus one theories and so on). 
  We will see in section \ref{sec:Konishi} that this leads directly to the well known factorization condition for SQCD.

\section{$\CN=1$ curve in $A_1$ theory}
\label{sec:A1}
  In this section we consider four-dimensional $\CN=1$ gauge theories obtained from the $A_{1}$ theory 
  in six dimensions (or compactification of two M5-branes).
  The gauge groups are always $\SU(2)$ in this case.
  From the generalized Hitchin system viewpoint (the first description in section \ref{subsubsec:1st}), 
  we have two Hitchin fields $\Phi$ and $\varphi$ where the spectral curves can be put in the form,
  in terms of $v=\lambda(\zeta)$ and $w$,
    \bea
    v^2
     =     f_2(t), ~~~~
    w^2
     =     V_2(t).
           \label{A1curves}
    \eea
  Nota that $v$ is identified with the coordinate $v=x_{4} + ix_{5}$ appearing in the M-theoretic geometry
  if we choose $\zeta = t \partial_{t}$ in the theory with the Type IIA brane realization as we will see below.
  Since the Hitchin fields commute, they must be proportional to each other (which is true only for the $A_{1}$ case). 
  This implies $\varphi=F(t) \Phi$ which gives us the condition $w=F(t)v$ or
    \begin{equation}\label{fac2}
    w^2
     =     F(t)^2v^2.
    \end{equation}
  
  Indeed, this condition can be obtained from the factorization condition in section \ref{subsec:factorization}.
  In this rank-one case the condition reads 
    \bea
    w
     =     \frac{A(t)v+B(t)}{C(t)v+D(t)}.
    \eea
  Since the two solutions of \eqref{A1curves} at fixed $t$ that we denote by ($v$, $\tilde{v}$) and ($w$, $\tilde{w}$)
  differ by a sign, we get
    \bea
    \frac{-A(t)v-B(t)}{C(t)v+D(t)}
     =   - w
     =     \tilde{w}
     =     \frac{A(t)\tilde{v}+B(t)}{C(t)\tilde{v}+D(t)}
     =     \frac{B(t)-A(t)v}{D(t)-C(t)v}.
           \label{around34}
    \eea
  We can consider the analogous sequence of equalities for the inverse relation 
  $v=(B(t)-D(t)w)/(C(t)- A(t)w)$.
  Equating the leftmost and rightmost terms we find $B(t)D(t)=A(t)C(t)v^2$ and $A(t)B(t)=C(t)D(t)w^2$. 
  These in turn imply that $w^{2}/v^{2}$ is a square of a function\footnote{If some of the polynomials are zero, 
    making these two equations trivial, it is easy to see that either $B(t)=C(t)=0$ or $A(t)=D(t)=0$ 
    and this leads directly to the conclusion in any case.} \eqref{fac2}.
  Thus we refer to \eqref{fac2} as factorization condition below.

  In this case we can turn on only mass terms for the adjoint chiral fields (quadratic superpotentials), 
  so the ratio $w/v$ tends to a constant (possibly zero) at all the punctures. 
  This is the boundary condition we have to impose for $F(t)$ at the punctures. 
  The value of the mass parameters for the adjoint fields in the various gauge groups (in a given weakly coupled limit)
  will be simply encoded in the difference between the values of $F(t)$ at the punctures. 
  Indeed, since $F(t)$ is not constant, it has poles. 
  To avoid singularities for $w$ away from the punctures, which indeed should not be there, 
  the poles should be located at the zeroes of $v$. 
  In order for this to be possible, $f_2(t)$ should have double zeroes. 
  This is the way in which the restriction on the Coulomb branch coordinates appears in this description. 

\subsection{$\SU(2)$ SQCD}
\label{subsec:SU(2)SQCD}
  We will now show how one can use the above results to determine the curve for $\CN=1$ theory
  obtained by the mass deformation of the adjoint chiral field 
  of $\CN=2$ $\SU(2)$ gauge theory with $N_{f}$ fundamental hypermultiplets.
  These theories have a brane realization in Type IIA string theory.
  While the existence of the brane realization is not needed for our approach,
  it is helpful to identify each vacuum of $\CN=1$ theory with the brane configuration.
  Thus, we shortly review the Type IIA brane set-up \cite{Witten:1997sc,Elitzur:1997fh,Elitzur:1997hc,Giveon:1998sr}.
  
\begin{figure}
  \centering
  \includegraphics[width=7cm]{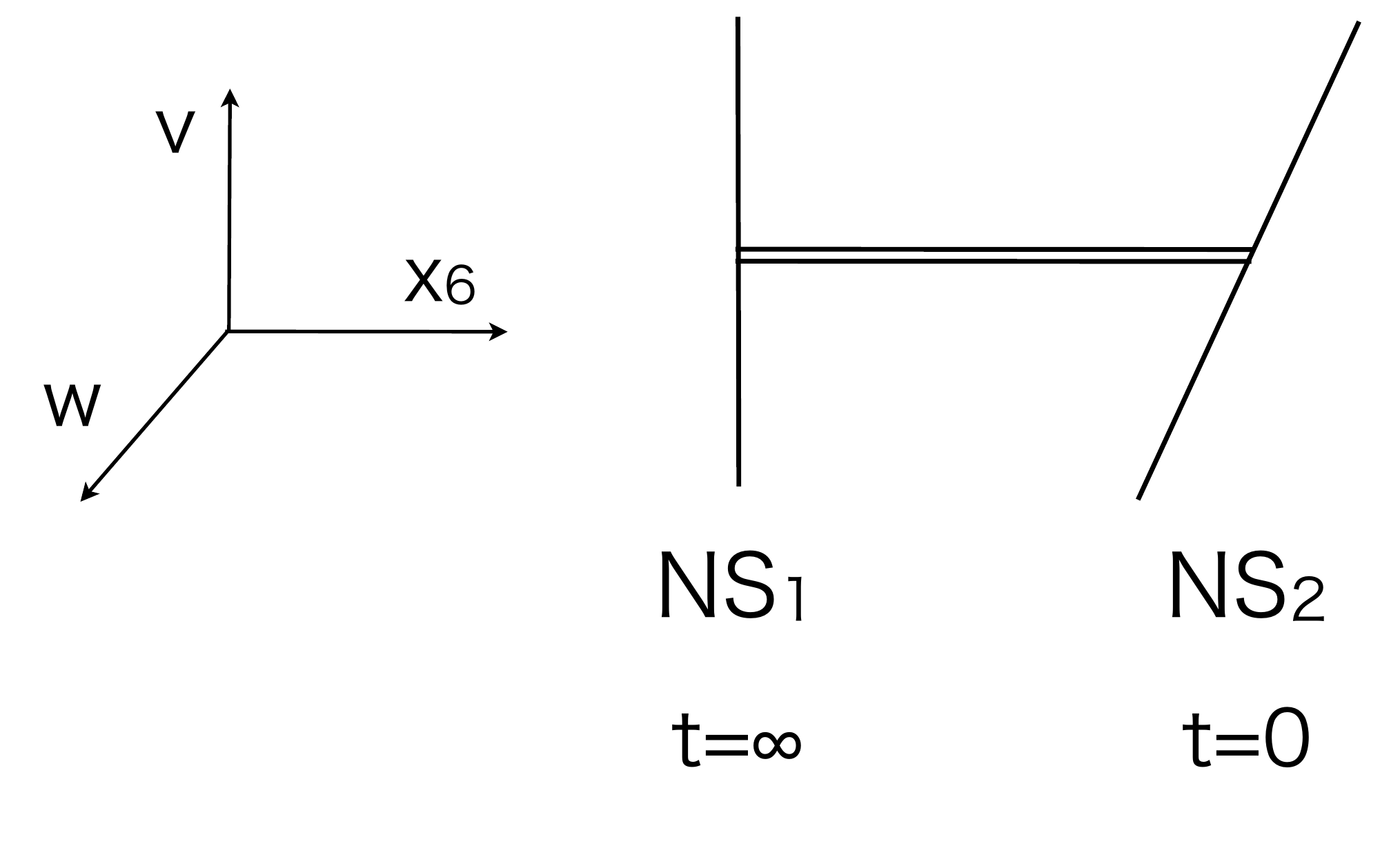}
  \caption{The rotation of the NS$_{2}$-brane. The coordinates are $v = x_{4} + i x_{5}$,
           $w = x_{8} + ix_{9}$ and $t = e^{-(x_{6}+ix_{10})}$.}
  \label{fig:SU(2)Nf0}
  \end{figure}

  We consider D4-branes extended along $x_{0},\ldots,x_{3},x_{6}$ directions
  and NS5-branes extended along $x_{0},\ldots,x_{3},x_{4},x_{5}$ directions. 
  Two D4-branes stretched between two NS5-branes induce at the low energy 
  four-dimensional $\CN=2$ $\SU(2)$ SYM theory on the worldvolume.
  In the following we denote the NS5-branes as NS$_{1}$ and NS$_{2}$ below, as in figure \ref{fig:SU(2)Nf0}.
  The addition of a D6-brane extended along $x_{0},\ldots,x_{3},x_{7},x_{8},x_{9}$ directions 
  between the NS5-branes in the $x_{6}$-direction
  corresponds to the inclusion of a fundamental hypermultiplet.
  We always move this D6-brane to the left of NS$_{1}$- or the right of NS$_{2}$-brane, 
  which produces the D4-branes stretched between the NS5- and D6-branes.
  Finally, in order to include the quadratic superpotential of the adjoint field we rotate one NS5-brane to $w$-direction,
  where $w=x_{8}+ix_{9}$.
  
  When $N_{f} \geq 2$, there is a Higgs branch in the corresponding $\CN=2$ theory.
  As in \cite{Argyres:1996eh}, generically the Higgs branch is divided into the non-baryonic branches 
  which are labeled by $r$ with $r=1,\ldots, [ N_{f}/2 ]$ and the baryonic branch.
  (The baryonic branch exists when $N_{f} \geq N$.)
  These branches intersect with the Coulomb branch at the (non-)baryonic branch roots
  which are submanifolds in the Coulomb branch.
  On these submanifolds there are loci where mutually-local massless particles appear.
  We will see that these become vacua of $\CN=1$ theory after the mass deformation.

\subsubsection{$\SU(2)$ SYM theory}
  The M-theory curve of $\CN=2$ SYM theory (which is identified with the SW curve \cite{Seiberg:1994rs}) is 
    \bea
    \Lambda^{2} t^{2} - P_{2}(v) t + \Lambda^{2}
     =     0,
           \label{N=2pureSU(2)}
    \eea
  where $P_{2} = v^{2} - u$ and $u$ is the Coulomb moduli parameter.
  The SW differential is given by $\lambda=v dt/t$.
  The curve can be written as 
    \bea
    v^2
     =     \Lambda^2t+u+\frac{\Lambda^2}{t},
    \eea
  as a double cover of the base sphere parametrized by $t$.
  The quadratic differential $\theta_{2} = f_{2}(dt/t)^{2}$ has two irregular singularities of degree $3$
  at $t=0$ and $t=\infty$.
  Note that our choice of the vector field is $\zeta = t\partial_{t}$ and $v = \lambda(\zeta) = xt$.
  
  Let us rotate the NS$_{2}$-brane at $t=0$. 
  This means that the $\mathcal{N}=1$ curve has the same behavior at $t=0$ 
  as the SW curve and subleading at infinity (in order to avoid a global rotation of the brane system). 
  In particular we require $w \rightarrow \mu v$ at $t=0$.
  This requirement leads to the curve 
    $w^2=\frac{(\mu \Lambda)^{2}}{t}+a$,
  where $a$ is a constant.
  In principle we can allow a constant term linear in $w$ but this can of course be eliminated by
  a redefinition of $w$ and $a$. 
  The curve is already in the form $w^2=V_2(t)$, so we can use (\ref{fac2}) 
    \bea
    \frac{w^2}{v^2}
     =     F^2(t)
     =     \frac{at+(\mu \Lambda)^{2}}{\Lambda^2t^2+ut+\Lambda^2}.
    \eea
  The above expression can be the square of a rational function only if $a=0$ and $u=\pm2\Lambda^2$. 
  Indeed, these are the points where the $\mathcal{N}=2$ curve degenerates.
  The $\CN=1$ curve is
    \bea
    w^{2}
     =     \frac{(\mu \Lambda)^{2}}{t},
           \label{N=1curvepure}
    \eea
  implying that $V_{2}(t)$ has a simple pole at $t=0$ where we have rotated the NS5-brane.
  This $\CN=1$ curve is sensitive only to the rotated puncture.
  The singularity of $V_{2}$ expresses that this puncture is obtained from the $\CN=2$ irregular puncture with degree $3$.
  This agrees with the one obtained in appendix \eqref{N=1curvepureHori} following the method in \cite{Hori:1997ab}.
  
  The Dijkgraaf-Vafa curve follows from these:
  by substituting \eqref{N=1curvepure} into the SW curve with $u=\pm 2\Lambda^{2}$,
  we obtain the curve $w^{2} - \mu v w + \mu S = 0$ where $S = \pm \mu \Lambda^{2}$. 
  This is identified with the DV curve: $S$ is the gluino condensate.
  We can see that by the one-loop matching this is indeed the correct value $S = \pm \Lambda_{\CN=1}^{3}$
  where $\Lambda_{\CN=1}$ is the dynamical scale of the theory after integrating out the massive adjoint fields.

\subsubsection{$\SU(2)$ $N_f=1$}
\label{subsubsec:SU(2)Nf1fac}
  We next consider the $N_{f}=1$ case
  which is realized by the addition of a D6-brane.
  Here we put the D6-brane to the right of the NS$_{2}$-brane.
  The SW curve is 
    \bea
    \label{WWc}
    \Lambda^{2} t^{2} - tP_{2}(v) + \Lambda (v + m)
     =     0,
    \eea
  where $m$ is the mass parameter of the quark field.
  This leads to 
    \bea
    v^2
     =     \frac{4\Lambda^2 t^3+4ut^{2} +4\Lambda m t+\Lambda^2}{4t^{2}}.
    \eea
  Note that we have shifted $v$ such that the linear term disappears.
  Thus the quadratic differential has singularities of degree $4$ at $t=0$ and of degree $3$ at $t= \infty$.
  
  Rotating the NS$_{2}$-brane at $t=0$ as before the ansatz of the $\CN=1$ curve is 
  $w^2= \frac{(\mu \Lambda)^{2}}{4t^2}+\frac{b}{t}+a$,
  where $a$ and $b$ are constants.
  We then require 
    $$F^2(t)=\frac{at^2+bt+(\mu \Lambda)^{2}}{4\Lambda^2 t^3+4ut^{2} +4\Lambda m t+\Lambda^2}.$$ 
  This equation imposes the constraint $a=0$, so the numerator becomes of the form $b(t+c)$ 
  where $bc=(\mu \Lambda)^{2}$, and implies that the denominator factorizes as $4\Lambda^2(t+c)(t+d)^2$. 
  We thus find $c=1/4d^{2}$ and $d$ is determined by the cubic equation $d^{3} - (m/\Lambda)d + 1/2 = 0$. 
  The Coulomb moduli is determined to be $u=\frac{\Lambda^{2}}{4d^{2}}(-3 + 8(m/\Lambda)d)$. 
  We thus find three vacua.
  When the mass parameter is set to zero, 
  the positions of the vacua in the Coulomb branch are $\mathbb{Z}_3$ symmetric, as expected. 
  Finally the $\CN=1$ curve is given by
    \bea
    w^{2}
     =     \frac{(\mu \Lambda)^{2}}{4t^{2}} + \frac{4d^{2} (\mu \Lambda)^{2}}{t},
           \label{N=1curveNf1}
    \eea
  which agrees with \eqref{N=1curveNf1Hori}.
  Note that the singularity of $V_{2}$ at the rotated puncture $t=0$ is a double pole, 
  and is different from that of $\SU(2)$ SYM case.
  This is because the rotated puncture is from the $\CN=2$ irregular one with degree $4$.
  
  To check that these are the correct points on the Coulomb branch, 
  we can e.g. rewrite the curve in the form \cite{Seiberg:1994aj,Argyres:1995wt,Hanany:1995na}
    \be\label{hyper}
    y^2
     =     (v^2-u)^2-4\Lambda^{4-N_f}\prod_i(v-m_i)
    \ee 
  (for $N_f=3$ instead of $v^2-u$ we have $v^2+\Lambda v-u$) and check that these are the only points 
  at which the discriminant vanishes (without setting $\Lambda=0$).
  At these points the curve degenerates to a sphere, corresponding to singularities of the Coulomb branch
  of $\CN=2$ theory which are not lifted by the $\CN=1$ deformation.
  In the massless case for example, the discriminant is $\Lambda^6(27\Lambda^6+16u^3)$, 
  which precisely vanishes at the three points found above setting $m=0$. 

  Since it is not trivial in this case, let us explain how one can extract the Dijkgraaf-Vafa curve. 
  The easiest way is to use the factorization condition to write $t$ in terms of $w$ and $v$, 
  and then plug the result in the $\mathcal{N}=1$ curve to get an equation in $w$ and $v$ only. 
  This procedure works, provided we use the $\mathcal{N}=2$ curve written in the parametrization as in \eqref{WWc} 
  (and redefine $w$ accordingly). 
  In the present case this is accomplished by the redefinitions 
    $$v'=v+\frac{\Lambda}{2t};\quad w'=w+\frac{\Lambda}{2t}.$$ 
  We then find
    \bea 
    w'^2-w'(\mu v-\frac{\mu\Lambda}{2d})+\mu^2\Lambda^2d
     =     0,
    \eea 
  which is precisely the curve we were looking for. 
  Without this redefinition we would have found a cubic equation in $w$.

\subsubsection{$\SU(2)$ $N_f=2$}
\label{subsubsec:SU(2)Nf2}
  Let us consider $\SU(2)$ gauge theory with $N_{f} = 2$.
  Note that there is Higgs branch in the corresponding $\CN=2$ theory:
  a non-baryonic branch and a baryonic branch.
  The latter exists only in the massless flavor case.
  There are three ways to construct this theory in Type IIA:
  the first is to place two D6-branes to the right of the NS$_{2}$-branes;
  the second is to put one D6 to the right of the NS$_{2}$ and one D6 to the left of the NS$_{1}$;
  the third is to put two D6-branes to the left of the NS$_{1}$.
  Though the gauge theory and $\CN=1$ vacua obtained from these should be the same,
  the $\CN=1$ curve looks different. 
  We will see these in the following. 

\paragraph{The first realization}
  The SW curve is given by
    \bea
    \Lambda^{2} t^{2} - P_{2}(v) t + (v+m_{1})(v+m_{2})
     =     0.
           \label{N=2curveSU(2)Nf21}
    \eea
  This leads to
    \bea
    v^{2}
     =     \frac{m_{+}^{2}}{(t-1)^{2}} + \frac{\Lambda^{2}t^{2} + ut + m_{1}m_{2}}{t-1},
    \eea
  where we defined $m_{\pm} = (m_{1}\pm m_{2})/2$.
  It is easy to see that the quadratic differential $\theta_{2} = f_{2}(dt/t)^{2}$ has singularities 
  of degree $2$ at $t=0,1$ and an irregular one at $t=\infty$ of degree $3$.
  We refer to the former singularity of degree less than or equal to $2$ as regular.
  The SW differential is $v\frac{dt}{t}$, 
  thus the residues at regular punctures $t=0$ and $t=1$ are $\pm m_{-}$ and $\pm m_{+}$ respectively.
  
\begin{figure}
  \centering
  \includegraphics[width=9.7cm]{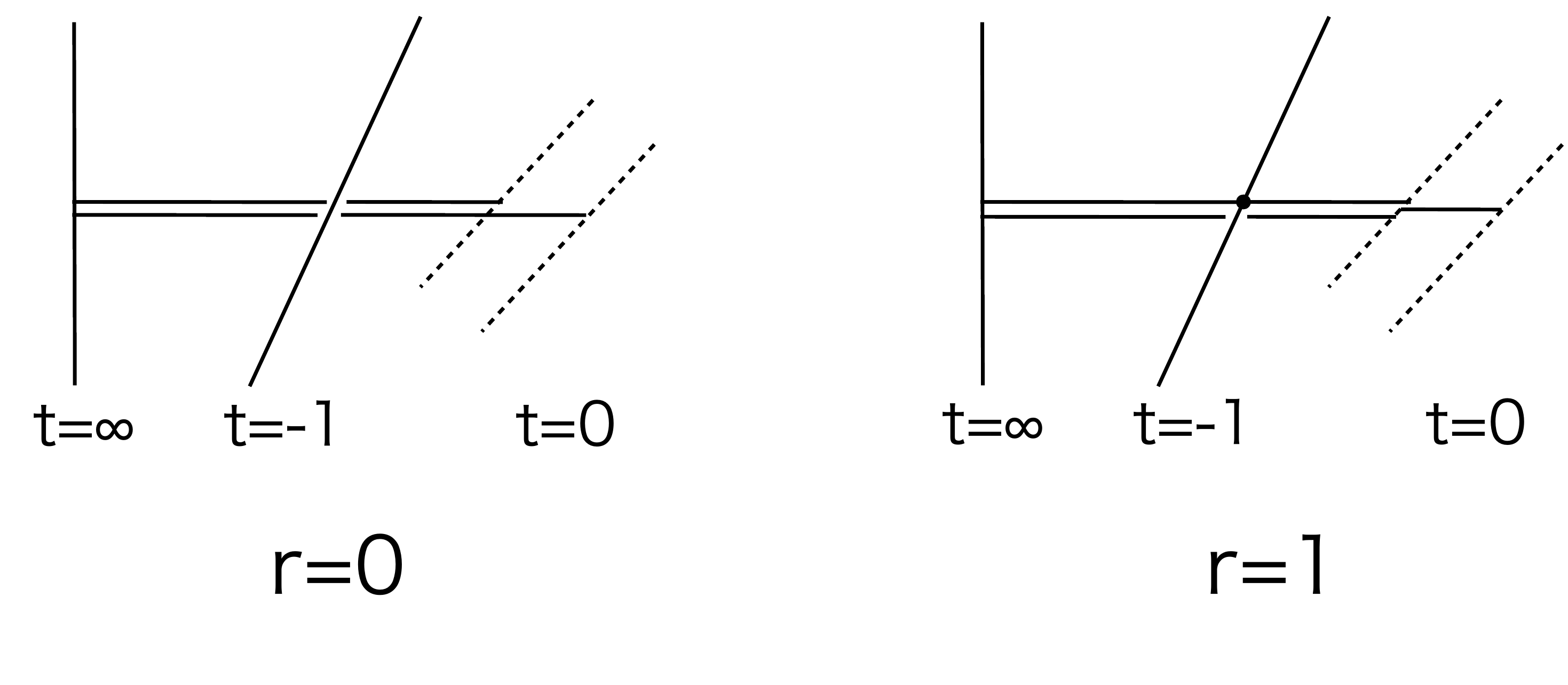}
  \caption{The first realization of $\SU(2)$ gauge theory with $N_{f} =2$. 
           Left: $r=0$ case (roots of baryonic branch). 
           The NS$_{2}$-brane is detached from the rest and is rotated to $w$ directions. 
           The dashed lines denote the D6-branes.
           Right: $r=1$ case (roots of non-baryonic branch). The NS$_{2}$-brane is still attached to D4-branes.}
  \label{fig:SU(2)Nf2}
  \end{figure}

  We will consider the rotation of the NS$_{2}$-brane as above.
  Let us set the mass parameter $m_{-}$ to be zero, namely $m_{1} = m_{2} =m_{+} =: m$ for simplicity.
  The ansatz of the $\CN=1$ curve is $w^{2} = \frac{\mu^{2} m^{2}}{(t-1)^{2}} + \frac{a}{t-1} + b$.
  Therefore we require
    \bea
    F(t)^{2}
     =   - \frac{(t-1)^{2}b+(t-1)a+\mu^{2} m^{2}}{t(\Lambda^{2}t^{2}(\Lambda^{2}-u)t-u+m^{2})}.
    \eea
  One solution is $a=\mu^{2}m^{2}$, $b=0$ and $u=-\Lambda^{2} \pm 2 \Lambda m$.
  The $\CN=1$ curve is written as 
    \bea
    w^{2}
     =     \frac{\mu^{2}m^{2} t}{(t-1)^{2}}.
    \eea
  At this locus of the Coulomb branch the SW curve degenerates.
  Thus this is the $r=0$ vacuum considered in \cite{Hori:1997ab}, as reviewed in appendix \ref{subsubsec:SU(2)Nf2Hori}.
  Note that when we set $m=0$, the $\CN=1$ curve is trivialized,
  signaling that the NS$_{2}$-brane is detached from the rest, as in figure \ref{fig:SU(2)Nf2}.
  Indeed the baryonic branch of the massless theory intersects with 
  the Coulomb branch at the locus obtained above. 
  The detachment happens when the residue of the SW differential at $t=1$ vanishes.
  We will see the same phenomenon later in the $\SU(2)$ with $N_{f}=4$ case.
  
  The other solution is $a=\mu \Lambda^{2}$, $b=0$ and $u=m^{2}$,
  where $F(t) = \mu$.
  The $\CN=1$ curve is written as 
    \bea
    w^{2}
     =     \frac{\mu^{2}m^{2}}{(t-1)^{2}} + \frac{\mu^{2} \Lambda^{2}}{t-1}.
           \label{N=1curveSU(2)Nf21-2}
    \eea
  See \eqref{N=1curveSU(2)Nf21-2Hori} for comparison.
  This corresponds to the $r=1$ vacuum,
  namely the deformation at the root of the non-baryonic branch.
  Indeed, one can easily see from the $\CN=1$ curve that
  when $t \rightarrow \infty$ we get $w = 0$,
  when $t \rightarrow -1$ we get $w \rightarrow \infty$, and
  when $t \rightarrow 0$ $w$ tends to two values $w = w_{\pm}$.
  These are explained by the figure \ref{fig:SU(2)Nf2}.
  In particular, the two values $w_{\pm}$ are interpreted as the positions of D4-branes in $w$-direction
  at $t=0$.
  Notice that the NS$_{2}$ cannot detach from the rest because of the $s$-rule.
  
  As a consistency check, one can check that the discriminant of the curve \eqref{hyper}
  with $N_{f}=2$ and $m_{1,2}=m$ has simple zeros at $u=-\Lambda^{2} \pm 2\Lambda m$ and a double zero at $u=m^{2}$.
  These are exactly the vacua we found above.

\paragraph{The second realization}
  Let us then consider the second realization.
  The SW curve is 
    \bea
    \Lambda (v+m_{1}) t^{2} + P_{2}(v) t + \Lambda (v+m_{2})
     =     0.
           \label{N=2curveSU(2)Nf22}
    \eea
  We here consider the massless theory for simplicity.
  This leads to
    \bea
    v^{2}
     =     \frac{\Lambda^{2}}{4t^{2}} (t^{4} + (2+4u/\Lambda^{2}) t^{2} + 1).
    \eea
  There are two irregular punctures of degree $4$ at $t = 0$ and $t=\infty$.
  
\begin{figure}
  \centering
  \includegraphics[width=10cm]{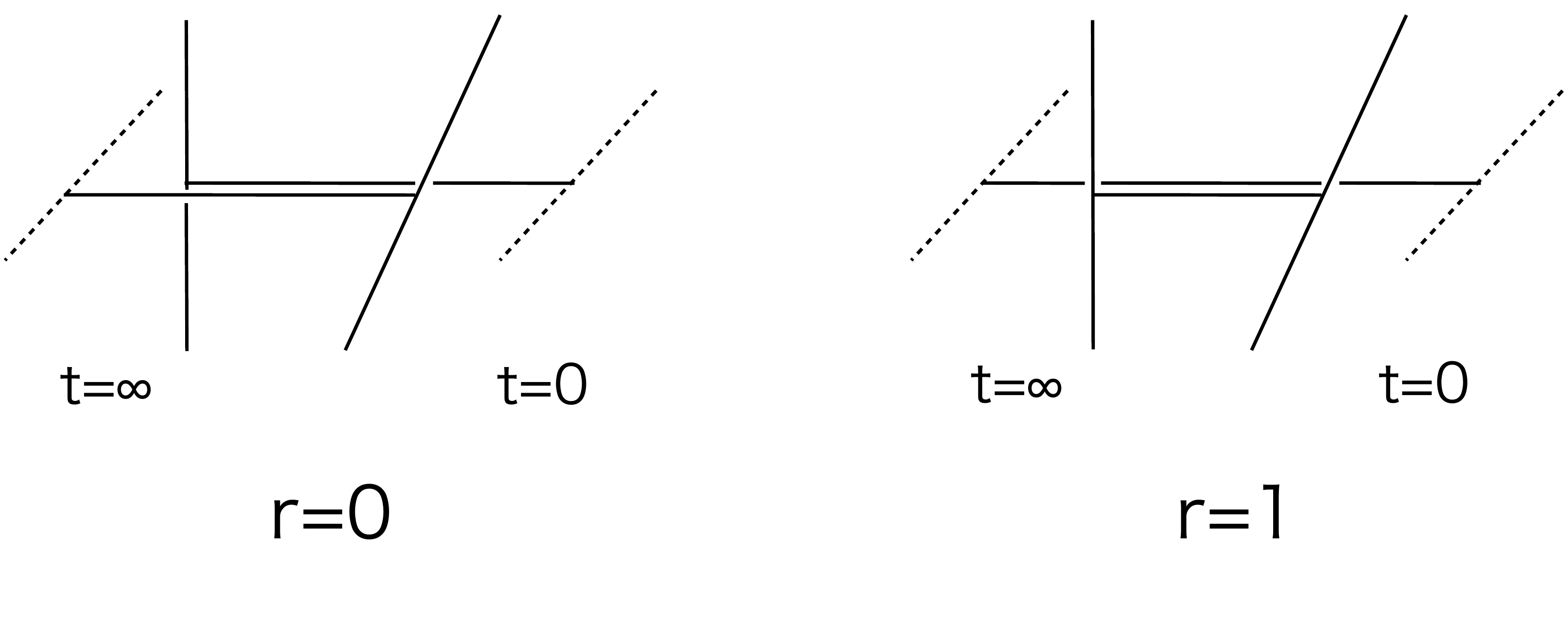}
  \caption{The second realization of $\SU(2)$ gauge theory with $N_{f}=2$.
           Left: $r=0$ case (roots of baryonic branch). 
           The NS$_{2}$-brane (and left D6-brane stretched by a D4-brane) is detached from the rest 
           and is rotated to $w$ directions. 
           Right: $r=1$ case (roots of non-baryonic branch). 
           The NS$_{2}$-brane is attached to a D4-brane.}
  \label{fig:SU(2)Nf22}
  \end{figure}

  We would like to rotate the brane at $t=0$.
  From that the boundary condition at $t=0$ is $w \rightarrow \mu v$ and $v^{2} \rightarrow \frac{\Lambda^{2}}{4t^{2}}$,
  the $\CN=1$ curve is of the following form
  $w^{2} = \frac{(\mu \Lambda)^{2}}{4t^{2}} + \frac{b}{t} + a$.
  Thus the factorization condition is
    \bea
    F(t)^{2}
     =     \frac{4}{\Lambda^{2}} \frac{a t^{2} + b t + (\mu \Lambda)^{2}/4}{t^{4} + (2 +4u/\Lambda^{2}) t^{2} + 1}.
    \eea
  There are two possibilities: one with the denominator factorized as 
  $(t-1)^{2}(t+1)^{2}$, meaning $u = -\Lambda^{2}$, and the numerator factorized as 
  $a (t \pm 1)^{2}$, meaning $a = (\mu \Lambda)^{2}/4$ and $b= \pm (\mu \Lambda)^{2}/2$.
  In this case, 
    \bea
    w^{2}
     =     \frac{(\mu \Lambda)^{2}}{4} (1 \pm 1/t)^{2}, ~~~~~
    F(t)
     =     \frac{\mu}{(\pm t + 1)}.
           \label{N=1curveSU(2)Nf22-1}
    \eea
  Note that the rhs of the $\CN=1$ curve is a square.
  The curve is identified with \eqref{SU(2)Nf2N=1curvesecond} (after the shift of $w$).
  Thus this solution of the factorization condition corresponds to the $r=0$ vacuum,
  and represents that the brane system is separated into two part as in figure \ref{fig:SU(2)Nf22}.
  
  The other possibility is $a=b=0$ and $u=0$ or $u=-\Lambda^{2}$.
  The latter corresponds to the previous solution
  after the shift of $w$-coordinate, thus we do not consider this.
  The $\CN=1$ curve is calculated as 
    \bea
    w^{2}
     =     \frac{(\mu \Lambda)^{2}}{4t^{2}}, ~~~~
    F(t)
     =     \frac{\mu}{t^{2} + 1} ~~{\rm or}~~
         - \frac{\mu}{t^{2} - 1}.
           \label{N=1curveSU(2)Nf22-2}
    \eea
  This agrees with the curve \eqref{SU(2)Nf2N=1curvesecond2}, namely the $r=1$ vacuum.
  See figure \ref{fig:SU(2)Nf22}.
  The singularity of $V_{2}$ of both \eqref{N=1curveSU(2)Nf22-1} and \eqref{N=1curveSU(2)Nf22-2} at $t=0$
  is indeed the same as that in section \ref{subsubsec:SU(2)Nf1fac}:
  the rotated puncture is of degree $4$.
  
\paragraph{The third realization}
  The SW curve is in this case
    \bea
    (v+m_{1})(v+m_{2}) t^{2} - P_{2}(v) t + \Lambda^{2}
     =     0.
           \label{N=2curveNf23}
    \eea
  Again we consider the massless case where
    \bea
    v^{2}
     =   - \frac{tu+\Lambda^{2}}{t (t-1)}.
    \eea
  There are two regular punctures at $t=1,\infty$ and an irregular one at $t=0$ of degree $3$. 
  
\begin{figure}
  \centering
  \includegraphics[width=10cm]{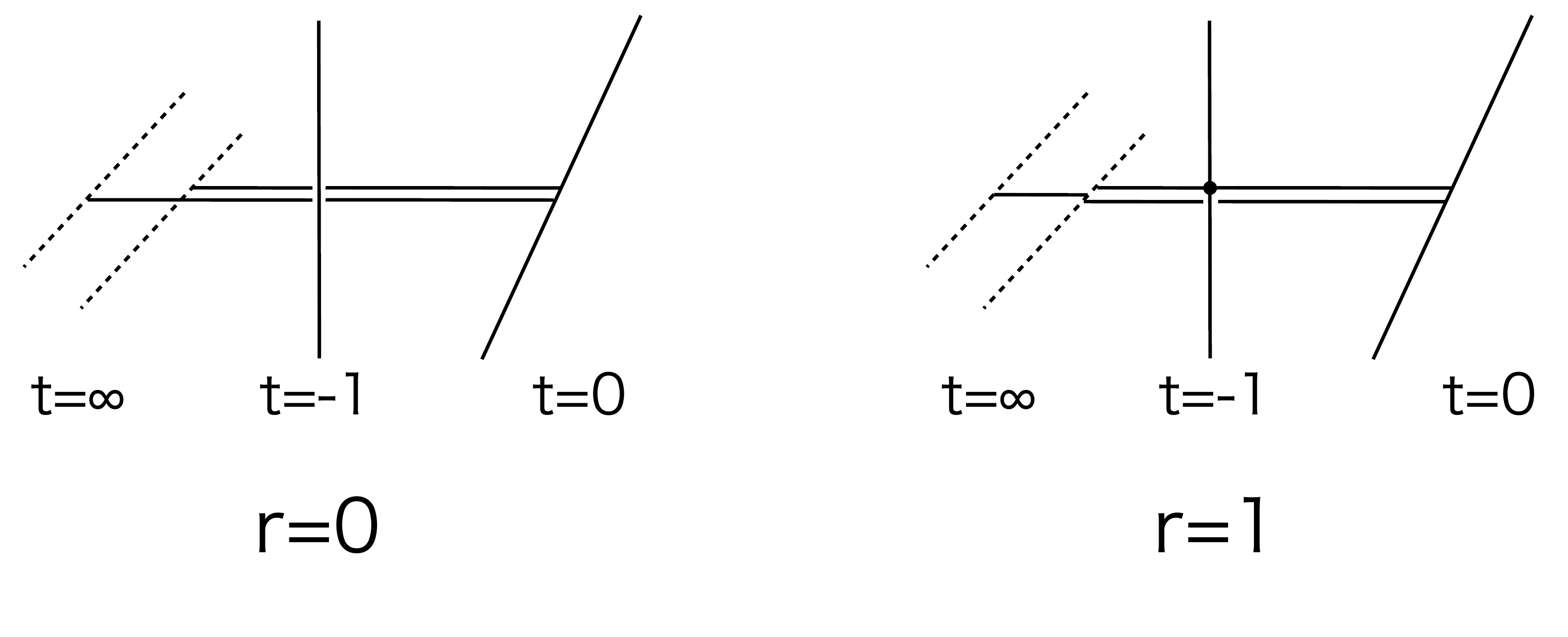}
  \caption{The third realization of $\SU(2)$ gauge theory with $N_{f}=2$.
           Left: $r=0$ case (roots of baryonic branch). 
           Right: $r=1$ case (roots of non-baryonic branch). }
  \label{fig:SU(2)Nf23}
  \end{figure}

  Let us rotate the NS$_{2}$-brane at $t=0$.
  The ansatz for the $\CN=1$ curve is $w^{2} = \frac{(\mu \Lambda)^{2}}{t} + a$.
  This is because the boundary condition at $t=0$ is $w \rightarrow \mu v$ and $v^{2} \rightarrow \Lambda^{2}/t$.
  Thus, we have
    \bea
    F^{2}
     =     \frac{(at + (\mu \Lambda)^{2})(t-1)}{tu+\Lambda^{2}}.
    \eea
  Solving the factorization condition leads to two solutions: 
  the one with $a=0$ and $u=-\Lambda^{2}$ corresponding to the $r=0$ vacuum
    \bea
    w^{2}
     =     \frac{(\mu \Lambda)^{2}}{t}, ~~~
    F(t)
     =     \mu.
    \eea
  the one with $a = -(\mu \Lambda)^{2}$ and $u=0$ corresponding to the $r=1$ vacuum
    \bea
    w^{2}
     =   - \frac{(\mu \Lambda)^{2}}{t} - (\mu \Lambda)^{2}, ~~~
    F(t)
     =     \mu (t+1).
    \eea
  By shifting $w$-coordinate, we can see that these two curves agree 
  with \eqref{SU(2)Nf2N=1curvethird1} and \eqref{SU(2)Nf2N=1curvethird2} respectively.
  See figure \ref{fig:SU(2)Nf23}.
  The singularity of $V_{2}$ at $t=0$ is a simple pole and coincides with that of $\SU(2)$ SYM case,
  since the rotated puncture is of degree $3$.
  
\subsubsection{$\SU(2)$ $N_{f}=3$}
\begin{figure}
  \centering
  \includegraphics[width=10cm]{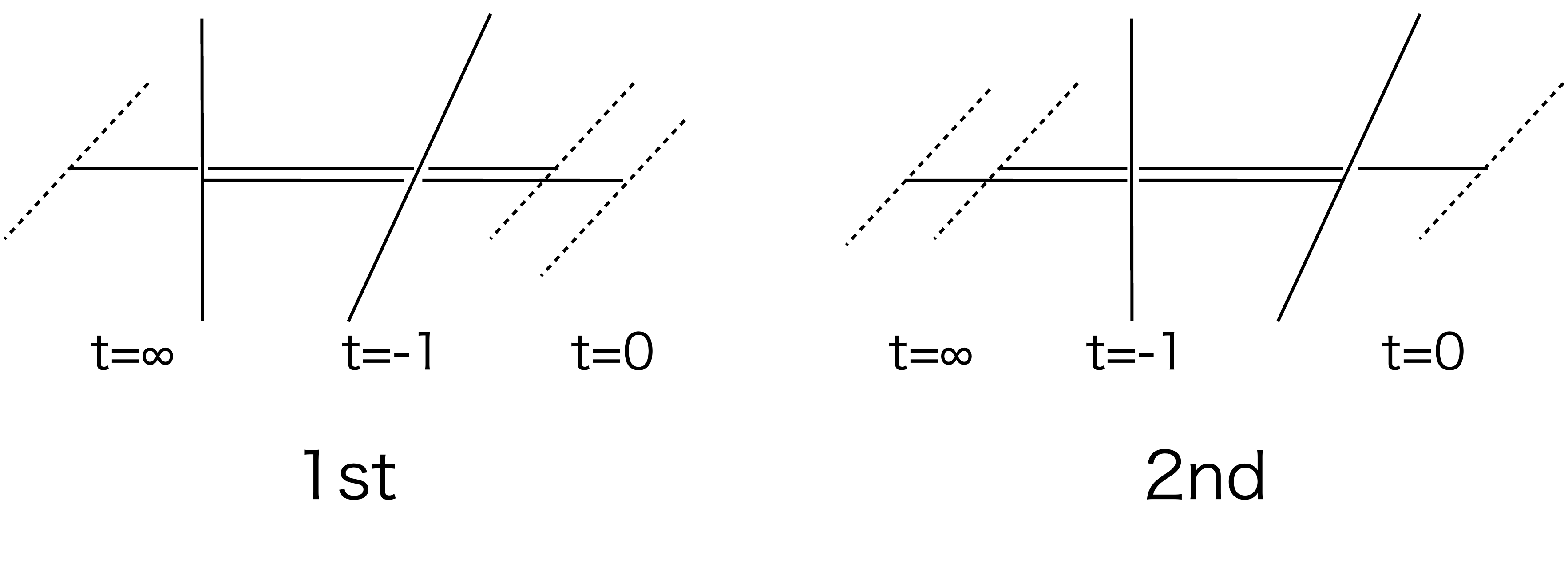}
  \caption{Left: the first realization of $\SU(2)$ gauge theory with $N_{f}=3$.
           The $r=1$ non-baryonic branch root is the same as the baryonic branch root.
           Right: the second realization.}
  \label{fig:SU(2)Nf3}
  \end{figure}

  Let us then turn to the $N_{f} = 3$ case.
  There are two realizations of this theory in Type IIA:
  the first is to put two D6-branes to the right of the NS$_{2}$
  and one D6-brane to the left of the NS$_{1}$,
  the second is the opposite to the first case, as in figure \ref{fig:SU(2)Nf3}.
  We will focus only on the second option below.
  
  Note that there is a non-baryonic branch with $r=1$ in the corresponding $\CN=2$ theory.
  In the massless theory the root of the non-baryonic branch is the same as that of the baryonic branch.
  
  The SW curve is given by
    \bea
    (v+m_{1})(v+m_{2}) t^{2} - P_{2}(v) t + \Lambda (v+m_{3})
     =     0.
           \label{N=2curveSU(2)Nf32}
    \eea
  We will for simplicity consider the case $m_1=-m_2=m$ and $m_3=0$, where the curve can be rewritten as 
    \bea
    v^2
     =     \frac{4m^2t^3-4ut^2+\Lambda^2t-\Lambda^2}{4t^2(t-1)}.
    \eea 
  In this parametrization the mass terms lead to a double pole at infinity. 
  There is also the irregular singularity of degree $4$ at $t=0$.
  
  Rotating the puncture at $t=0$, 
  the $\mathcal{N}=1$ curve is of the form: $w^2=\frac{at^2+bt+c}{4t^2}$. 
  We thus find the factorization condition
    $$F^2(t)=\frac{(at^2+bt+c)(t-1)}{4m^2t^3-4ut^2+\Lambda^2t-\Lambda^2}.$$ 
  One solution corresponds to the fact that the denominator has the form $4m^2(t-\alpha)^2(t-\beta)$.
  The constants $\alpha$ and $\beta$ are determined by 
  $4m^{2}\alpha^{3} - \Lambda^{2} \alpha + 2 \Lambda^{2} = 0$ and $\beta = \Lambda^{2}/4m^{2} \alpha^{2}$.
  In terms of $\alpha$, we find $u=(\Lambda^{2} + 8m^{2}\alpha^{3})/4\alpha^{2}$.
  We thus get three vacua.
  The above factorization of the denominator in turn implies that $at^2+bt+c=a(t-\beta)(t-1)$.
  By considering the asymptotics at zero as before, we can fix $a=4m^{2} \mu^{2} \alpha$.
  Note that in the massless case $\alpha=2$, $u=\Lambda^{2}/16$, $a=0$, $c=-b = \mu^{2} \Lambda^{2}$.
  That is the $\CN=1$ curve is
    \bea
    w^{2}
     =     \frac{\mu^{2} \Lambda^{2}}{4t^{2}} - \frac{\mu^{2} \Lambda^{2}}{4t}.
           \label{N=1curveSU(2)Nf32-2}
    \eea
  The $V_{2}$ has a double pole at $t=0$ representing the behavior of the rotated puncture of degree $4$.
  This is the same as that in subsection \ref{subsubsec:SU(2)Nf1fac}.
  Also the curve agrees with \eqref{N=1curveSU(2)Nf32-2Hori}.
  Therefore this denotes the $r=0$ vacuum. 
  
  The other solution can be obtained by requiring the denominator to be a multiple of $t-1$, 
  which imposes the constraint $u=m^2$. 
  $F^2(t)$ then reduces to $\frac{at^2+bt+c}{4m^2t^2+\Lambda^2}$. 
  This can be the square of a rational function only if the numerator is a multiple of the denominator. 
  Thus we obtain $a=4m^2\mu^2$, $b=0$ and $c=\mu^2\Lambda^2$, namely the curve is
    \bea
    w^{2}
     =     \frac{\mu^2\Lambda^2}{4t^{2}} + m^2\mu^2.
    \eea
  Notice that in the massless case the factorization condition and asymptotics 
  at the NS5-branes do not allow to fix all the parameters in the $\mathcal{N}=1$ curve. 
  Once the mass parameters are turned on, this is no longer the case.
  This may corresponds to the $r=1$ vacuum as depicted in figure \ref{fig:SU(2)Nf3}.

  As a consistency check we can rewrite the curve \eqref{hyper} as $y^2=(v^2+\Lambda v-u)^2-4\Lambda(v^3-m^2v)$ 
  and check that the discriminant has degree five in $u$, a double zero at $u=m^2$ 
  and other three simple zeros located precisely at the points described above. 
  When $m$ is sent to infinity the vacuum at $u=m^2$ disappears 
  and the other three are at $u_i\sim \omega_3^i(m^2\Lambda)^{2/3}$, 
  where $\omega_3$ is the third root of unity. This matches precisely the behavior of the theory with $N_{f}=1$. 
  Note that $m^2\Lambda$ is precisely proportional to the cube of the dynamical scale of the theory with $N_{f}=1$
  by the one-loop matching.

\subsubsection{$\SU(2)$ $N_f=4$}
\label{subsubsec:SU(2)Nf4}
  The SW curve of this theory is  
    \bea
    (v+m_{1})(v+m_{2})t^{2} - (1+q) P_{2}(v) t + q(v+m_{3})(v+m_{4})
     =     0,
           \label{N=2curveSU(2)Nf4}
    \eea
  where we have chosen the coefficients in such a way that the punctures are at $t=0, q, 1$ and $\infty$.
  All the punctures are regular.
  For simplicity we analyze the case $m_1=-m_2$ and $m_3=-m_4$. 
  We can write the SW curve as 
    \bea
    v^2
     =     \frac{m_2^2t^2-(1+q)tu+qm_4^2}{(t-1)(t-q)}.
           \label{N=2curveSU(2)Nf42}
    \eea
  With this choice of mass parameters the quadratic differential $\theta_{2} = f_{2}(dt/t)^{2}$ has 
  simple poles at 1 and $q$ and double poles at zero and infinity
  whose coefficients are $m_{4}^{2}$ and $m_{2}^{2}$. 
  We can rotate either at $q$ or at $1$. 
  Let us consider the first option, so the $\mathcal{N}=1$ curve has the form $w^2=\frac{at+b}{t-q}$. 
  The factorization condition is
    $$F^2(t)  =  \frac{(at+b)(t-1)}{m_2^2t^2-(1+q)tu+qm_4^2}.$$ 
  Notice that $q$ and $m_i$'s are parameters and should not be tuned. 
  The only parameter we can constrain is the coordinate on the Coulomb branch.
  
  Requiring the denominator to be a square we find the two solutions $u= \pm 2\sqrt{q}m_{2}m_{4}/(1+q)$. 
  We are then forced to set $b=-a$.
  Imposing then $w\sim\mu v$ at $t=q$ we find 
    \bea
    a
     =     \frac{\mu^2}{(q-1)^2}(m_2 \pm \sqrt{q} m_4)^2.
    \eea 
  The $\pm$ depends on which solution we choose for $u$. 
  Notice that $a$ diverges in the limit $q\rightarrow1$. 
  
  We can also require that the denominator is a multiple of $t-1$. 
  This implies then $$u=\frac{m_2^2+qm_4^2}{q+1};\quad a=m_2^2\mu^2;\quad b=-q\mu^2m_4^2.$$
  This corresponds to detaching from the brane system the brane located at $t=1$.
  This is signaled by the vanishing of the residue at $t=1$ of the quadratic differential of the SW curve.
  
  There is another solution: $a=b=0$. 
  As in the previous solution, we now see that this corresponds to detaching the NS5-brane at $t=q$.
  Since in this case the rotated brane is detached from the rest, the coordinate $w$ will be trivial except at $t=q$. 
  We thus expect the $\mathcal{N}=1$ curve to develop two branches and become of the form $(t-q)w^2=0$.
  Indeed, this is precisely what we get if we multiply the curve written before by $t-q$ on both sides 
  and then set $a=b=0$. 
  However, this solution will be consistent only if the residue of the pole at $t=q$ 
  in the quadratic differential vanishes, which is true only for $$u=\frac{qm_2^2+m_4^2}{q+1}.$$  
  This solution must be included whenever we can eliminate the corresponding pole in the differential 
  with a suitable choice of point in the Coulomb branch. 
  It is thus the structure of the $\mathcal{N}=2$ curve which dictates when this solution is allowed.
  This phenomenon will occur in rank one theories with regular punctures and 
  we have already encountered it in the study of the $N_f=2$ theory.
  
  Rewriting the SW curve in the form 
    $$y^2=(v^2-u)^2-\frac{4q}{(q+1)^2}(v^2-m_2^2)(v^2-m_4^2),$$ 
  we can test as before our result checking that the above solutions are the only points in the Coulomb branch 
  where the discriminant of the rhs vanishes. 
  It turns out that it is a polynomial of degree six in $u$, whose roots are precisely the four values found above. 
  The last two solutions are actually double roots. 
  With a more general choice of mass parameters they split. 
  For large $m_2$ and $m_4$ only the first two solutions remain finite and correspond to the vacua of $\SU(2)$ SYM 
  (once we take into account the one-loop matching condition). 
  The other four run away to infinity as all the matter fields become infinitely massive. 
  In the massless limit all the vacua merge at $u=0$, 
  in agreement with the expectation that the theory flows to an IR fixed point in this limit \cite{Leigh:1995ep}. 

\paragraph{The S-dual frame}
\label{frame}
  As we have just seen, the $\mathcal{N}=1$ curve diverges when $q\rightarrow1$. 
  To approach this limit the most natural procedure is to change S-duality frame 
  and consider a description of the theory in which the gauge group becomes weakly coupled. 
  More precisely what we want to do is to analyze the mass deformation of the $\mathcal{N}=2$ dual theory. 
  How can we do that in the present framework? 
  Indeed the first step is to rewrite properly the SW curve. We can e.g. consider the reparametrization of the sphere 
  that interchanges the punctures at $t=1,\infty$ leaving $t=0$ fixed by $t \rightarrow t/(t-1)$. 
  This is accompanied by the redefinition $q\rightarrow q/(q-1)$. 
  The resulting curve is 
    \bea
    \theta_{2}
     =     \frac{m_2^2(1-q)t+(t-1)[qm_4^2-(1+q)tU]}{t^2(t-1)^2(t-q)}(dt)^2.
    \eea
  Note that the $q$ here differs from the original $q$.
  Notice also that in bringing the original curve \eqref{N=2curveSU(2)Nf42} to this form 
  we have redefined the Coulomb branch coordinate as 
    \bea
    (1+q)U
     =     m_4^2q+u(1-2q)+m_2^2(q-1).
    \eea
  The mass parameters are the same but now $m_2$ is the residue at $t=1$. 
  The limit $q\rightarrow1$ now corresponds to a weak coupling limit. 
  
  The important point is that this reparametrization should be accompanied by a change in the vector field: 
  instead of $t\partial_t$ we have to use $\zeta=t(t-1)\partial_t$, which fixes the points $t=0,1$. 
  The SW curve written in terms of $t$ and $y=v(t-1)=\lambda(\zeta)$ reads 
    $$y^2=\frac{m_2^2(1-q)t+(t-1)[qm_4^2-(1+q)tU]}{t-q}.$$
  
  Rotating at $q$ as before, we find the $\mathcal{N}=1$ curve $w^2=\frac{a}{t-q}+b$. 
  The factorization condition now reads $$F^2(t)=\frac{a+b(t-q)}{m_2^2(1-q)t+(t-1)[qm_4^2-(1+q)tU]}.$$ 
  One can now determine the $\mathcal{N}=1$ points as before. 
  The analysis is very similar to the previous case and we will not repeat it. 
  Indeed, the set of points in the Coulomb branch which are not lifted coincides with that found before 
  once we take into account all the redefinitions, as it should be. 
  The relevant case is the analogue of the first solution studied previously, 
  when the denominator of $F^2(t)$ becomes a square. 
  This implies $b=0$ and 
    $$U=\frac{(\sqrt{q}m_4\pm\sqrt{q-1}m_2)^2}{q+1}.$$ 
  Imposing the constraint $w\sim\mu v$, we get the $\mathcal{N}=1$ curve 
    \bea
    w^2
     =     \mu^2\frac{((q-1)\sqrt{q}m_4\pm q\sqrt{q-1}m_2)^2}{t-q}.
    \eea
  Now the divergence at $q=1$ has disappeared! 
  The divergent limit is instead $q\rightarrow\infty$, the weak coupling limit of the original description. 
  
  Using the recipe discussed before we can now easily extract the DV curve 
  (write $t$ in terms of $v,w$ using $F(t)$ and plug it in the $\mathcal{N}=1$ curve). 
  We find as expected $w^2-\mu wv+\mu S=0$, where $S$ is the gluino condensate in the present duality frame 
    $$S=\mu(\sqrt{q}m_4\pm\sqrt{q-1}m_2)((q-1)\sqrt{q}m_4\pm q\sqrt{q-1}m_2).$$ 
  Notice that it vanishes for $q=1$, confirming that the gauge group becomes weakly coupled in this limit. 
  
  From this example it should be clear how to approach the $\mathcal{N}=1$ deformation of various duality frames 
  in more general cases as well (at least on the sphere): 
  the recipe is simply to rewrite the $\mathcal{N}=2$ curve in the proper way and change 
  at the same time the location of the zeroes for the vector field. 
  The systematic analysis of this problem will be done elsewhere.

\subsection{$\SU(2)^{2}$ quiver}
\begin{figure}
  \centering
  \includegraphics[width=10cm]{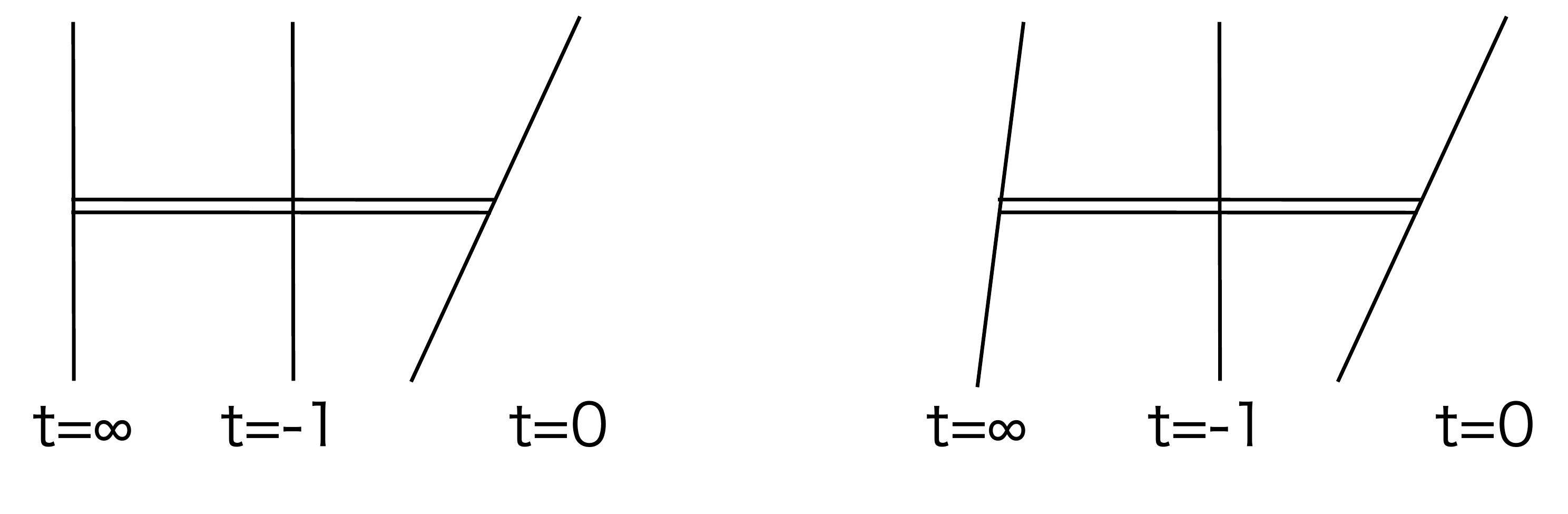}
  \caption{Left: the rotation of right NS5-brane (NS$_{3}$-brane).
           Right: the rotation of the left and right NS5-branes (NS$_{1}$ and NS$_{3}$-branes).}
  \label{fig:SU2x2fac}
  \end{figure}

\label{subsec:SU(2)^{2}}
  As a next example, let us consider $\CN=2$ $\SU(2) \times \SU(2)$ gauge theory coupled 
  to a bifundamental hypermultiplet.
  Here we focus on the massless hypermultiplet.
  The Seiberg-Witten curve is
    \bea
    \Lambda_{1}^{2} t^{3} + P_{2,1}(v) t^{2} + P_{2,2}(v) t + \Lambda_{2}^{2}
     =     0,
           \label{N=2curveSU(2)^{2}}
    \eea
  where $P_{2,i} = v^{2} - u_{i}$ and $u_{i}$ are the Coulomb moduli parameters of two $\SU(2)$ gauge groups.
  The curve can be written as
    \bea
    v^{2}
     =   - \frac{\Lambda_{1}^{2} t^{3} + u_{1}t^{2} + u_{2}t + \Lambda_{2}^{2}}{t(t+1)}.
    \eea
  There are two irregular punctures of degree $3$ at $t=0$ and $t= \infty$,
  and one regular puncture at $t=-1$.
  This theory can be realized by three NS5-branes stretched by two-D4-branes, as in figure \ref{fig:SU2x2fac}.
  We refer to NS5-branes at $t = \infty, -1, 0$ as NS$_{1}$-, NS$_{2}$-, NS$_{3}$-branes respectively.
  
  Let us first consider the case where only the puncture corresponding to NS$_{3}$-brane is rotated, 
  as in figure \ref{fig:SU2x2fac}. 
  This induces the mass term of the adjoint field of one $\SU(2)$ gauge group
  (coming from the right two D4-branes).
  The boundary condition fixes
  $w^{2} = - \frac{(\mu \Lambda_{2})^{2}}{t} + a$.
  Thus the factorization condition is
    \bea
    F(t)^{2}
     =     \frac{(t+1)(-at + (\mu \Lambda_{2})^{2})}{\Lambda_{1}^{2} t^{3} + u_{1}t^{2} + u_{2}t + \Lambda_{2}^{2}}.
    \eea
  The above expression will be a square only if $a=0$ and the denominator is factorized as
  $(t+1)(\Lambda_{1}t \pm \Lambda_{2})^{2}$.
  This means $u_{1} = \Lambda_{1}^{2} \pm 2 \Lambda_{1}\Lambda_{2}$ 
  and $u_{2} = \Lambda_{2}^{2} \pm 2 \Lambda_{1}\Lambda_{2}$.
  Therefore
    \bea
    w^{2}
     =   - \frac{(\mu \Lambda_{2})^{2}}{t}, ~~~~
    F(t)
     =     \frac{\mu \Lambda_{2}}{\pm \Lambda_{1}t + \Lambda_{2}}.
    \eea
  Again $V_{2}$ has a simple pole at $t=0$ representing the rotation of the irregular puncture of degree $3$.
  
  We can easily get the DV curve from these.
  Due to the factorization condition the Seiberg-Witten curve can be written as
  $v^{2} = - (\Lambda_{1}t \pm \Lambda_{2})^{2}/t$.
  By substituting the $\CN=1$ curve, we get
    \bea
    v^{2}
     =     \left(  \frac{-\Lambda_{1}\Lambda_{2}^{2} \mu^{2}/w^{2} \pm \Lambda_{2}}{\mu \Lambda_{2}/w} \right)^{2}.
    \eea
  This implies
    \bea
    w^{2} (w - \mu v) \mp \mu^{2} \Lambda_{1} \Lambda_{2} w
     =     0.
    \eea
  This is the form of the DV curve which we will see in section \ref{sec:Konishi}
  with $f_{0} = \mp \mu^{2} \Lambda_{1} \Lambda_{2}$ and $f_{1} = 0$.
  
  Next, we turn on the superpotential for both groups by rotating the NS$_{1}$ and NS$_{3}$-branes.
  We choose the boundary conditions $w \rightarrow \mu_{1} v$ at $t=\infty$
  and $w \rightarrow \mu_{2} v$ at $t=0$. This corresponds to turning on the superpotentials 
  $W_1=-(\mu_1/2)\Tr\Phi_1^2$ and $W_2=(\mu_2/2)\Tr\Phi_2^2$.
  Thus, the ansatz of the $\CN=1$ curve is 
  $w^{2} = - t (\mu_{1} \Lambda_{1})^{2} - \frac{(\mu_{2} \Lambda_{2})^{2}}{t} - a$.
  Therefore the factorization condition is 
    \bea
    F^{2}
     =     \frac{(t+1)( (\mu_{1} \Lambda_{1})^{2} t^{2} + a t + (\mu_{2} \Lambda_{2})^{2})}
           {\Lambda_{1}^{2} t^{3} + u_{1}t^{2} + u_{2}t + \Lambda_{2}^{2}}.
    \eea
    
  One solution is that the numerator becomes $(t+1)^{2} ((\mu_{1} \Lambda_{1})^{2} t + (\mu_{2} \Lambda_{2})^{2})$,
  which in turn implies that the denominator should include 
  a factor $(\mu_{1} \Lambda_{1})^{2} t + (\mu_{2} \Lambda_{2})^{2}$.
  We thus find 
    \bea
    u_1
     =     \frac{\mu_2^2}{\mu_1^2}\Lambda_2^2 + 2\frac{\mu_1}{\mu_2} \Lambda_1^2,\quad ~
    u_2 
     =     \frac{\mu_1^2}{\mu_2^2}\Lambda_1^2 + 2 \frac{\mu_2}{\mu_1} \Lambda_2^2.
    \eea 
  Notice that this vacuum runs away to infinity if we set to zero one of the two mass parameters. 
  Consequently, it exists only if we turn on both superpotentials. The $\mathcal{N}=1$ curve is then 
  \beq\label{newvac}w^2=-\frac{(t+1)(\mu_1^2\Lambda_1^2t+\mu_2^2\Lambda_2^2)}{t};\quad \frac{w}{v}=F(t)=
  \frac{\mu_1\mu_2(t+1)}{\mu_2t+\mu_1}.\eeq We can now easily extract the DV curve: 
  using the second equation we can write $t$ in terms of $v$ and $w$ 
  and plugging this back into the $\mathcal{N}=1$ curve we find a cubic equation for $w$. 
  After the redefinition $y=w-(\mu_1+\mu_2)v/3$, it becomes precisely the curve we will find in section \ref{sec:Konishi}
  with $g_1(v)=g_2(v)=0$ and 
    $$f_1(v)=-\mu_1S_1=(\mu_2-\mu_1)\frac{\Lambda_1^2\mu_1^2}{\mu_2};\quad 
    f_2(v)=\mu_2S_2=(\mu_1-\mu_2)\frac{\Lambda_2^2\mu_2^2}{\mu_1},$$
  where $S_1$ and $S_2$ are the two gluino condensates. 
  
  When $\mu_{1} = \mu_{2} =: \mu$, 
  namely the NS$_{1}$ and NS$_{3}$-branes are aligned, the $\CN=1$ curve becomes
    \bea
    w^{2}
     =   - t (\mu \Lambda_{1})^{2} - \frac{(\mu \Lambda_{2})^{2}}{t}
         - \mu^{2} (\Lambda_{1}^{2} + \Lambda_{2}^{2}), ~~~~
    F(t)
     =     \mu.
           \label{SU(2)^2sol1}
    \eea
  This solution corresponds to the locus on the Coulomb branch where the SW curve degenerates to
    \bea
    v^{2}
     =     \frac{(t+1)(\Lambda_{1}^2 t + \Lambda_{2}^2)}{t},
    \eea
  but is still of genus-one. With this choice of mass parameters the NS$_2$ brane is detached.
  
  We have another solution imposing that the numerator factorizes as $(t+1)(\mu_{1} \Lambda_{1} t \pm \mu_{2} \Lambda_{2})^{2}$.
  Consequently the denominator should be $(t+1)(\Lambda_{1} t \pm \Lambda_{2})^{2}$,
  which means $u_{1} = \Lambda_{1}^{2} \pm 2 \Lambda_{1} \Lambda_{2}$
  and $u_{2} = \Lambda_{2}^{2} \pm 2 \Lambda_{1} \Lambda_{2}$.
  The $\CN=1$ curve is in this case
    \bea
    w^{2}
     =   - t (\mu_{1} \Lambda_{1})^{2} - \frac{(\mu_{2} \Lambda_{2})^{2}}{t}
         \mp 2\mu_{1} \mu_{2} \Lambda_{1} \Lambda_{2}, ~~~~
    F(t)
     =     \frac{\mu_{1} \Lambda_{1} t \pm \mu_{2} \Lambda_{2}}{\Lambda_{1} t \pm \Lambda_{2}}.
           \label{SU(2)^2sol2}
    \eea
  This solution reduces to the one found previously when we just rotated one puncture if we set to zero one of the mass 
  parameters (in our case $\mu_1$). In this case the SW curve completely degenerates to a sphere,
  indicating that the theory is in the confining phase.
  
  We saw that there are two different kinds of solutions above.
  In particular when we choose the mass parameters as $\mu_{1} = \mu_{2}$, 
  the theory below this energy scale is just $\CN=1$ $\SU(2) \times \SU(2)$ gauge theory 
  with two bifundamental chiral multiplets which was studied in \cite{Intriligator:1994sm}.
  Note that by integrating out the massive adjoints we get $(B\tilde{B})^{2}$ terms where $B$ is the bifundamental field.
  But these terms are cancelled each other by the above choice of the masses.
  There is a Coulomb branch parametrized by the moduli consisting of the bifundamental fields.
  The curve on the Coulomb branch was found in \cite{Intriligator:1994sm} to be
    \bea
    w^{2}
     =   - t \Lambda_{1}'^{2} - \frac{\Lambda_{2}'^{2}}{t} - v_{2},
    \eea
  where $v_{2}$ is the moduli parameter and $\Lambda_{i}' := \mu_{i} \Lambda_{i}$ ($i=1,2$) 
  are the dynamical scales of the theory after integrating out the adjoint chirals by one-loop matching
  (see \cite{Tachikawa:2011ea} for convention).
  This clarifies what is the physical meaning of our $\mathcal{N}=1$ curve: the two solutions just described precisely 
  coincide with the curve of Intriligator and Seiberg, for special choices of the moduli parameter as expected.
  We see that the second solution \eqref{SU(2)^2sol2} is a special point on the Coulomb branch
  where the genus-one $\CN=1$ curve degenerates.

\subsection{$\SU(2)$ theories in class $\CS$}
\label{subsec:}
  Let us now give another general argument to find curves of $\CN=1$ mass deformation of $\CN=2$ class $\CS$ theory, 
  making use of the second perspective given in section \ref{subsubsec:2nd}.
  This will be applied to analyze $\CN=1^{*}$ theory.
  
  The Hitchin field $\Phi$ is a section of $K_{\cal C}\otimes {\rm End}{\cal E}$ 
  with ${\cal E}=K_{\cal C}^{-1/2}\oplus K_{\cal C}^{1/2}$.
  This is specified by a meromorphic quadratic differential $\theta_2$ on ${\cal C}$, 
  with singularities at the punctures and with a suitable gauge choice reduces to
    $$\Phi=\left( \begin{array}{cc}
    0 & 1 \\ 
   \theta_2 & 0
    \end{array}
   \right) \ \ .
    $$
  In the ${\cal N}=1$ set-up, the equations \eqref{curvas} will take a quite simple universal form.
  Indeed, both $\phi_i$ are aligned in the Lie algebra. 
  This implies that one can specify the curve 
    \bea
    x_1^2
     =     R, ~~~~
    x_1 x_2
     =     RP
    \eea
  where $R=\frac{1}{2}{\rm Tr}\phi_1^2$ is a section of $K_{\cal C}^2{\cal L}^{-2}$ 
  and $RP=\frac{1}{2}{\rm Tr}\phi_1\phi_2$ is a section of $K_{\cal C}$.
  Let us choose then a gauge where
    $$\phi_1=\left( \begin{array}{cc}
    0 & \sigma^{-1} \\ 
    \sigma R & 0
    \end{array}
    \right)$$
  is a section of $K_{\cal C}{\cal L}^{-1}$ in the adjoint of ${\cal E}$.
  The relation between $\phi_1$ and the Hitchin field is then simply $\Phi=\sigma \phi_1$ which sets $\theta_2=\sigma^2R$.
  
  The ${\cal N}=2$ spectral curve has therefore the form 
    $$x^2=\sigma^2 R,$$
  which means that some of the zeros of the quadratic differential $\theta_2$ get constrained to pair in couples
  to reproduce the square of the section $\sigma$.
  This implies the lowering of the genus of the SW curve and generates the ${\cal N}=1$ vacua
  associated to the corresponding massless dyonic sectors.
  
  The complete canonical transformation to get the rotated system would read\footnote{A more general one would be 
    $x=\sigma x_1$ and $w= \tau x_2 + \rho(x_1,t)$, but the last is a global shift on $\varphi$ which can we reabsorbed.}
    $$x=\sigma x_1\quad {\rm and }\quad w= \tau x_2,$$
  where $\sigma\tau=\partial_tz$, $z$ being an appropriate coordinate on ${\cal C}$. 
  The complete ${\cal N}=1$ curve is then given by
    $$x^2=\sigma^2 R \quad {\rm and}\quad xw=PR\partial_tz.$$
  The usual form of the latter as a generalized Konishi anomaly equation is obtained then 
  by introducing the scalar variable $v=i_\zeta \lambda$ -- 
  where $\zeta$ is a suitable vector field setting the S-duality frame as discussed, for a particular example, 
  in section \ref{subsubsec:SU(2)Nf4} --
  and by resolving the explicit $t$ dependence using the degenerate SW curve. 
  It is not granted that this last step can always be performed in a closed form.

\subsubsection{${\cal N}=1^*$ SU(2) gauge theory}
\label{subsubsec:N=1*}
  As an example of the above construction, let us consider the ${\cal N}=1^*$ $\SU(2)$ gauge theory.
  The base curve is therefore elliptic, namely a complex one dimensional torus with a distinguished point. 
  The canonical bundle of the torus $T^2_\tau$ is trivial and admits three square roots, namely the three spin structures.
  Let ${\cal L}_\nu$, with $\nu=1,2,3$ denote the three corresponding line bundles.
  A meromorphic section of ${\cal L}_\nu$ is ${\tt s}_\nu(z)=\sqrt{{\cal P}(z)-{\cal P}(e_\nu)}$, 
  where ${\cal P}(z)$ is the Weierstrass ${\cal P}$-function 
  and $\{e_\nu\}_{(\nu=1,2,3)}=\left(1/2,\tau/2,(1+\tau)/2\right)$ are the three Weierstrass points in $T^2_\tau$.
  The local CY space can be chosen therefore in three ways, 
  being the total space of ${\cal L}_\nu\oplus{\cal L}_\nu$ on $T^2_\tau$.
  
  The quadratic differential of the ${\cal N}=2^*$ $\SU(2)$ gauge theory is given by
  $\theta_2(z)=m^2{\cal P}(z)+u$, where $m$ is the mass of the adjoint hypermultiplet and $u$ is the Coulomb modulus.
  Due to the triviality of the canonical bundle and imposing that the rotated geometry is not more singular 
  than the unrotated one, the ${\cal N}=1$ curve is given by
    $$
    x_1^2=m^2\,\, , \quad x_1x_2=\mu m,
    $$
  where $\mu$ is a further mass scale related to the ${\cal N}=1$ breaking 
  and we fixed $R=m^2$ and $RP=\mu m$ to be constant sections of the trivial bundle on the elliptic curve.

  Therefore, the equation for the quadratic differential $\theta_2$ reads
    \bea
    \theta_2
     =     m^2{\tt s}_\nu(z)^2,
    \eea
  which fixes the Coulomb modulus to $u=u_{\nu}=-m^2{\cal P}(e_\nu)$.
  These are the well known three ${\cal N}=1^*$ vacua of the $\SU(2)$ gauge theory.

\section{$\CN=1$ curve in $A_{N-1}$ theory}
\label{sec:AN}
  In this section we consider four-dimensional $\CN=1$ gauge theories obtained from the $A_{N-1}$ theory.
  In this case the factorization condition \eqref{facct} cannot be simplified enough,
  thus we do not have a general approach to fix the moduli of the curve.
  We will see some examples where it is possible to find $\CN=1$ curves.

\subsection{$\SU(N)$ SYM theory}
\label{subsec:SU(N)pure}
  Let us first consider $\SU(N)$ pure SYM theory.
  The $\CN=1$ breaking by brane rotation can be set for a generic choice of the superpotential, 
  while we will discuss the breaking via complex structure rotation on a case by case basis.
  The SW curve is given by an $N$-sheeted cover of the twice punctured sphere 
  ${\cal C}={\mathbb P}^1\setminus\{0,\infty\}$. 
  Its equation is 
    \be
    \Lambda^N\left(t+\frac{1}{t}\right)
     =     P_N(xt),
    \label{n2}
    \ee
  where $t$ is a stereographic coordinate on the sphere, $v=xt$, $\lambda=xdt$ and $P_N(v)=v^N+\sum_{i=2}^N v^{N-i}u_i$.
  $u_{i}$ are the Coulomb moduli parameters.

  Let us fix the boundary condition for $\varphi$ to be 
    \bea
    \varphi\sim 0 \quad 
    &{\rm as}& \quad t\sim 0,\nonumber \\ 
    \varphi\sim W'(V) \quad 
    &{\rm as}& \quad t\sim \infty,
    \label{bcpureN}
    \eea
  where $\zeta=t\partial_t$ is the non-singular vector field fixing both the punctures and $V=i_\zeta\Phi$.
  Notice that $\zeta$ is non singular and therefore we can fix the boundary conditions ambiguity we discussed above 
  by imposing vanishing boundary condition at the puncture located at $t=0$.
  
  We can now discuss how the curve (\ref{gs}) is affected by the above boundary conditions.
  From the algebraic system we can eliminate the $t$ variable and obtain an algebraic equation 
  involving the variables $v=tx$ and $w$ only. 
  This can be partially fixed by the boundary conditions as follows:
  Consider the function $w(w-W'(v))/W'(v)$ which is vanishing at the two punctures and independent of $t$. 
  This implies that
    \be
    w(w-W'(v))
     =     f(v),
           \label{n1}
    \ee
  where $f(v)$ is a polynomial of degree less than the degree of $W'$ 
  (independent of $t$ by definition and of $w$ since it would then be too divergent).
  This fixes the curve as the projection of the system (\ref{gs}) in the two fibers directions.
  To proceed further we have to specify the superpotential $W$.

\subsubsection{The massive deformation}
  Let us consider in this section the pure massive deformation $W(v)=\mu v^2/2$. 
  In this case the equation (\ref{n1}) reads
    \be
    w^2-\mu w v-f_0
     =     0
           \label{nnn1}
    \ee
  which implies that
    \bea
    V
     =     \frac{1}{\mu}\left(\varphi-f_0\varphi^{-1}\right),
           \label{Vpure}
    \eea
  where $f_0$ is a constant to be determined.
  The last expression has to satisfy equation (\ref{n2}) and this fixes uniquely $P_N$ 
  to be the Chebyshev polynomial because of the boundary conditions for $\varphi$.
  By introducing the parameter $s$ such that $s = w/\mu \Lambda$ and $s+s^{-1} = xt/\Lambda$,
  this means the SW curve is now
    \bea
    s^N+s^{-N}
     =     t+t^{-1},
    \eea
  The lhs is equal to $2T_{N}(\frac{s+s^{-1}}{2})$, where $T_{N}$ is the Chebyshev polynomial.
  This fixes the corresponding value of the condensate to $f_0=-\Lambda^2\mu^2\omega_i$,
  where $i=1, \ldots, N$ and $\omega_{i}^{N}=1$.
  Thus one finds $N$ vacua.
  The $\CN=1$ curve is thus written as
    \bea
    w^{N}
     =     \mu^{N} \Lambda^{N} t.
    \eea
  When $N=2$ this reproduces the curve \eqref{N=1curvepure} (with $t \rightarrow 1/t$) in the previous section.
  
  We conclude that the eigenvalues of the Hitchin field are
  $x_i=\frac{\Lambda}{t}\left(t^{1/N}\omega_i+t^{-1/N}\omega_i^{-1}\right)$.
  The SW curve is given by
    \bea
    0
     =     \det \left(v\1-V_i\right),
    \eea
  where $V_{i}$ is \eqref{Vpure} with $f_{0}$ being the $i$-th vacuum value.
  Correspondingly one has 
    \be
    \Phi=\frac{\Lambda}{t}\left[\Omega+\omega_i \Omega^{-1}\right], ~~~
    \varphi=\mu\Lambda\Omega
    \label{bfbf}
    \ee
  with 
    $$\Omega=\left(\begin{matrix}
    0 &  1  & \ldots & \ldots & 0\\
    0 & 0  &  1  & \ldots & 0 \\
    \vdots & \vdots & \ddots & \ddots & \vdots\\
    0 &  0  & \ldots & 0 & 1\\
    t  &   0  &\ldots     &\ldots & 0
    \end{matrix}\right)
    $$
  satisfying $\Omega^N=t {\bf 1}_N$.
  
  Let us split now the Hitchin field in the two components, 
  namely the one which corresponds to the rotated brane and the one corresponding to the fixed one.
  This means that we write
    $$\Phi
     = \sigma_1\phi_1  +\sigma_2\phi_2 \,\,~~ {\rm and}\quad 
    \varphi=\tau\phi_2,
    $$
  where in this case ${\cal L}={\cal K}^{1/2}_{P^1}={\cal O}(-1)$ and we choose $\sigma_2=\Lambda/t$, 
  $\sigma_1=1/\Lambda\mu t$ and $\tau=\mu\Lambda$ in the patch around $t=0$.
  Therefore, by \eqref{bfbf} we obtain
    \bea
    \phi_1
     =     \omega_i \Lambda^{2}\mu\Omega^{-1}, ~~~
    \phi_2
     =     \Omega,
    \eea
  and $\varphi = \Lambda\mu\phi_2$.
  The explicit canonical transformation is 
    \be
    x
     =     \frac{1}{\Lambda\mu t}x_1+\frac{\Lambda}{t}x_2,~~~\quad 
    w
     =     \Lambda\mu x_2,
           \label{hhh}
    \ee
  and the $z$ parameter satisfies $\partial_tz=\sigma_1\tau=\frac{1}{t}$, that is $t=e^z$.

  This implies that the ${\cal N}=1$ curve of the rotated system is described by the equations
    \be
    x_2^N
     =     e^z,\qquad 
    x_1^N
     =     (\mu\Lambda^{2})^{N}e^{-z}\quad {\rm and}~~\,\,
    x_1 x_2
     =     \mu\Lambda^2\omega_i
           \label{nn1}
    \ee
  which follows from the fact that $\phi_1\phi_2=\omega_i \Lambda^{2}\mu{\bf 1}_N$.
  This is precisely equivalent to the curve given in \cite{Hori:1997ab}. 
  If we now decouple the multiplet in the adjoint sending $\mu$ to infinity, 
  we flow to $\mathcal{N}=1$ $\SU(N)$ SYM theory provided we send to zero $\Lambda$ as required by the one-loop matching 
  keeping fixed $\mu\Lambda^2=\Lambda_{\mathcal{N}=1}^{3}$. 
  In this limit the above curve becomes the one proposed in \cite{Witten:1997ep} 
  if we identify $\Lambda_{\mathcal{N}=1}^{3}$ with the $\zeta$ parameter appearing in that paper.
  
  The above system of equations \eqref{nn1} is indeed equivalent to \eqref{n2} and \eqref{nnn1}: 
  using
    $$v=x_2\Lambda+\frac{x_1}{\Lambda\mu},$$
  the projection of the curve on the $(v,t)$ plane gives \eqref{n2}.
  In particular the dependence on $\mu$ has to disappear. 
  With our identification
    $$t+\frac{1}{t}=x_{2}^{N}+\frac{x_{1}^{N}}{\mu^N\Lambda^{2N}}=x_{2}^{N}+\frac{1}{x_{2}^{N}}.$$ 
  Using the Chebyshev polynomial it is easy to see that the rhs is precisely 
  $2T_N(v/2\Lambda)$, with $v$ as above. 
  We thus get the SW curve at the points where the curve can be rotated.
  Eq.~\eqref{nnn1} can be deduced as follows
    $$vw=\left(x_2\Lambda+\frac{x_1}{\Lambda\mu}\right)\Lambda\mu x_2=\frac{w^2+\mu^2\Lambda^2\omega_i}{\mu}.$$
  
  When $v$ tends to infinity either $x_1$ or $x_2$ will go to infinity 
  (while the other goes to 0 as long as $\mu$ is different from zero). 
  In the first case we get $x_1/\mu\Lambda\rightarrow v$, so it describes the unrotated brane 
  (and indeed $x_2=w/\mu\Lambda\rightarrow0$). 
  In the second one we find $x_2\Lambda\mu=w\rightarrow\mu v$, reproducing the boundary condition (\ref{bcpureN})
  (so $x_2$ describes the rotated brane).

\paragraph{Factorization method}
  Let us consider another way to get the $\CN=1$ curve which makes use of the factorization condition given in 
  section \ref{subsec:factorization}.
  We write the SW curve in the form
    \bea
    v^N
     =     \sum_{k=2}^{N-1}v^{N-k}u_k+\Lambda^Nt+u_N+\frac{\Lambda^N}{t}.
    \eea
  In the case with the quadratic superpotential, we impose that the $\mathcal{N}=1$ curve has the same asymptotic behavior
  as the SW curve at infinity and subleading at zero as in \eqref{bcpureN}. 
  We thus get 
    \bea
    w^N-\sum_{k=2}^{N}w^{N-k}a_k
     =     bt,
    \eea
  where $a_{k}$ and $b$ are constants. 
  In this case the factorization condition can be written as 
  $v=P_{N-1}(w,t)/Q_{N-1}(w,t)$, where $P$ and $Q$ are polynomials in $w$ of degree (at most) $N-1$, 
  with coefficients polynomial in $t$. 
  On the other hand, the form of the $\mathcal{N}=1$ curve tells us that $t$ can be written as a polynomial 
  in $w$ and substituting in the factorization condition we find that $v$ must be a rational function of $w$ only: 
    \beq
    \label{facym}
    v
     =     \frac{\alpha_0w^k+\dots+\alpha_k}{w^{k-1}+\dots+\beta_{k-1}},
    \eeq
  for some $k$.
  Note that the exponents of the first terms in the denominator and the numerator are fixed by
  the boundary condition $w \rightarrow \mu v$ when $w \rightarrow \infty$.
  
  We can now proceed as follows: the SW curve tells us that $v$ diverges at $t=0$. 
  This behavior is reproduced by \eqref{facym} only if the denominator vanishes there and its roots must clearly coincide
  with the roots of the lhs of the $\mathcal{N}=1$ curve. 
  We can now evaluate $v^N$ using \eqref{facym}. 
  The resulting expression has necessarily poles of order at least $N$ in $w$. 
  Plugging instead the $\mathcal{N}=1$ curve in the SW curve we find an object with poles of order at most $N$ in $w$ 
  (depending on the multiplicity of the roots of the lhs of the $\mathcal{N}=1$ curve). 
  Since the ratio of these two quantities must tend to one at $t=0$ (just because $v$ diverges there), 
  we conclude that both must have a pole of order exactly $N$. The asymptotic behavior of $v$ at $t=0$ dictated 
  by the SW curve and by the factorization condition are thus compatible only if $k=2$ in \eqref{facym}, so 
    \bea
    v
     =     \frac{\alpha_0 w^2 + \alpha_1 w + \alpha_2}{w+\beta_{1}},
    \eea 
  and the $\mathcal{N}=1$ curve reduces to $(w+\beta_{1})^N=bt$. 
  By shifting $w$ we can then set $a=0$ and the boundary condition $w\sim\mu v$ at infinity 
  implies $\beta_{1}=\mu^N\Lambda^N$. 
  Plugging now the $\mathcal{N}=1$ curve in the SW curve we can easily see that the factorization condition reduces 
  to $\mu v=w+\omega_{i} \mu^2\Lambda^2w^{-1}$ 
  and the Coulomb branch coordinates are restricted to be at the Chebyshev points, as above.

\subsubsection{Argyres-Douglas points}
\label{subsubsec:AD1}
  Let us now consider the higher order superpotential $W(v)=g v^N/N$. 
  Eq.~(\ref{n1}) now reads
    $$w^2-g w v^{N-1}-f_{N-2}(v)=0.$$
  Let us solve the boundary conditions and the commutation relation via the ansatz $\varphi=a(t) g V^{N-1}$, 
  where $a$ is a meromorphic function in $t$ with a zero at $t=0$ and regular at $t\sim\infty$. 
  The $\CN=1$ curve then implies that
    $$(a^2-a)g^2 V^{2N-2}-f_{N-2}(V)=0,$$
  which is compatible with (\ref{n2}) only if $V^N=k(t)\1$ and $(a^2-a)k(t)=c$ is a constant.
  This fixes $f_{N-2}=cg^2v^{N-2}$.
  The Coulomb moduli are then fixed as $P_N(v)=v^N-u_N$ and $k(t)=\Lambda^N(t+1/t)-u_N$.
  The condition $(a^2-a)k=c$ has then two solutions, namely $u_N=\mp 2\Lambda^N$ and $a=\frac{t}{t\mp 1}$ 
  with $c=\pm\Lambda^N$ and therefore $f_{N-2}(v)=\pm\Lambda^N g^2 v^{N-2}$.
  This value of $u_{N}$ denotes the maximal conformal fixed point \cite{Argyres:1995jj,Eguchi:1996vu}.
  Notice that $a(t)$ is consistent with the requirements we put on it few lines ago.
  The SW curve at this point then reads
    $$\Lambda^{N}\left(t+\frac{1}{t}\right)=v^{N}\pm2\Lambda^{N}$$ 
  (for definiteness we will choose the minus sign in the following). 

  The Hitchin field is then $\Phi=(\Lambda/t)\Omega_{ad}$, where $\Omega_{ad}^N=(t+1/t+2){\bf 1}$. 
  More specifically we can choose
    $$\Phi=\frac{\Lambda}{t}\left(
    \begin{matrix}
    0 & \ldots & 0 & 1+\frac{1}{t}\\
    1 & 0  & \ldots & 0 \\
    0 & \ddots & 0 & \vdots\\
    0 & \ldots & t+1 & 0\\
    \end{matrix}\right),$$ 
  where the diagonal dots stand for 1's. 
  The SW curve as written above can be seen as the spectral curve of the field $V=i_\zeta\Phi$, 
  namely $\det(v{\bf 1}-V)=0$.
  The field $\varphi$ now reads
    $$\varphi=(t+1)g\Lambda^{N-1}\Omega_{ad}^{-1}=
    g\Lambda^{N-1}\left(\begin{matrix}
    0 & 1+\frac{1}{t} & 0 & \ldots\\
    0 & 0  & \ddots & 0 \\
    0 & \ldots & 0 & \frac{1}{t}\\
    1 & 0 & \ldots & 0\\ 
    \end{matrix}\right),$$ 
  where the diagonal dots stand for $1+1/t$'s. 
  Plugging this in (\ref{curvas}) we find the system
    \bea 
    x^N
    &=&     (\Lambda/t)^N\left(t+\frac{1}{t}+2\right),
            \nonumber \\
    w^N
    &=&     g^N\Lambda^{N(N-1)}t\left(t+1\right)^{N-2},
            \nonumber \\
    xw
    &=&     g\Lambda^N\left(1+\frac{1}{t}\right).
    \eea

  The description of the vacuum at the conformal fixed point in the rotated complex structure is provided 
  by the canonical transformation ${\cal O}(-2)\oplus{\cal O}(0) \to {\cal O}(-1)\oplus{\cal O}(-1)$ given by
    $$
    x=\sigma_1 x_1\,\,~~{\rm and}\quad w=\tau x_2,
    $$
  where $\sigma_1=\Lambda/t$ and $\tau=1/\Lambda$ in the patch around $t=0$. 
  The ${\mathbb C}^*$ parameter is fixed by $\sigma_1\tau=\partial_t z=1/t$.
  Since now $\phi_1=\Omega_{ad}$ and $\phi_2=(t+1)\left(\Lambda^N\right)\Omega_{ad}^{-1}$,
  in the new coordinates the spectral curve reads
    \bea
    x_1^N
    &=&    (t+1)^2/t,\\
    x_1 x_2
    &=&    g\Lambda^N(t+1),\\
    x_2^N
    &=&    (g\Lambda^N)^N t(t+1)^{N-2}.
    \eea
  Notice that in the rotated geometry only the combination $\hat g=g\Lambda^N$ appears.

\subsection{$\SU(2)$ gauge theory coupled to SCFTs of Argyres-Douglas type}
\label{subsubsec:Anm}
  We here consider the adjoint mass deformation of
  $\CN=2$ $\SU(2)$ gauge theory coupled to the $D_n$ and $D_m$ SCFTs \cite{Cecotti:2011rv,Bonelli:2011aa},
  by gauging the diagonal part of two $\SU(2)$ flavor symmetries of SCFTs,
  though this is not higher rank theory.
  The $D_{n}$ theory is an $\CN=2$ SCFT of Arygres-Douglas type which can be obtained as a superconformal fixed point
  on the Coulomb branch of $\CN=2$ $\SU(n-1)$ gauge theory with two flavors.
  We will use the approach in section \ref{subsec:SU(N)pure} to obtain the $\CN=1$ curve.
  
  The SW curve of this theory is
    \be
    x^2
     =     \frac{\Lambda^2}{t^{n+2}}+\ldots + \frac{u}{t^2} + \ldots + \Lambda^2 t^{m-2},
           \label{Amn}
    \ee
  where $u$ and $\Lambda$ are the Coulomb modulus and the dynamical scale of the $\SU(2)$ gauge group 
  (which we consider fixed).
  $t$ is the coordinate on the base sphere, and the SW differential is $\lambda = x dt$.
  The dots denote the moduli and the couplings of the (generically) strongly coupled theory 
  which will not enter the ${\cal N}=1$ curve.
  In terms of the variable $v=x t$ the curve reads
    \be
    v^2
     =     \frac{\Lambda^2}{t^{n}}+\ldots + u + \ldots + \Lambda^2 t^{m}.
           \label{Amnv}
    \ee
  The only single trace superpotential is a pure massive deformation for the adjoint scalar in the vector multiplet
  $W(v)=\frac{\mu}{2} v^2$ and therefore, as in the case of the $\SU(N)$ SYM theory, 
  we rotate at one puncture and stay with the curve which depends on the number and type of punctures
    \beq
    w^n(w-\mu v)^m + \sum_{i=0}^{n+m-2} w^i f_i(v) = 0,
    \label{zio}
    \eeq
  where $f_i(v)$ are polynomials of degree $n+m-2-i$.
  
  The complete analysis of the vacua of this class of theories boils down to find the solutions 
  of the factorization condition.
  In the $n,m=1,2$ cases the theory has a Lagrangian description and corresponds 
  to $\SU(2)$ gauge theory with $N_f=0,1,2$ already discussed in the previous section.
  A class of vacua for $n=m$ can be identified as follows.
  The curve \eqref{zio} can be reduced to
    \beq
    w(w-\mu v)=g_0
    \eeq
  by setting $f_i=0$ $i>0$ and $f_0=g_0^N$ with $g_0$ a constant.
  Correspondingly $V=\frac{1}{\mu}\left(\varphi+f_0\varphi^{-1}\right)$,
  where $\varphi$ has vanishing boundary condition at $t\sim\infty$ and is diverging at $t\sim 0$.
  Therefore, from eq.~(\ref{Amnv}), we get the matrix equation
    $$
    \frac{1}{\mu^2}\left(\varphi^2+ 2 f_0 + f_0^2\varphi^{-2}\right) = \1 \left(
    \frac{\Lambda^2}{t^{n}}+\ldots + u + \ldots + \Lambda^2 t^{n} \right),
    $$
  which has to be solved under the above boundary conditions.
  Henceforth one finds
    \bea
    \frac{\varphi^2}{\mu^2}&=&\frac{\Lambda^2}{t^{n}}+\ldots + u_+, \\
    \frac{f_0^2\varphi^{-2}}{\mu^2}&=&u_-+\ldots \Lambda^2 t^n, \\
    2f_0&=&u-u_+-u_-,
    \eea
  where $u_\pm$ are constants. 
  In such a case one has two vacua $f_0=\pm \mu^2\Lambda^2$ and the ${\cal N}=2$ moduli are fixed at
    $$
    v^2
     = \Lambda^2\left(t^n \pm 2 + t^{-n}\right).
    $$
  Correspondingly 
    $$
    \Phi=\frac{\Lambda}{t}(t^n\pm 1)\left( \begin{matrix}
    0 & 1\\
    \frac{1}{t^n} & 0\\
    \end{matrix}
    \right),~~~
    \varphi=\mu\Lambda \left( \begin{matrix}
    0 & 1\\
    \frac{1}{t^n} & 0\\
    \end{matrix}
    \right).
    $$
  Fixing $\sigma=\Lambda/t$ and $\tau=1/\Lambda$ as in the case in subsection \ref{subsubsec:AD1}, we stay with 
    $$
    \phi_1=(t^n\pm 1)\left(\begin{matrix}
    0 & 1\\
    \frac{1}{t^n} & 0\\
    \end{matrix}
    \right),~~~
    \phi_2=\mu\Lambda^2\left(\begin{matrix}
    0 & 1\\
    \frac{1}{t^n} & 0\\
    \end{matrix}
    \right)
    $$
  and the rotated curve
    \bea
    x_1^2
    &=&    \frac{(t^n\pm 1)^2}{t^n},
           \\
    x_1x_2
    &=&    \mu\Lambda^2 \frac{(t^n\pm 1)}{t^n},
           \\
    x_2^2
    &=&    \frac{(\mu\Lambda^2)^2}{t^n},
    \eea
  which describes the theory in the ${\cal O}(-1)\oplus{\cal O}(-1)$ complex structure.

\subsection{$T_N$ theories coupled to $\CN=1$ vector multiplets}
  We will now see how our technique allows to predict the form of the $\mathcal{N}=1$ curve 
  found in \cite{Maruyoshi:2013hja} for the $T_N$ theories with two $\SU(N)$ flavor symmetries gauged. 
  Let us start from a linear quiver of the form $\SU(N)^n$ with bifundamental hypermultiplets. 
  The SW curve can be written as $$\Lambda_1^Nt^{n+1}+t^nP^1_N(v)+\dots+\Lambda_2^N=0,$$ 
  where $\Lambda_i$ represent the dynamical scales of the two gauge groups at the ends of the quiver. 
  We can bring it to the $N$-sheeted cover form: 
    $$v^N=\sum_{k=2}^{N}v^{N-k}Q_k(t),$$ 
  where $Q_k(t)$ (for $k<N$) have simple poles at $n-1$ points (the position of the NS5-branes) 
  and $Q_N(t)$ has two other poles at zero and infinity, associated with the irregular singularity. 
  If we turn on a quadratic superpotential for the first and last gauge groups with the mass parameters $\mu_{1}$
  and $\mu_{2}$, we can impose vanishing boundary conditions (for the $w$ coordinate) at all the simple punctures. 
  The $\mathcal{N}=1$ curve can then be parametrized as follows: 
    \bea
    w^N
     =     \sum_{k=2}^{N-1}a_kw^{N-k}+\mu_1^N\Lambda_1^Nt+a_N+\alpha\frac{\mu_2^N\Lambda_2^N}{t},
    \eea
  where $\alpha$ is a constant depending on the gauge coupling constants of $\CN=2$ gauge groups
  in the linear quiver.
  This is the most general expression compatible with the absence of poles at the simple punctures 
  and reproducing the asymptotics of $v$ at zero and infinity, no matter how many simple punctures we introduce. 
  If we have at least $N-1$ simple punctures, with a suitable choice of the coupling constants 
  (i.e.~considering the collision limit) we can degenerate the UV curve to a sphere with two irregular punctures, 
  one maximal regular (described by a Young tableau with a single row of length $N$) 
  and possibly other regular punctures.
  In this way we obtain the curve describing $\mathcal{N}=1$ gaugings of the $T_N$ theory
  (possibly coupled to other matter sectors). 
  Note here that the theory has adjoint multiplets with masses $\mu_{1,2}$.
  When we set the mass parameters to be equal this recovers the set-up in \cite{Maruyoshi:2013hja}.

  In \cite{Maruyoshi:2013hja} it was argued that the above curve captures the matrix of ${\rm U}(1)$ gauge couplings 
  also for $T_N$ theories coupled together by $\mathcal{N}=2$ vector multiplets, 
  provided we gauge two $\SU(N)$ flavor symmetries by $\CN=1$ vector multiplets. 
  The associated SW curve in this case has two irregular punctures (that we can set at $t=0,\infty$) 
  and a number of full punctures that can be thought of as arising from the collision limit of $N$ simple punctures 
  \cite{Chacaltana:2010ks}. 
  Since the form of the $\mathcal{N}=1$ curve derived above does not depend on the number of simple punctures, 
  we directly recover this result.

  If we now turn off the quadratic superpotential for one of the two gauge groups (sending e.g. to zero $\mu_2$), 
  the $\mathcal{N}=1$ curve reduces to 
    $$w^N=\sum_{k=2}^{N-1}a_kw^{N-k}+\mu_1^N\Lambda_1^Nt+a_N.$$ 
  Indeed this curve is still valid also when we send $\Lambda_2$ to zero, thus decoupling the second gauge group. 
  In this way we get the $\mathcal{N}=1$ curve for the $T_N$ theory with a single $\SU(N)$ gauged.
  
  Notice that for $\Lambda_2\neq0$ we can actually fix precisely the moduli of the curve: 
  the above formula is equal to the curve we found previously when we analyzed the massive deformation of SYM theory. 
  In that case we argued that all the parameters $a_k$ are actually zero. 
  This exploited the form of the factorization condition (combined with the fact that $t$ could be written 
  as a function of $w$) and the behavior of $v$ at infinity ($v\sim t$) and at zero ($v\sim1/t$). 
  All these properties hold in the present case as well, (at least as long as $\Lambda_2$ is nonzero) 
  so the conclusion is the same. We thus get the $\mathcal{N}=1$ curve $w^N=\mu_1^N\Lambda_1^Nt.$
  
  We will now provide a recipe to extract the $\mathcal{N}=1$ curve for the $T_N$ theory 
  coupled to three $\SU(N)$ gauge groups. 
  The corresponding $\mathcal{N}=2$ curve is (we denote with $\lambda$ the SW differential)
    $$\lambda^N=\sum_{k=2}^{N}\lambda^{N-k}\theta_k,$$ 
  where $\theta_k$ are the meromorphic differentials. In the present case they have the form 
    \be
    \label{mmdd}
    \theta_k
     =     \frac{P_k(t)}{t^k(t-1)^k}(dt)^k\quad (k<N);\quad
    \theta_N
     =     \frac{P_{N+3}(t)}{t^{N+1}(t-1)^{N+1}}(dt)^N,
    \ee 
  where we have set the three punctures to be at $t=0,1,\infty$.
  So far we have mainly discussed theories with Lagrangian descriptions, 
  whose $\mathcal{N}=1$ curve can be extracted just analyzing the behavior at the punctures of the SW curve 
  written in terms of the coordinate $v$ introduced by Witten (the SW differential reads $\lambda=vdt/t$). 
  This can be extracted considering the vector field on the sphere that preserves the points $t=0,\infty$: 
  $\zeta=t\partial_t$. 
  The $v$ coordinate is then obtained acting on $\zeta$ with the SW differential $\lambda(\zeta)=v$. 
  For the theory under consideration, we must use a vector field which preserves the three irregular singularities. 
  We thus impose that it has a zero at $t=1$ as well. 
  This will of course introduce a pole, that we locate at $t=a$. The vector field then reads 
    $$\zeta'=\frac{t(t-1)}{t-a}\partial_t.$$ 
  Acting on it with the SW differential we find the coordinate 
    $$y=\lambda(\zeta')=\frac{v(t-1)}{t-a}.$$ 
  Rewriting the SW curve in terms of $t$ and $y$ we get 
    $$y^N=\sum_{k=2}^{N}y^{N-k}\vartheta_k,$$ 
  where $\vartheta_k$ are meromorphic functions of the form 
    $$\vartheta_k=\frac{P_k(t)}{(t-a)^k}\quad(k<N);\quad\vartheta_N=\frac{P_{N+3}(t)}{t(t-1)(t-a)^N}.$$ 
  The boundary conditions for the $\mathcal{N}=1$ curve at the three punctures should be extracted from the SW curve 
  in this set of coordinates. We thus get a curve of the form 
    $$w^N=\sum_{k=2}^{N-1}w^{N-k}a_k+at+a_N+\frac{b}{t}+\frac{c}{t-1}.$$ 
  More in general, notice that a simple zero for the vector field decreases the order of the pole of the 
  $k$-th differential by $k$. 
  Consequently, the pole structure of the $\mathcal{N}=1$ for all the theories 
  considered in \cite{Maruyoshi:2013hja} can be recovered in the present setting 
  imposing that the vector field has a zero at all the irregular punctures.

\subsection{An example with higher degree superpotential}

  We will now illustrate with an example how one can determine the $\mathcal{N}=1$ curve 
  when higher degree superpotentials are turned on. 
  We consider an $\SU(3)\times \SU(3)$ gauge theory (with a massless bifundamental hypermultiplet 
  as the only matter field) with quadratic superpotential for one gauge group $W'_1(\Phi)=\mu\Phi$ 
  and cubic for the second gauge group $W'_2(\Phi)=g\Phi^2-\mu\Phi$. We can write the SW curve in the form
    $$v^3=v\frac{tu_1-u_2}{t-1}+\frac{\Lambda_2^{3}t^3+t^2v_1-tv_2+\Lambda_1^3}{t(t-1)}.$$ 
  The $\mathcal{N}=1$ curve has the by now familiar form 
    $$w^3=wV_2(t)+V_3(t).$$ 
  Given the above superpotentials, the meromorphic functions $V_2(t)$ and $V_3(t)$ can be found imposing 
  the boundary conditions 
    $$\frac{w}{v}\rightarrow0\;\; (\text{at}\; t=0)\quad \frac{w}{v}\rightarrow\mu\;\; (\text{at}\; t=1)\quad 
    \frac{w}{gv^2}\rightarrow1\;\; (\text{at infinity}).$$ 
  Let us analyze the asymptotics at each puncture: 
    \begin{itemize}
    \item {\bf At t=0:} $v^3\sim u_2v-\Lambda_1^3/t$ so $V_2(t)$ and $V_3(t)$ must be bounded.
    \item {\bf At t=1:} $$v^3\sim\frac{u_1-u_2}{t-1}v+\frac{\Lambda_1^3+\Lambda_2^3+v_1-v_2}{t-1},$$ 
                        consequently 
                        $$V_2(t)\sim\frac{\mu^2(u_1-u_2)}{t-1},\quad 
                        V_3(t)\sim\frac{\mu^3(\Lambda_1^3+\Lambda_2^3+v_1-v_2)}{t-1}.$$
    \item {\bf At infinity:} $$v^3\sim u_1v+\Lambda_2^3t,\quad \text{so}
                        \quad w^3\sim(\alpha t+\beta)w+g^3t^2+\gamma t+\delta.$$ 
                        We can allow a pole of degree one for $V_2(t)$, since this does not affect 
                        the asymptotic behavior of $w$ (at leading order).
    \end{itemize}
  We thus find the following answer: 
    \bea
    \begin{aligned}
    V_3(t)&=\frac{g^3(t-1)^3+a(t-1)^2+b(t-1)+\mu^3(\Lambda_1^3+\Lambda_2^3+v_1-v_2)}{t-1},\\
    V_2(t)&=c(t-1)+d+\frac{\mu^2(u_1-u_2)}{t-1}.
    \end{aligned}
    \eea 
  The parameters $a$, $b$, $c$ and $d$ and the Coulomb branch coordinates can then be fixed 
  imposing the factorization condition. 
  
  In general, in order to write down the $\mathcal{N}=1$ curve if we have to impose the boundary condition $w\sim v^k$ 
  at a puncture, the most convenient procedure is to consider the Newton polygon associated to the puncture. 
  In \cite{Gaiotto:2009we} it is explained how to construct it for the $\mathcal{N}=2$ theory: 
  one marks the point in the plane with coordinates $(k,i)$ if the $k$-th differential has a pole of degree $i$ 
  at the puncture. 
  One then considers the convex envelope of all the marked points. 
  The slope of the edges of the polygon encode the leading behavior of the $N$ roots at the puncture. 
  One has to repeat the same procedure in this case, but using the pole structure of the meromorphic functions 
  appearing in the SW curve written in terms of the coordinate identified by the choice of vector field. 

  The above boundary condition simply means that we have to multiply by $k$ the slope of all the edges, 
  or equivalently that we have to multiply by $k$ the ``height'' of all the vertices. 
  All the dots contained in the interior of the new Newton polygon are allowed and the corresponding term 
  in the meromorphic functions $V_k(t)$'s should be included. 
  In the above example the $\mathcal{N}=2$ curve has a simple pole at infinity, 
  so the associated Newton polygon is the triangle with vertices $(0,0)$, $(3,0)$ and $(3,1)$. 
  The leading behavior of the three roots at infinity is $v\sim t^{1/3}$. 
  Since we want to impose the boundary condition $w\sim v^2$, 
  the three roots of the $\mathcal{N}=1$ curve should be $w\sim t^{2/3}$. 
  This is done multiplying by $2$ the height of the third vertex. 
  The $\mathcal{N}=1$ Newton polygon is thus the triangle with edges $(0,0)$, $(3,0)$ and $(3,2)$. 
  We can easily see that the point $(2,1)$ lies in the interior of the triangle and is thus allowed. 
  In fact we included in $V_2(t)$ a term linear in $t$. 
  This algorithm can be used in principle to write down the curve in the general case.

\section{Linear quivers and the Dijkgraaf-Vafa curve}
\label{sec:Konishi}
  In this section we will provide another method to write down the curve 
  for $\CN=1$ theories with Lagrangian descriptions that admit a brane construction in Type IIA string theory, 
  focusing especially on linear quivers. 
  For this class of theories there is a globally well defined projection of the curve on the $(w,v)$ plane, 
  that can be identified with the Dijkgraaf-Vafa curve. 
  In this section we will make use of the generalized Konishi anomaly equations 
  to determine the curve explicitly in some examples.

\subsection{$\SU(N)$ SYM theory}

  In \cite{deBoer:2004he} it was recognized that for $\CN=1$ SYM theory the M-theory curve can be written 
  in terms of the Dijkgraaf-Vafa curve. 
  We will now review that argument and see how one can fix all the parameters of the curve 
  using the generalized anomaly equations, once the superpotential and the point in the Coulomb branch are given. 

  The SW is written as \eqref{n2}.
  If we break extended supersymmetry adding a polynomial superpotential (of degree $n+1$) 
  that we denote by $W(\Phi)$ for the chiral multiplet in the adjoint representation, 
  the curve describing our $\mathcal{N}=1$ theory will now be embedded in a threefold. 
  The system defining the $\mathcal{N}=1$ curve can be conveniently written as \cite{deBoer:2004he} 
    \bea
    \Lambda^{N}\left(t+\frac{1}{t}\right)
    &=&    P_N (v),
           \nonumber \\
    w^2-w W'(v) +f_{n-1}(v)
    &=&    0,
           \label{n=1}
    \eea
  where $f_{n-1}(v)$ is a polynomial of degree $n-1$. 
  The second equation is implied by the boundary conditions at the punctures 
  \cite{deBoer:2004he}\footnote{The analysis of \cite{deBoer:2004he} applies to ${\rm U}(N)$ SYM theory. 
    In the present context this statement is slightly modified. 
    This will not be important in this section and we postpone a detailed discussion of this point to the next section,
    where the $\mathcal{N}=1$ deformation of SQCD is considered.}
  :
    \be\label{bc}
    \begin{array}{l}
    \text{for}\quad t\rightarrow0\qquad w\rightarrow W'(v),\quad v\simeq t^{-1/N},\\
    \text{for}\quad t\rightarrow\infty\qquad w\rightarrow0,\quad v\simeq t^{1/N},
    \end{array}
    \ee
  and is referred to as DV curve.
  As seen in the previous sections, the brane cannot be rotated at a generic point of the moduli space. 
  The submanifold at which the curve can be rotated is defined by a certain factorization equation 
  (we will see in the next subsection that it precisely coincides with the factorization condition introduced 
  in this paper), which implies that the curve degenerates to a genus (at most) $n-1$ surface \cite{deBoer:2004he}. 

  So far we have determined the general form of the DV curve. 
  However, we still do not know the precise form of $f_{n-1}$ in a given vacuum. 
  It can actually be determined explicitly in several cases, as we will see.
  In order to determine the polynomial $f$, let us recall that the chiral condensates 
  in such a theory can be computed explicitly using the generalized Konishi anomaly studied in \cite{Cachazo:2002ry}: 
  if we introduce the generating functions
    $$ T(v)=\left\langle {\rm Tr}\frac{1}{v-\Phi}\right\rangle,\; R(v)=\frac{-1}{32\pi^2}\left\langle{\rm Tr}
    \frac{\W_{\alpha}\W^{\alpha}}{v-\Phi}\right\rangle,$$
  where $\W_{\alpha}$ denotes the field strength superfield, we can determine all chiral condensates simply 
  expanding $R(v)$ and $T(v)$ in powers of $1/v$, once we know them explicitly. 
  These can in turn be determined recognizing that they satisfy the following Ward-identities \cite{Cachazo:2002ry}:
    \begin{equation}\label{angen}
    \begin{aligned}
    & W'(v)T(v) +c_{n-1}(v) = 2R(v)T(v),\\
    & W'(v)R(v) - f_{n-1}(v) = R^{2}(v).
    \end{aligned}
    \end{equation}
  Notice that $R(v)$ satisfies the same equation as $w$ in (\ref{n=1}). 
  This motivates the identification $R(v)=w$ proposed in \cite{deBoer:2004he}. 
  Furthermore, it was shown in \cite{Cachazo:2003yc} that $T(v)dv=dlog(2t)=dt/t$. 

  The $f$ polynomial appearing in the above equation is precisely the $f_{n-1}$ entering in the DV curve 
  and $c$ is another polynomial in $v$. 
  From the second equation we get immediately 
    \be\label{mres}
    R(v)=\frac{1}{2}\left(W'(v)-\sqrt{W'(v)^{2}-4f_{n-1}(v)}\right).
    \ee

  Let us first of all determine $c_{n-1}(v)$: 
  from the first Ward identity in (\ref{angen}) we find 
    $$c_{n-1}(v)=\left(W'(v)T(v)\right)_+,$$ 
  where $(s)_+$ stands for the polynomial part of $s$. 
  Since $T(v)=N/v+O(v^{-3})$, for quadratic or cubic superpotentials (or whenever $T(v)=N/v+O(v^{-n-1})$, 
  which will be true only on a submanifold of the Coulomb branch) 
  we can keep only the leading term
     \be
     \label{ciao}
     c_{n-1}(v)=-N\frac{W'(v)}{v}.
     \ee 
  In this section we will discuss the cases of quadratic superpotential and 
  $\mathcal{N}=1$ deformation of Arygres-Douglas points only, so this approximation can be used. 
 
  With a similar argument we can also determine $f_{n-1}(v)$: since $T(v)dv=dt/t$ we have $T(v)=\frac{1}{t}\frac{dt}{dv}$.
  Plugging this into the first Ward-identity in (\ref{angen}) and using (\ref{ciao}) we get 
    $$R(v)=\frac{1}{2}\left(W'(v)-N\frac{t}{v}\frac{dW(v)}{dt}\right).$$ 
  Comparing now with (\ref{mres}) we then find 
    \be\label{poly}
    f_{n-1}(v)
     =     \frac{1}{4}\left[W'(v)^{2}-N^2\frac{t^2}{v^2}\left(\frac{dW(v)}{dt}\right)^2\right],
    \ee
  thinking of $W(v)$ as a function of $t$, determined by the SW curve. 
  At a generic point in the Coulomb branch this formula will not give a polynomial of degree $n-1$. 
  However, this will be true at all points at which the NS5-brane can be rotated. 

  Let us analyze explicitly the two cases we have studied before: $\SU(N)$ SYM at the Arygres-Douglas point 
  and the case of quadratic superpotential. 
  We will now see that using (\ref{poly}) we can recover the same result found before with the help of the Hitchin field.

\paragraph{$\SU(N)$ SYM at the Argyres-Douglas point}
  The SW curve at the Argyres-Douglas point is (\ref{n2}) with $P_N(v)=v^N\pm 2\Lambda^{N}$, thus
  $v^N=\Lambda^N\left(t+\frac{1}{t}\mp 2\right)$. 
  It is known that this point of the moduli space is not lifted if we turn on a superpotential of the form $v^N$. 
  Let us set $W(v)=(g/N)v^N$. 
  From (\ref{poly}) we immediately find 
    $$f_{N-2}(v)=\frac{g^2}{4}\left[v^{2N-2}-\frac{t^2}{v^2}\left(\frac{dv^N}{dt}\right)^2\right].$$ 
  Using the SW curve we can rewrite $v^N$ in terms of $t$. 
  We then find
    $$\frac{t^2}{v^2}\left(\frac{dv^N}{dt}\right)^2
     =\frac{\Lambda^{2N}}{v^2}\left(t-\frac{1}{t}\right)^2=v^{2N-2}\pm4\Lambda^Nv^{N-2},$$
  where we have used the identity 
  $\left(t-\frac{1}{t}\right)^2=\left(t+\frac{1}{t}\mp2\right)^2\pm4\left(t+\frac{1}{t}\mp2\right)$. 
  The final result is then $f_{N-2}(v)=\mp g^2\Lambda^Nv^{N-2}$. 
  The second equation defining the $\mathcal{N}=1$ curve consequently is 
    \be\label{curva2}w^2-gv^{N-1}w\mp g^2\Lambda^Nv^{N-2}
     =     0.
    \ee 
  This agrees with the equation obtained in section \ref{subsubsec:AD1}.
  
  We can make a simple consistency check on our formula for $f_{N-2}$: 
  using (\ref{angen}) and (\ref{mres}) it is straightforward to determine $T(v)$ explicitly
    $$T(v)\equiv\sum_{i}\frac{\langle\Tr\Phi^i\rangle}{v^{i+1}}
     =    \frac{N}{v\sqrt{1\pm4(\Lambda/v)^N}}
     =     \frac{N}{v}\mp\frac{2N\Lambda^N}{v^{N+1}}+\dots$$ 
  We can now compute $P_N(v)$ using this formula and check that we get precisely our starting point.
  In order to do this the simplest way is to introduce $U_{i}=\frac{1}{i}\langle\Tr\Phi^i\rangle$ 
  and then use the formula given in \cite{Cachazo:2003yc} which relates the SW curve to these quantities:
    \begin{equation}\label{curva}
    P_{N}(x)
     =     \left(x^{N}e^{-\sum_{i\geq1}\frac{U_i}{x^i}}\right)_{+}, 
    \end{equation}
  It is easy to check that this equation gives precisely $P_N(v)=v^{N}\pm2\Lambda^{N}$.

\paragraph{Quadratic superpotential}
  We now turn to the analysis of the massive deformation $W(v)=(\mu/2)v^2$. 
  In this case the points which are not lifted can be determined using Chebyshev polynomials: 
  $P_{N}(v)=2\Lambda^{N}T_{N}(v/2\Lambda)$. 
  If we introduce the variable $s$ defined as $v/\Lambda=s+1/s$, then $2T_N(v/2\Lambda)=s^N+s^{-N}$.
  In this case $f_{n-1}$ is simply a constant and from (\ref{poly}) we find 
    $$f_{0}=\frac{\mu^2}{4}\left[v^2-N^2t^2\left(\frac{dv}{dt}\right)^2\right]
     =    \frac{\mu^2\Lambda^2}{4}\left[(s+s^{-1})^2-N^2t^2\left(\frac{d(s+s^{-1})}{dt}\right)^2\right].$$
  In order to evaluate the last term we exploit the equation coming from the SW curve: $s^N+s^{-N}=t+t^{-1}$. 
  If we differentiate with respect to $t$ this equation we get 
    $$1-\frac{1}{t^2}=\frac{N}{s}(s^{N}-s^{-N})\frac{ds}{dt}.$$ 
  Taking the square and multiplying by $t^2$ the lhs becomes
    $\left(t-\frac{1}{t}\right)^2
     =    \left(t+\frac{1}{t}\right)^2-4
     =    s^{2N}+s^{-2N}-2=(s^N-s^{-N})^2.$ 
  The rhs becomes instead $(s^N-s^{-N})^2\left(\frac{Nt}{s}\frac{ds}{dt}\right)^2$. 
  Equating these two we finally get $\left(\frac{Nt}{s}\frac{ds}{dt}\right)^2=1$. 
  Using this formula the equation for $f_0$ reduces to 
    $$f_0=\frac{\mu^2\Lambda^2}{4}\left[(s+s^{-1})^2-(s-s^{-1})^2\right]=\mu^2\Lambda^2.$$ 
  We thus find the $\mathcal{N}=1$ curve 
    \be\label{mass}
    w^2-\mu vw + \mu^2\Lambda^2
     =     0.
    \ee
  This correctly reproduces the result found previously.

\subsection{The curve for SQCD}
  We have seen that the $\mathcal{N}=1$ deformation of $\SU(N)$ SYM theory can be described by a curve 
  whose form is dictated by the SW curve of the parent $\mathcal{N}=2$ theory 
  and the DV curve, as noticed in \cite{deBoer:2004he}. 
  This recipe can be easily extended to the theory with matter fields in the fundamental. 
  The matrix model curve can be easily derived from the generalized anomaly equations: 
  considering the variation $\delta\Phi=\frac{\W_{\alpha} \W^{\alpha}}{z-\Phi}$, 
  we obtain the Ward identity which is the same as the one in \eqref{angen}
  where $W(v)$ is now the superpotential of the theory with the matter fields \cite{Seiberg:2002jq}.
  
  The above variation indeed changes the trace part of $\Phi$, which is not a problem if the gauge group is ${\rm U}(N)$.
  Since we are dealing with $\SU(N)$ gauge theories, we have to slightly modify the above variation 
  in order to preserve the traceless condition. 
  This can be done simply setting \cite{DiPietro:2011za}
    $$\delta\Phi=\frac{\W_{\alpha}\W^{\alpha}}{v-\Phi}+\frac{32\pi^2}{N}R(v)I.$$
  The resulting Ward identity is then 
    \be
    \label{mcurve}
    R^{2}(v)
     =     \left(W'(v)-\left\langle W_{\Phi}' \right\rangle\right)R(v)-f_{n-1}(v),
    \ee 
  where $W_{\Phi}'$ is the trace of the derivative of the superpotential with respect to $\Phi$ 
  (times the prefactor $1/N$). 
  This is automatically zero for ${\rm U}(N)$ gauge theories because of the equations of motion, 
  but it is not zero in the present case. 
  The DV curve is then obtained just identifying $R(v)$ with the complex coordinate $w$ as before. 
  Notice that this procedure is equivalent to using the formula valid for ${\rm U}(N)$ gauge theories 
  with a modified superpotential which includes a linear term whose coefficient is equal to $W_{\Phi}'$.

  We should note that the shift in the variation in principle changes the result of the SYM theory 
  in the previous subsection.
  However, at least for the vacua considered there, the effect of the shift is irrelevant:
  for the quadratic superpotential $\tr W'(\Phi) = \tr \Phi = 0$;
  at the Argyres-Douglas points, $\langle \tr \Phi^{N-1} \rangle =0$.

  Note also that we are not forced to restrict to deformations of the $\mathcal{N}=2$ theory 
  that include only a polynomial superpotential for $\Phi$. 
  Superpotential terms of the form $\tilde{Q}P(\Phi)Q$ with $P$ a generic polynomial can be easily incorporated. 
  The only way in which they affect the curve is by modifying the term $\langle W_{\Phi}'\rangle$. 
  The curve (\ref{mcurve}) can be derived as in the previous sections from the boundary conditions for $w$: 
  at the punctures corresponding to the presence of the NS5-branes the $v$ coordinate\footnote{In this section 
    when we mention the SW curve we always refer to the curve in the parametrization with $v$ and $t$.} 
  diverges and the presence of the superpotential $W(\Phi)$ is encoded in the difference 
  between the boundary conditions imposed at the two punctures. 
  We are free to impose the condition $w\rightarrow0$ at one puncture and consequently 
  $w\rightarrow W'(v)-\langle W_{\Phi}'\rangle$ at the second one. 
  At this puncture $w$ tends to $W'(v)$ in the sense that their ratio tends to one, 
  the difference being a constant given by the vev of the chiral operator $W_{\Phi}'$.
  
  Let us comment further on the boundary conditions for $w$. 
  As is well known and seen also in subsection \ref{subsec:SU(2)SQCD}, 
  the Riemann surface associated to $\SU(N)$ SQCD with $N_f$ flavors, 
  in which the M5-branes are compactified, is a sphere with two, three or four punctures 
  depending on the number of flavors and on the Type IIA realization of the theory, 
  that is on the number of semi-infinite D4 branes inserted on each side. 
  Whenever there are less than $N$ D4-branes on both sides the theory is described by a sphere 
  with two irregular punctures, where the $v$ coordinate tends to infinity. 
  In this case the situation is very similar to the SYM case analyzed before, so there is no need to discuss it further.
  If $N_f\geq N$, we can indeed put $N$ semi-infinite D4 branes
  (or D4-branes whose ends are attached to D6-branes) on the same side. 
  In this case there are one irregular puncture at let's say $t=\infty$ (provided $N_f<2N$),
  one maximal regular puncture at $t=0$ and one minimal puncture at $t=1$. 
  In this case the punctures associated to the ``transverse'' NS5-branes are those at $t=1$ and infinity, 
  where $v$ diverges. 
  The puncture at the origin is instead related to the $N$ D4-branes and the $v$ coordinate remains finite in this limit.
  The $N$ asymptotic values for $v$ just correspond to the mass parameters of the associated flavors \cite{Witten:1997sc}.
  The boundary conditions for $w$, which describe the rotation of the NS5-branes, should then be imposed 
  at $t=1$ and $t=\infty$. 
  The third puncture simply does not play any role in this discussion. 

  If $N_f=2N$ the $\mathcal{N}=2$ theory is conformal and the underlying base sphere has four regular punctures: 
  two minimal and two maximal. 
  The punctures associated to the NS5-branes are the two minimal ones 
  and this is where the boundary conditions for $w$ should be imposed. 
  Again, at the maximal punctures $v$ does not diverge and its limit can be identified with the various mass parameters. 
  This discussion generalizes in a straightforward way to quiver theories: 
  a linear quiver with $n$ gauge groups is associated to a sphere with $n+1$ minimal punctures 
  and this is where we have to impose the boundary conditions for $w$. 
  At one of them we can impose the condition $w=0$, 
  since an overall rotation of the brane system just corresponds to a reparametrization of the curve. 
  The other $n$ boundary conditions precisely encode the information about the various superpotentials 
  for the chiral superfields in the adjoint. 
  We thus see how the general picture gradually emerges. 
  We will discuss in more detail quiver theories in the following subsection.

\subsubsection{The factorization condition}
  Our proposal for the M-theory curve associated to softly broken $\mathcal{N}=2$ SQCD is then
    \bea
    t^2-tP_N(v)+\Lambda^{2N-N_f}\prod_{i}(v-m_i)
    &=&    0,
           \nonumber \\
    w^2-w\widetilde{W}'(v)+f_{n-1}(v)
    &=&    0,
    \label{SQCD}
    \eea
  where $\widetilde{W}'(v)$ stands for $W'(v)-\left\langle W_{\Phi}'\right\rangle$.
  As we mentioned before, the restriction on the Coulomb branch coordinates is encoded in a factorization condition 
  which relates the above two equations. 
  It proves convenient in order to define it, to introduce the variables $y=2t - P_N(v)$ and $x=2w - \widetilde{W}'(v)$. 
  The above equations then become 
    $$y^2=P_N^{2}(v)-4\Lambda^{2N-N_f}\prod_{i}(v-m_i);\quad 
    x^2=(\widetilde{W}'(v))^{2}-4f_{n-1}(v).$$ 
  The factorization condition then reads as follows: 
    $$y^2= H_{N-k}^{2}(v)F_{2k}(v);\quad x^2=Q_{n-k}^2(v)F_{2k}(v),$$ 
  where $F$ has only roots of multiplicity one. 
  This equation encodes the well-known fact that a point in the $\mathcal{N}=2$ Coulomb branch is not lifted 
  only if there are at least $N-n$ massless monopoles. 

  We will now see how this factorization condition is related to our constraint \eqref{facct}. 
  In the special case of SQCD, we can rewrite \eqref{facct} as follows: 
    \be\label{fac}
    w
     =     \frac{A(v)t+B(v)}{C(v)t+D(v)},
    \ee
  where $A,B,C$ and $D$ are polynomials in $v$ 
  (any higher power of $t$ can be eliminated using the SW curve which in the present case is quadratic in $t$). 
  This is the equation that should be added to (\ref{SQCD}).
  Clearly, modulo a redefinition of the four polynomials, 
  the same equation holds for the variables $y$ and $x$ introduced before. 
  It is convenient to work in terms of them because the two solutions at fixed $v$ 
  (that we indicate with $x$, $\tilde{x}$, $y$ and $\tilde{y}$) differ just by a sign. 
  Now the argument proceeds as in the analysis of the rank one case around \eqref{around34}
  which leads to $B(v)D(v)=A(v)C(v)y^2$ and $A(v)B(v)=C(v)D(v)x^2$. 
  These in turn imply\footnote{If some of the polynomials are zero, making these two equations trivial, 
    it is easy to see that either $B(v)=C(v)=0$ or $A(v)=D(v)=0$ and this leads directly to (\ref{fac1}) in any case.} 
    \be\label{fac1}\frac{A^2(v)}{D^2(v)}=\frac{x^2}{y^2}; \quad x^2y^2=\frac{B^2(v)}{C^2(v)}.
    \ee
  It is easy to see that both these equations are equivalent to the factorization condition. 
  We have thus established that from the brane perspective the constraint which selects the points 
  in the moduli space which are not lifted by the $\mathcal{N}=1$ deformation is equivalent to the requirement 
  that the M-theory curve is an $N$-sheeted cover of the base sphere.

\subsubsection{Comparison with previous results}
  Specializing this construction to the case of quadratic superpotential 
  $W=\frac{\mu}{2} {\rm Tr} \Phi^2+ \sum_{i=1}^{N_{f}}\sqrt{2} \widetilde{Q}^i\Phi Q_i
   + \sum_{i=1}^{N_{f}} m_{i} \widetilde{Q}^{i}Q_{i}$, 
  we recover the class of models studied in \cite{Hori:1997ab}. 
  We will see that our curve allows to reproduce the results presented in that paper. 
  Our curve is given by (\ref{SQCD}) with 
    $$\widetilde{W}'(v)=\mu v-M;\qquad M\equiv\frac{\sqrt{2}}{N}\langle \widetilde{Q}^iQ_i\rangle.$$ 
  The polynomial $f_{n-1}(v)$ in this case is just a constant, equal to $\mu$ times the gluino condensate $S$. 
  We then immediately see that we can rewrite the matrix model curve as follows 
    \be\label{vfun}
    v
     =     \frac{w^2+wM+\mu S}{\mu w}=\frac{(w-w_+)(w-w_-)}{\mu w}.
    \ee
  We have seen before that the coordinate $w$ can be identified with $R(v)$ which encodes all the chiral condensates 
  $\langle {\rm Tr} \Phi^n\W_{\alpha}\W^{\alpha}\rangle$. 
  If one considers as in \cite{Cachazo:2003yc} the transformation $\delta Q^i=\frac{1}{v-\Phi}Q^i$, 
  the corresponding anomalous Ward identity reads as (for fixed flavor index $i$) 
    $$\sqrt{2}\langle \widetilde{Q}^i\frac{1}{v-\Phi}Q_i\rangle
     =     \frac{R(v)}{v+m_i}-\frac{R(v_I)}{v+m_i},$$ 
  where with $v_I$ $(I=1,2)$ we indicate the two preimages on the DV curve of the point $v=-m_i$ 
  (see \cite{Cachazo:2003yc} for a detailed explanation of this point). 
  We thus see that $R(v_I)$ (or equivalently the two solutions for $w$ at $v=-m_i$) is naturally interpreted 
  as $-\sqrt{2}\langle \widetilde{Q}^iQ_i\rangle$. 
  It was shown in \cite{Hori:1997ab} that the matrix of meson vev's is diagonal 
  with at most two different eigenvalues $M_{\pm}$.
  This is explained in the present framework by the possibility of choosing $v_I$ in the above formula 
  (clearly different choices correspond to different vacua of the theory). 

  Let us now focus for simplicity on the theory with equal bare masses ($m_i=m$ $\forall i$). 
  From the above argument we can compute the meson condensates simply setting $v=-m$ in the matrix model curve. 
  This in turn implies that the parameters $w_{\pm}$ become precisely proportional to the meson vevs 
  ($w_{\pm}=-\sqrt{2}M_{\pm}$), after a simple redefinition which brings the curve to the form 
    $$v+m=\frac{w^2+wM+\mu S}{\mu w}=\frac{(w-w_+)(w-w_-)}{\mu w}.$$  
  This coincides with the one found in \cite{Hori:1997ab}.
  
  Another consequence of the analysis performed in \cite{Hori:1997ab} is that $w$ is a global coordinate 
  on the M-theory curve describing the massive deformation of $\mathcal{N}=2$ SQCD. 
  This fact is a simple consequence of the factorization condition: we can invert (\ref{fac}) to obtain 
    $$t=\frac{B(v)-D(v)w}{C(v)w-A(v)}.$$ 
  Since we have just seen that $v$ is a function of $w$ only, 
  it is enough to substitute (\ref{vfun}) in the previous expression. 
  The explicit form of the function $t(w)$ clearly depends on the boundary conditions we choose to impose for $w$. 
  Notice that the DV curve only tells us that $W'(v)$ can be expressed as a function of $w$. 
  We thus see that for a generic choice of superpotential it will not be true that $t$ and $v$ are functions of $w$ only.

\subsection{Linear quivers}
  In order to extend what we have done so far to quiver gauge theories, 
  we will again use the generalized Konishi anomaly equations as a guidance. 
  We expect the curve to be described by the SW curve of the underlying $\mathcal{N}=2$ theory, 
  together with the DV curve which encodes the boundary conditions for the coordinate $w$ 
  (or equivalently which carries the information associated to the various superpotentials) 
  and a constraint similar to (\ref{fac}), encoding the restrictions on the Coulomb branch coordinates.

\subsubsection{Generalized anomaly equations for linear quivers}
  The analysis of \cite{Cachazo:2002ry,Seiberg:2002jq,Cachazo:2003yc} has been generalized to linear unitary quivers
  with gauge group $\prod_{k=1}^{n} \SU(N_{k})$ 
  in \cite{Naculich:2003cz,Casero:2003gf,Casero:2003gr}. 
  For each chiral multiplet in the adjoint representation $\Phi_k$ we can define the generating functional 
    $$R_k(v)=-\frac{1}{32\pi^2}\left\langle{\rm Tr}\frac{\W_{k \alpha} \W_{k}^{\alpha}}{v-\Phi_k}\right\rangle.$$ 
  Considering the variations 
    $$\delta\Phi_k=\frac{\W_{k \alpha}\W_{k}^{\alpha}}{v-\Phi_k}+\frac{32\pi^2R_k(v)}{N_k}I,$$
  we find the Ward identities (compare with equations (5.2)-(5.4) of \cite{Casero:2003gf}): 
    \be\begin{aligned} 
    R_k^2(v)=&\left(W'_k(v)-\langle W'_{\Phi_k}\rangle\right)R_k(v)-
    f_k(v)-\frac{1}{32\pi^2}\left\langle{\rm Tr}\widetilde{X}\frac{\W_{k \alpha}\W_{k}^{\alpha}}{v-\Phi_k}X\right\rangle\\
    &+\frac{1}{32\pi^2}\left\langle {\rm Tr} X\frac{\W_{k \alpha}\W_{k}^{\alpha}}{v-\Phi_k}\widetilde{X}
    \right\rangle\qquad (k=2,\dots,n-1),\\
    R_1^2(v)=&\left(W'_1(v)-\langle W'_{\Phi_1}\rangle\right)R_1(v)-
    f_1(v)-\frac{1}{32\pi^2}\left\langle{\rm Tr}\widetilde{X}\frac{\W_{1 \alpha}\W_{1}^{\alpha}}{v-\Phi_1}X\right\rangle,
    \\
    R_n^2(v)=&\left(W'_n(v)-\langle W'_{\Phi_n}\rangle\right)R_n(v)-
    f_n(v)-\frac{1}{32\pi^2}\left\langle{\rm Tr} X\frac{\W_{n \alpha}\W_{n}^{\alpha}}{v-\Phi_n}\widetilde{X}\right\rangle,
    \\
    \end{aligned}
    \ee 
  where $X$ denotes the bifundamentals, $\langle W'_{\Phi_k}\rangle$ is the vev of 
  the ${\rm U}(N_{k})$ equation of motion as explained in the previous section (divided by $N_k$ as before) and 
    $$f_k(v)=\frac{1}{32\pi^2}\left\langle\frac{W'_k(\Phi_k)-W'_k(v)}{v-\Phi_k}\W_{k \alpha}\W_{k}^{\alpha}\right\rangle.
    $$ 
  The further complication is the presence of the generating functionals 
  $\left\langle{\rm Tr}\widetilde{X}\frac{\W_{k \alpha}\W_{k}^{\alpha}}{v-\Phi_k}X\right\rangle$, 
  which can be evaluated considering more general variations as explained in detail in \cite{Casero:2003gf}. 
  Combining all the chiral ring equations one can show that the $R_k(v)$'s functionals satisfy constraints 
  of degree $k=2,\dots,n+1$ (where $n$ is the number of gauge groups), which in principle allow to determine them. 
  In \cite{Casero:2003gf} the quadratic and cubic constraints have been determined explicitly. 
  From now on we will denote the quantity $W'_k(v)-\langle W'_{\Phi_k}\rangle$ simply with $W'_k(v)$, 
  hoping that no confusion arises. 
  This is the analogue of the redefinition of the superpotential discussed in the previous section. 
  With this modification the equations given in \cite{Casero:2003gf} apply in our case as well. 
  For example, in the present notation the quadratic constraint reads: 
  \be\label{quadratic}\sum_{k=1}^{n-1}R_kR_{k+1}=\sum_{k=1}^{n}(R_k^2-W'_kR_k+f_k).\ee

\subsubsection{The $\mathcal{N}=1$ curve}
  From the analysis of SQCD in the previous section, we expect the right boundary condition for $w$ at the $i$-th puncture
  to be of the form 
    $$w\rightarrow\sum_{k<i}W'_k(v),$$ 
  where $i=1,\ldots,n+1$ and $W'_k(\Phi_k)$ stands for the derivative of the ``modified'' superpotential 
  associated to the $k$-th gauge group. 
  Clearly we are free to impose vanishing boundary conditions at one puncture. 
  This leads to the equation 
    $$\frac{w(w-W'_1(v))(w-W'_1(v)-W'_2(v))\dots(w-\sum_{i=1}^{n}W'_i(v))}{(\sum_{i=1}^{n}W'_i(v))^n}=0,$$ 
  or equivalently 
    \be
    \label{curvalin}
    w(w-W'_1(v))(w-W'_1(v)-W'_2(v))\dots(w-\sum_{i=1}^{n}W'_i(v))+\sum_{j=2}^{n+1}w^{n+1-j}f_{(j-1)N-1}(v)=0,
    \ee 
  for some polynomials $f_{(j-1)N-1}(v)$ of degree $(j-1)N-1$ 
  (notice that by assuming they are polynomials we impose the constraint that $w$ does not diverge for $v$ finite, 
  which is indeed the expected behavior). 
  In the above formula $N=\text{max}_i\{\text{deg}W'_i\}$. 

  In order to make contact with the generalized anomaly equations, we should now understand the relationship 
  between the $w$ coordinate and the generating functionals $R_k(v)$. 
  Let us then reconsider the SQCD case: as we have seen, the boundary conditions lead to the curve $w(w-W'(v))+f(v)=0$. 
  We then get the two solutions $w_1(v)$ and $w_2(v)=W'(v)-w_1(v)$. 
  One of them coincides with $R(v)=\langle{\rm Tr}\frac{\W_{\alpha}\W^{\alpha}}{v-\Phi}\rangle$ 
  (which one clearly depends on our choice of boundary conditions). 
  With a shift in $w$ we can eliminate the linear term finding the equation 
    \be
    \label{one}
    y^2-\frac{W'(v)^2}{4}+f(v)=0.
    \ee 
  The fact that $R(v)$ (defined as a power series in $1/v$) coincides 
  with one of the two roots $w_{1,2}$ can then be rewritten as follows 
    \be\label{two}(y-R(v)+\frac{W'(v)}{2})(y+R(v)-\frac{W'(v)}{2})=0.
    \ee 
  Equating (\ref{one}) and (\ref{two}), we then get a quadratic constraint for $R(v)$: $R^2(v)-W'(v)R(v)+f(v)=0$, 
  which reproduces precisely the anomalous Ward identity introduced previously. 
  Our goal is then to generalize (\ref{two}).
  
  We can easily eliminate from (\ref{curvalin}) the term of degree $n$ in $w$ with a shift. 
  This leads to the equation 
    \be\label{curlin}\prod_{i=1}^{n+1}\left(y+\sum_{k=i}^{n}\frac{n+1-k}{n+1}W'_{k}(v)
    -\sum_{k=1}^{i-1}\frac{k}{n+1}W'_{k}(v)\right)+\sum_{j=2}^{n+1}y^{n+1-j}g_{(j-1)N-1}(v)=0.
    \ee
  By analogy with the SQCD case, we expect the $n+1$ solutions in $y$ (at fixed $v$) to be proportional 
  to various linear combinations of $R_k(v)$'s. 
  More precisely, we expect to be able to rewrite (\ref{curlin}) in the form
    \be\label{resolvents}\prod_{i=1}^{n+1}\left(y+\sum_{j=1}^{n}c_{ij}R_j(v)
    +\sum_{k=i}^{n}\frac{n+1-k}{n+1}W'_{k}(v)-
    \sum_{k=1}^{i-1}\frac{k}{n+1}W'_{k}(v)\right)=0.
    \ee 
  Equating the terms with the same power of $y$ in (\ref{curlin}) and (\ref{resolvents}), leads to $n$ equations 
  in the $R_k(v)$'s which should reproduce the various constraints coming from the generalized Konishi anomaly equations. 
  
  Imposing that the terms proportional to $y^{n-1}$ reproduce the quadratic constraint (\ref{quadratic}), 
  we find the solution (setting $R_0(v)=R_{n+1}(v)=0$) 
    $$\sum_{j=1}^{n}c_{ij}R_j(v)=R_{i-1}(v)-R_{i}(v).$$
  The curve \eqref{resolvents} with this choice of $c_{ij}$ already appeared in the literature: 
  it coincides with the matrix model curve for linear quivers found in \cite{Dijkgraaf:2002vw}. 
  This naturally generalizes our findings for SQCD. 
  As shown in \cite{Casero:2003gf}, one can check that the above curve correctly reproduces 
  the anomaly cubic constraint as well.
  
  The above discussion shows that the coefficients of the polynomials $g_{jN-1}(v)$'s appearing 
  in (\ref{curlin}) are proportional to the various chiral condensates of the theory, exactly as for SQCD. 
  In the simplest case of two gauge groups we can be completely explicit: 
  in this case (\ref{curlin}) reduces to 
    $$\begin{aligned}&\left(y+\frac{2}{3}W'_1(v)+\frac{1}{3}W'_2(v)\right)
    \left(y-\frac{1}{3}W'_1(v)+\frac{1}{3}W'_2(v)\right)\left(y-\frac{1}{3}W'_1(v)-\frac{2}{3}W'_2(v)\right)\\
    &+g_{N-1}(v)y+g_{2N-1}(v)=0.\end{aligned}$$
  Using (\ref{resolvents}) and comparing with the explicit expressions for the quadratic and cubic constraints 
  given in \cite{Casero:2003gf}, we find 
    \bea
    g_{N-1}(v)
    &=&    \sum_{k=1,2}f_k(v)=\frac{1}{32\pi^2}\sum_{k=1,2}
           \left\langle {\rm Tr}\frac{W'_k(\Phi_k)-W'_k(v)}{v-\Phi_k}\W_{k \alpha}\W_{k}^{\alpha}\right\rangle,
           \nonumber \\
    g_{2N-1}(v)
    &=&    f_2(v)\left(\frac{2W'_1}{3}+\frac{W'_2}{3}\right)
         - f_1(v)\left(\frac{W'_1}{3}+\frac{2W'_2}{3}\right)+g_1(v)+g_2(v),
    \eea 
  where 
    $$g_1(v)=\frac{1}{32\pi^2}\left\langle{\rm Tr}\frac{W'_1(\Phi_1)-W'_1(v)}{v-\Phi_1}
    \W_{1 \alpha}\W_{1}^{\alpha}X\widetilde{X}\right\rangle,$$
    $$g_2(v)=\frac{1}{32\pi^2}\left\langle{\rm Tr} \frac{W'_2(\Phi_2)-W'_2(v)}{v-\Phi_2}
    \W_{2 \alpha}\W_{2}^{\alpha}\widetilde{X}X\right\rangle.$$

  The final point we have to discuss is the factorization condition. 
  In the case of linear quivers the $\mathcal{N}=2$ curve is a polynomial equations of degree $n+1$ in $t$. 
  The argument of the previous section for SQCD then implies that the factorization condition reads: 
    $$w=\frac{A_1(v)t^n+\dots+A_{n+1}(v)}{B_1(v)t^n+\dots+B_{n+1}(v)}.$$

\section{Discussion}
\label{sec:discussion}
  In this paper we have proposed a general framework to study the $\mathcal{N}=1$ breaking of class $\cal S$ theories. 
  We provided for a large class of models a recipe for constructing the M-theory curve of the theory 
  and we have seen that the condition determining whether a point on the Coulomb branch is lifted or not arises 
  in a simple way from the geometric set-up. 
  Our construction exploits the choice of a vector field and we have seen in the case of $\SU(2)$ SQCD 
  that changing this choice allows to study the $\mathcal{N}=1$ breaking in S-duality frames. 
  There are several directions worth to be further pursued.

  In this paper we focused on $A_{N-1}$ theories, 
  it would be interesting to analyze along the same lines other classical groups. 

  Another aspect which deserves to be deepened is the analysis of the mathematical properties 
  of the generalized Hitchin system, with particular emphasis on its moduli space 
  and its possible relation with integrable systems and their quantization 
  \cite{Nekrasov:2009rc,Bonelli:2009zp,Mironov:2009uv,Maruyoshi:2010iu,Nekrasov:2011bc,Bonelli:2011na,
  Tan:2013tq,Vartanov:2013ima}.

  An intriguing related question is whether one can find a two-dimensional field theory description 
  encoding the infrared properties of the ${\cal N}=1$ four dimensional gauge theory.
  For the subset of theories obtained via breaking of theories in ${\cal S}$-class the chiral condensates 
  can be computed via instanton counting by fixing the Coulomb moduli to the vacua of the superpotential 
  \cite{Fucito:2005wc}. 
  From the AGT viewpoint \cite{Alday:2009aq}, this amounts to fix the momentum of the internal channel 
  to some particular values, 
  which correspond to the stationary points of the Hitchin system's Hamiltonians \cite{Dorey:2001qj}. 
  
  This has a resemblance with the insertion of surface operators in the four-dimensional gauge theory 
  and the supersymmetry breaking induced by these defects. 
  For example a relation between surface operators and DV curves has been recently studied in \cite{Gaiotto:2013sma}. 
  
  The geometrical framework we studied should help to elucidate and deepen our understanding of ${\cal N}=1$ dualities. 
  There are a number of questions that should be investigated with respect to this issue, 
  for example to which deformation the marginal parameters correspond to 
  and how to read the superpotential of the dual theory. 
  Presumably the choice of vector field plays an important role in this respect.

  An interesting class of ${\cal N}=1$ four-dimensional gauge theories has been introduced recently 
  in \cite{Franco:2012mm,Heckman:2012jh,Xie:2012mr}.
  These are associated to bypartite graphs on bordered Riemann surfaces. 
  It would be interesting to investigate whether these models have a connection with the ones described 
  by the generalized Hitchin system.  

  Finally the analysis of ${\cal N}=1$ theories on more general four-dimensional manifolds could be considered 
  in an extension of the geometrical framework considered in this paper.
 \\

\noindent{\bf Note added}:
While finalizing this paper, it appeared in the arXiv \cite{Xie:2013gma} where a similar geometric set-up is considered.

\section*{Acknowledgments}
It is a pleasure to thank Francesco Benini, Michele Del Zotto, Hirosi Ooguri, Sara Pasquetti, Jaewon Song, 
Yuji Tachikawa, Futoshi Yagi, Kazuya Yonekura and Peng Zhao
for many stimulating discussions.
KM would like to thank Particle physics group, SISSA and Simons Summer Workshop 2013 
in Mathematics and Physics, where this work was partly done, for hospitality.
SG would like to thank the Kavli Institute for the Physics and Mathematics of the Universe, University of Tokyo 
for hospitality during the completion of this work.
This research was partly supported 
by the INFN Research Project PI14 ``Nonperturbative dynamics of gauge theory'', 
by the INFN Research Project TV12, by PRIN ``Geometria delle variet`a algebriche'', 
by MIUR-PRIN contract 2009-KHZKRX,
and by a Simons Investigator award from the Simons Foundation to Hirosi Ooguri.
The work of KM is supported by a JSPS postdoctoral fellowship for research abroad.

\appendix
\section{Comparison with IIA brane theory perspective}
\label{sec:Hori}
  In this appendix we follow the method in \cite{Hori:1997ab,Giveon:1997sn} and compute the $\CN=1$ curve.
  We see the agreement with the results in section \ref{sec:A1}
  using the factorization condition from the generalized Hitchin viewpoint.

\subsection{$\SU(2)$ SQCD}
  Let us begin with $\SU(2)$ gauge theory with $N_{f} \leq 4$ flavors.
  By the mass deformation of the adjoint chiral field, the loci on the Coulomb branch of corresponding $\CN=2$ theory
  where massless particles appear become the vacua of the $\CN=1$ theory.
  In this case it was argued in \cite{Hori:1997ab} that 
  only if the $\CN=2$ curve is completely factorized to a sphere, the curve can be rotated to $x^{7,8,9}$ direction.
  Since $w$ and $v$ are related by $w = \mu v$ ($v \rightarrow \infty$), we have \cite{Hori:1997ab}
    \bea
    v
     =     \frac{(w - w_{+})(w - w_{-})}{\mu w}.
    \eea
  Writing in the form 
    \bea
    w^{2} - (\mu v w + c) + f_{0}
     =     0,
    \label{DV}
    \eea
  which is identified with the DV curve,
  this implies $c= w_{+}+w_{-}$ and $f_{0} = w_{+}w_{-}$.
  The curve is always of this form in $\SU(2)$ gauge theory with $N_{f}$ flavors.
  This is also true in $\SU(N)$ gauge theory with $N_{f}$ flavors.

\subsubsection{Pure SYM}
\label{subsubsec:pureSU(2)Hori}
  The brane configuration of this theory is described by two D4-branes suspended between two NS5-branes.
  Again we denote the NS5-branes as NS$_{1}$ and NS$_{2}$ below, as in figure \ref{fig:SU(2)Nf0}.
  The $\CN=2$ Seiberg-Witten curve is \eqref{N=2pureSU(2)}.
  Let us consider the rotation of the NS5$_{2}$-brane (at $t=0$) 
  which corresponds to the inclusion of the mass term of the adjoint chiral field. 
  The boundary conditions are
    \bea
    {\rm NS}_{1}:~
    w 
    &\rightarrow& 0 ~~{\rm as}~~ t= v^{2}/\Lambda^{2} ~~(v \rightarrow \infty),
    \nonumber \\
    {\rm NS}_{2}:~
    w 
    &\rightarrow& \mu v ~~{\rm as} ~~ t= \Lambda^{2}/v^{2} ~~(v \rightarrow \infty).
    \eea
  This completely fixes the $\CN=1$ curve as follows.
  By using the DV curve we have $v \sim w/\mu$ when $w \rightarrow \infty$
  and $v \sim (w_{+}w_{-}/\mu \Lambda) w^{-1}$ when $w \rightarrow 0$.
  Thus the boundary conditions can be written as
    \bea
    {\rm NS}_{1}:~
    w 
    &\rightarrow& 0 ~~{\rm as}~~ t= \left(w_{+} w_{-}/\mu \Lambda \right)^{2} w^{-2},
    \nonumber \\
    {\rm NS}_{2}:~
    w 
    &\rightarrow& \mu v ~~{\rm as} ~~ t= (\mu \Lambda)^{2} w^{-2}.
    \label{boundary2}
    \eea
  This means simply
    \bea
    t
     =     \frac{(\mu \Lambda)^{2}}{w^{2}},
           \label{2ndHitchinpureSU(2)}
    \eea
  with $w_{+} w_{-} = (\mu \Lambda)^{2}$.
  In other words, this is expressed as
    \bea
    w^{2}
     =     \frac{(\mu \Lambda)^{2}}{t},
           \label{N=1curvepureHori}
    \eea
  Notice that we have unknown constants $w_{+}$ and $w_{-}$ in the curve.
  This is determined by substituting \eqref{DV} and \eqref{2ndHitchinpureSU(2)} 
  into the $\CN=2$ curve \eqref{N=2pureSU(2)} and expanding by $w$, which leads to
  $w_{+} + w_{-} = 0$ and $w_{+} w_{-} = (\mu \Lambda)^{2}$.

\subsubsection{$N_{f} = 1$}
\label{subsubsec:SU(2)Nf1}
  There are two ways to construct $\CN=2$ $\SU(2)$ gauge theory with one flavors in the IIA and M-theory.
  One is to place the D6-brane to the right of the NS$_{2}$-brane.
  The other is to the left of the NS$_{1}$-brane.
  Let us consider these in order.
  
  For the first case, the $\CN=2$ curve is \eqref{WWc}.
  Let us below focus on the massless case $m = 0$.
  The massive case does not change the form of the final curve.
  Then, we consider the rotation of the NS$_{2}$-brane.
  The boundary conditions are
    \bea
    {\rm NS}_{1}:~
    w
    &\rightarrow& 0 ~~{\rm as}~~ t= v^{2}/\Lambda^{2} = (w_{+}w_{-}/\mu \Lambda)^{2} w^{-2} ~~(v \rightarrow \infty),
    \nonumber \\
    {\rm NS}_{2}:~
    w
    &\rightarrow& \mu v ~~{\rm as}~~ t= \Lambda/v = \mu \Lambda /w ~~(v \rightarrow \infty).
    \eea
  where we have used the DV curve as in \eqref{boundary2}.
  These determine 
    \bea
    t 
     =     (\mu \Lambda) \frac{w-w_{-}}{w^{2}},
    \eea
  where the root of the numerator was obtained by the fact that $t=0$ leads to $v=0$.
  This can be rewritten as
    \bea
    w^{2} - \frac{\mu \Lambda}{t} w + \frac{\mu \Lambda w_{-}}{t}
     =     0.
    \eea
  By eliminating the linear term in $v$, we get
    \bea
    w^{2}
     =     \frac{(\mu \Lambda)^{2}}{4t^{2}} - \frac{\mu \Lambda w_{-}}{t},
           \label{N=1curveNf1Hori}
    \eea
  implying that $V_{2}$ has double poles at $t=0$.
  By checking the consistency with the $\CN=2$ curve as above we get
  $w_{+} + 2w_{-} = 0$ and $w_{+}^{2} w_{-} = - (\mu \Lambda)^{3}$.

  Let us then consider the second realization of this theory.
  The SW curve is
    \bea
    \Lambda (v + m) t^{2} - P_{2} (v) t + \Lambda^{2}
     =     0.
    \eea
  Again we set $m=0$ for simplicity.
  The boundary conditions are
    \bea
    w
    &\rightarrow& 0 ~~{\rm as}~~ t= v/\Lambda = (w_{+}w_{-}/\mu \Lambda) w^{-1} ~~(v \rightarrow \infty),
    \nonumber \\
    w
    &\rightarrow& \mu v ~~{\rm as}~~ t= \Lambda^{2}/v^{2} = (\mu \Lambda)^{2} w^{-2} ~~(v \rightarrow \infty).
    \eea
  This gives the $\CN=1$ curve:
    \bea
    t
     =     (\mu \Lambda)^{2} \frac{1}{w(w-w_{+})}.
    \eea
  The constants $w_{+}$ and $w_{-}$ are determined by the same condition as the first case.
  The $\CN=1$ curve is written as
    \bea
    w^{2} - w_{-} w - \frac{(\mu \Lambda)^{2}}{t}
     =     0.
    \eea
  The singularity structure is the same as that of the pure case in subsection \ref{subsubsec:pureSU(2)Hori}.
  Indeed, this should be the case since we have rotated the $\CN=2$ puncture of degree $3$ at $t=0$.

\subsubsection{$N_{f}=2$}
\label{subsubsec:SU(2)Nf2Hori}
  As in subsection \ref{subsubsec:SU(2)Nf2}, we see three different realizations of this theory.  
  
\paragraph{the first realization}
  The SW curve is (\ref{N=2curveSU(2)Nf21}).
  We would like to rotate the NS$_{2}$-brane corresponding to the regular puncture at $t=1$.
  The boundary conditions are
    \bea
    w
    &\rightarrow& 0 ~~{\rm as}~~ t= v^{2}/\Lambda^{2} = (w_{+}w_{-}/\mu \Lambda)^{2} w^{-2} ~~(v \rightarrow \infty),
    \nonumber \\
    w
    &\rightarrow& \mu v ~~{\rm as}~~ t= 1 ~~(v \rightarrow \infty).
    \eea
  Thus the $\CN=1$ curve is
    \bea
    t 
     =     \frac{(w-w_{+})^{2-r}(w-w_{-})^{r}}{w^{2}}
    \eea
  with $r=0,1$.
  $w_{\pm}$ are determined by $w_{+}^{r} w_{-}^{2-r} = (\mu \Lambda)^{2}$ 
  and $(2-r)w_{+} + r w_{-} = 0$,
  which are the same as eq.~(4.7) and (4.8) in \cite{Hori:1997ab}.
  Thus the interpretation of ``$r$'' is the same as in \cite{Hori:1997ab}:
  $r=1$ corresponds to non-baryonic branch and $r=0$ corresponds to a locus where the $A$-cycle collapses.
  In this $\SU(2)$ with $N_{f} = 2$ example, the $r=0$ locus is just the root of the baryonic branch.
  
  When $r=0$, $w_{+}=0$.
  Therefore the DV and the $\CN=1$ curves get simplified to
    \bea
    w 
     =     \mu v + w_{-}, ~~~
    t 
     =     1.
    \eea
  Indeed, on the root of the baryonic branch, the NS$_{2}$-brane is detached from the other branes.
  Thus the NS$_{2}$ is rotatable, meaning a flat NS5-brane with $t=1$ with $w = \mu v$. 
  (See figure \ref{fig:SU(2)Nf2}.) 
  Let us notice that the $\CN=2$ curve is factorized as
    \bea
    0
     =     \Lambda^{2} t^{2} - P_{2}(v) t + v^{2}
     =   - (t-1) (-\Lambda^{2} t + v^{2}).
    \eea
  Thus, it is easy to see that the $\CN=2$ Coulomb moduli is fixed to be $u=-\Lambda^{2}$. 
  
  When $r=1$, the curves are
    \bea
    w^{2} - \mu v w + (\mu \Lambda)^{2}
     =     0, ~~~
    w^{2} 
     =     \frac{(\mu \Lambda)^{2}}{t-1}.
           \label{N=1curveSU(2)Nf21-2Hori}
    \eea
  The $\CN=1$ curve agrees with \eqref{N=1curveSU(2)Nf21-2}.
  
\paragraph{the second realization}
  The $\CN=2$ curve is \eqref{N=2curveSU(2)Nf22}.
  The boundary conditions are
    \bea
    {\rm NS}_{1}:~
    w
    &\rightarrow& 0 ~~{\rm as}~~ t=  (w_{+}w_{-}/\mu \Lambda) w^{-1} ~~(v \rightarrow \infty),
    \nonumber \\
    {\rm NS}_{2}:~
    w
    &\rightarrow& \mu v ~~{\rm as}~~ t= (\mu \Lambda) w^{-1} ~~(v \rightarrow \infty).
    \eea
  Thus, we get
    \bea
    t 
     =     (\mu \Lambda) \frac{(w-w_{-})^{r-1}}{w (w - w_{+})^{r-1}},
    \eea
  with $r=0,1$. 
  $w_{\pm}$ are determined by the same equations as above.
  
  When $r=0$, the curves again get simplified:
    \bea
    w 
     =     \mu v + w_{-}, ~~~
    w - w_{-} + \frac{\mu \Lambda}{t}
     =     0.
           \label{SU(2)Nf2N=1curvesecond}
    \eea
  The $\CN=1$ curve denotes that when $t \rightarrow \infty$ we have $w = w_{-}$,
  and when $t \rightarrow 0$ we have $w = \infty$.
  This precisely describes the rotation of the NS$_{2}$ on the root of the baryonic branch 
  as depicted in figure \ref{fig:SU(2)Nf22}.
  We interpret $w_{-}$ as the position in $w$-direction of the D4-brane
  stretched between the NS$_{2}$ and D6-branes.
  This agrees with the result obtained in \eqref{N=1curveSU(2)Nf22-1}.
  Indeed the $\CN=1$ curve \eqref{SU(2)Nf2N=1curvesecond} can be written as
  $w^{2} - \mu \Lambda (1/t \pm 1) w=0$, since $w_{-} = \pm \mu \Lambda$.
  Thus by shifting $w$-coordinate, we have $\tilde{w}^{2} = (\mu \Lambda/2)^{2} (1/t \pm 1)^{2}$,
  which is \eqref{N=1curveSU(2)Nf22-1}.
    
  When $r=1$,
    \bea
    w^{2} - \mu v w + (\mu \Lambda)^{2}
     =     0, ~~~
    w - \frac{(\mu \Lambda)}{t}
     =     0.
           \label{SU(2)Nf2N=1curvesecond2}
    \eea
  The $\CN=1$ curve simply denotes that when $t \rightarrow \infty$, $w = 0$
  and when $t \rightarrow 0$, $w \rightarrow \infty$,
  which is depicted in figure \ref{fig:SU(2)Nf22}.
  The argument as above shows that this agrees with \eqref{N=1curveSU(2)Nf22-2}.

\paragraph{the third realization}
  The $\CN=2$ curve is \eqref{N=2curveNf23}.
  We only show the results here.
  The DV curves for $r=0,1$ are the same as above. 
  The $\CN=1$ curves are as follows: 
  when $r=0$
    \bea
    w^{2} - 2w_{-} w + w_{-}^{2} - \frac{(\mu \Lambda)^{2}}{t}
     =     0,
           \label{SU(2)Nf2N=1curvethird1}
    \eea
  with $w_{-} = \pm \mu \Lambda$,
  meaning that when $t \rightarrow \infty$ the equation has double zero at $w_{+}$
  which is the position of D4-branes at $t=\infty$ in $w$-direction (see figure \ref{fig:SU(2)Nf23});
  when $r=1$
    \bea
    w^{2} + w_{+}w_{-} - \frac{(\mu \Lambda)^{2}}{t}
     =     0,
           \label{SU(2)Nf2N=1curvethird2}
    \eea
  meaning that when $t \rightarrow \infty$ the equation has two zeros at $w_{\pm}$ ($w_{+} + w_{-} =0$).
  This denotes the non-baryonic branch as in figure \ref{fig:SU(2)Nf23}.
  The NS$_{2}$-brane cannot detach from the D4-branes because of the $s$-rule.

\subsubsection{$N_{f} = 3$}
  We consider two realizations depicted in figure \ref{fig:SU(2)Nf3} in order.

\paragraph{the first realization}
  The SW curve is
    \bea
    \Lambda (v+m_{1}) t^{2} - P_{2}(v) t + (v+m_{2}) (v+m_{3})
     =     0.
    \eea
  There are two regular punctures at $t=0$ and $t=-1$ and one irregular puncture of degree $4$ at $t=\infty$.
  As before we consider the massless case.
  
  The boundary conditions after the rotation of the NS$_{2}$-brane are
    \bea
    {\rm NS}_{1}:~
    w
    &\rightarrow& 0 ~~{\rm as}~~ t= v/\Lambda = (w_{+}w_{-}/\mu \Lambda) w^{-1} ~~(v \rightarrow \infty),
    \nonumber \\
    {\rm NS}_{2}:~
    w
    &\rightarrow& \mu v ~~{\rm as}~~ t= 1 ~~(v \rightarrow \infty).
    \eea
  Thus,
    \bea
    t 
     =     \frac{(w - w_{+})^{2-r}}{(w - w_{-})^{1-r}}
    \eea
  with $r=0,1$.
  It is easy to see $w_{+}^{-1+r} w_{-}^{2-r} =- \mu \Lambda$ and $(2-r)w_{+} + (-1+r)w_{-} = 0$.
  
  When $r=0$, the curves are
    \bea
    w^{2} - (3 w_{+} + \mu v) + w_{+} w_{-}
     =     0, ~~~~
    w^{2} + \frac{\mu \Lambda}{2} w - \frac{(\mu \Lambda)^{2}}{4(t-1)}
     =     0.
    \eea
  This corresponds to the locus in the Coulomb branch where the $A$-cycle collapses.
  This also expresses the behavior of the rotation of the $\CN=2$ regular puncture:
  the behavior at singularity is just the same as that of the irregular puncture of degree $3$.
    
  When $r=1$, the curves are
    \bea
    w 
     =    \mu v + w_{-}, ~~~~
    t
     =    1.
    \eea
  This describes the root of the baryonic branch, as can be seen in figure \ref{fig:SU(2)Nf3}.
  Notice that the configuration in the figure can also be considered as the root of the non-braryonic branch
  by reconnecting the NS$_{2}$-branes to the D4-brane stretched between the NS$_{1}$ and D6-branes.
  Contrary to the first and the third realizations of the $N_{f}=2$ case, no $s$-rule is applied.
  
\paragraph{the second realization}
  The SW curve is \eqref{N=2curveSU(2)Nf32}.
  Let us present only the result of the massless case here.
  The DV curves for $r=0,1$ are the same as above.
  The $\CN=1$ curves are
    \bea
    w^{2} - \left( \frac{\mu \Lambda}{t} + 2w_{-} \right) w + w_{-}^{2} + \frac{\mu \Lambda w_{+}}{t}
     =     0
    \eea
  for $r=0$, and
    \bea
    w - w_{-} - \frac{\mu \Lambda}{t}
     =     0
    \eea
  for $r=1$.
  In the former expression, we can shift $w$ to get
    \bea
    w^{2}
     =     \frac{\mu^{2} \Lambda^{2}}{4t^{2}} - \frac{\mu \Lambda w_{+}}{t}.
           \label{N=1curveSU(2)Nf32-2Hori}
    \eea
  where $w_{-} = - \mu \Lambda/4$.
  This agrees with \eqref{N=1curveSU(2)Nf32-2}.
  The latter denotes the brane system consisting of the NS$_{2}$, D4 and D6-branes (figure \ref{fig:SU(2)Nf3}).
  
\subsubsection{$N_{f} = 4$}
  As the final example of $\SU(2)$ SQCD we consider the $N_{f}=4$ case.
  The SW curve is \eqref{N=2curveSU(2)Nf4}.
  Again we consider the massless case.
  The boundary conditions are when $v \rightarrow \infty$, $t=q, 1$; 
  $w \rightarrow \mu v$ as $v \rightarrow \infty$, $t = q$.
  The solution is
    \bea
    t
     =     q (w - w_{+})^{r-2} (w - w_{-})^{2-r},
    \eea
  where $r=0,1,2$.
  By checking the consistency with the SW curve we see that $w_{+} = w_{-}$.
  Therefore the $\CN=1$ curve is trivially $t =q$.
  In this theory all the vacua with $r=0,1,2$ are at $u = 0$.
  Indeed, only the locus where the SW curve factorizes is at $u=0$.
  However, this denotes the point where superconformal symmetry is restored.
  The adjoint chiral mass deformation of this point leads to $\CN=1$ superconformal fixed point 
  in the IR \cite{Leigh:1995ep}. 
  Therefore, our curve system is related to this $\CN=1$ fixed point.

\subsection{$\SU(2)^{2}$ quiver}
  Let us then consider $\SU(2)^{2}$ quiver gauge theory.
  The SW curve is \eqref{N=2curveSU(2)^{2}}.

\subsubsection{Rotation of an NS5-brane}
\label{subsec:SU(2)^2}
  The $\CN=1$ deformation of this quiver has been studied in \cite{Giveon:1997sn}.
  Let us first study the case where one NS5-brane is rotated.

\begin{figure}
  \centering
  \includegraphics[width=10cm]{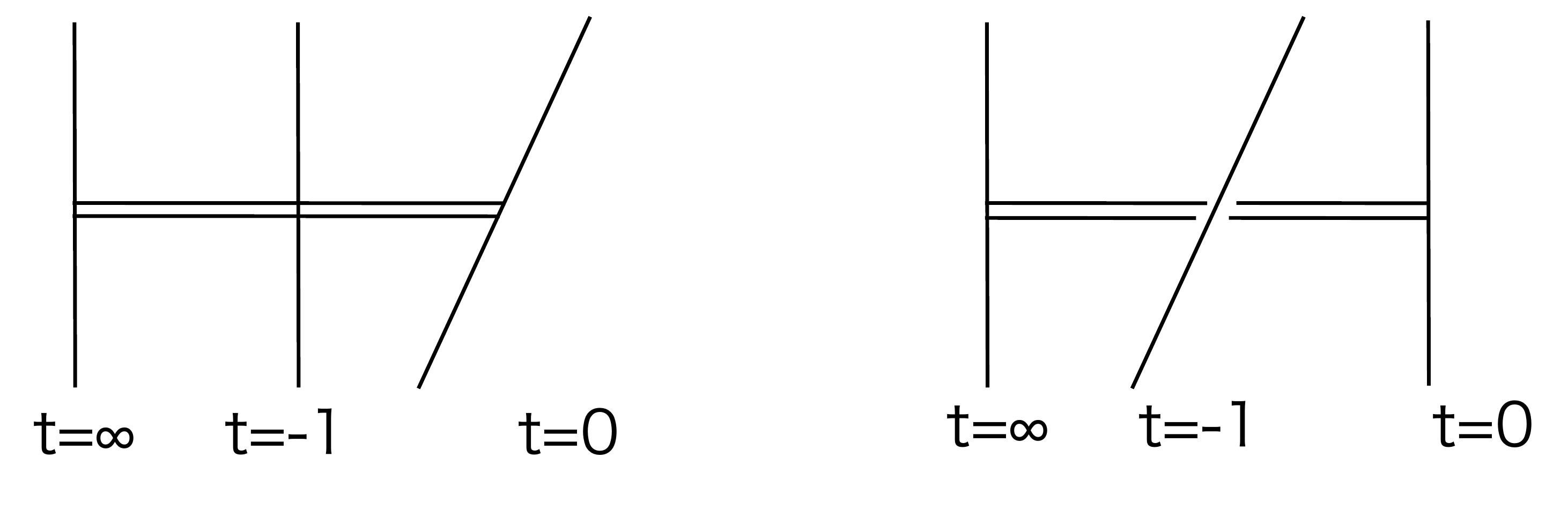}
  \caption{Left: the rotation of NS$_{3}$.
           Right: the rotation of NS$_{2}$.
           The NS5-branes can be detached from the rest.}
  \label{fig:SU(2)x21}
  \end{figure}

\paragraph{rotation of the NS$_{3}$}
  Let us first consider the mass deformation of the adjoint chiral of the second $\SU(2)$ vector multiplet by $\mu$.
  This is described by the rotation of the right NS5-branes as in figure \ref{fig:SU(2)x21}.
  We refer to three NS5-branes as NS$_{1}$, NS$_{2}$ and NS$_{3}$ from the left.
  The boundary conditions are
    \bea
    {\rm NS_{1}}: 
    w
    &\rightarrow& 0 ~~{\rm as}~~ t= - v^{2}/\Lambda_{1}^{2} ~~(v \rightarrow \infty),
    \nonumber \\
    {\rm NS_{2}}:
    w
    &\rightarrow& w_{2} ~~{\rm as}~~ t= -1 ~~(v \rightarrow \infty),
    \nonumber \\
    {\rm NS_{3}}:
    w
    &\rightarrow& \mu v ~~{\rm as}~~ t= - \Lambda_{2}^{2}/v^{2} ~~(v \rightarrow \infty),
    \eea
  where $w_{2}$ is an unknown constant.
  The DV curve is easily determined:
  since $w \rightarrow \mu v$, $v$ is written as $v = \frac{P_{3}(v)}{\mu R_{2}(v)}$
  where $P_{3}$ and $R_{2}$ are polynomials of degree $3$ and $2$. 
  The boundary conditions at NS$_{1}$ and NS$_{2}$ means $R_{2} = w (w - w_{2})$.
  Thus
    \bea
    v
     =     \frac{P_{3}(v)}{\mu w (w - w_{2})}.
    \eea
  We will later see the relation between this and the DV curve obtained in the previous note.
  
  Let us then write the $\CN=1$ curve.
  Let $P_{3}$ be $w^{3} + p_{1}w^{2} + p_{2} w + p_{3}$.
  Since the DV curve behaves $v \sim w/\mu$ at $w \rightarrow \infty$ and $v \sim - p_{3}/\mu w_{2} w$ at $w =0$,
  so $t = - (p_{3}/\mu \Lambda_{1} w_{2})^{2} w^{-2}$ and $t = - (\mu \Lambda_{2})^{2} w^{-2}$
  at these points.
  Thus, the curve is simply
    \bea
    w^{2} + \frac{(\mu \Lambda_{2})^{2}}{t}
     =     0
    \eea
  with $p_{3} = \mu^{2} \Lambda_{1} \Lambda_{2} w_{2}$.
  This also determines $w_{2} = \pm \mu \Lambda_{2}$.
  Indeed, this curve has the expected singularity of the rotated puncture at $t=0$ 
  of $\CN=2$ irregular one of degree $3$, seen in subsection \ref{subsubsec:pureSU(2)Hori}.
  This suggests that this behavior around the puncture is also true in the more generic theories
  and will help the analysis drastically.
  
  The other coefficients in $P_{3}$ are obtained by substituting these two equations into $\CN=2$ curve 
  and by checking the orders $w^{-1}$, $w^{-5}$ and $w^{-6}$ terms
    \bea
    p_{1}
     =   - w_{2}, ~~~
    p_{2}
     =   - p_{3}/w_{2}.
    \eea
  Now the DV curve can be written as, after the shift $w \rightarrow w + w_{2}/2$,
    \bea
    w^{2} (w - (\mu v - w_{2}/2)) + w f_{0} + f_{1}(v)
     =     0,
    \eea
  where 
    \bea
    f_{0}
     =   - \mu^{2} \Lambda_{1} \Lambda_{2}, ~~~
    f_{1}
     =     \frac{\mu w_{2}^{2}}{4} v + \frac{\mu^{2} w_{2}}{2} (\Lambda_{1} \Lambda_{2} - \Lambda_{2}^{2}/4).
    \eea
  This is indeed the form of the DV curve obtained in the previous note with $W'_{1} = 0$
  and $W'_{2} = \mu v - w_{2}/2$.

\paragraph{rotation of the NS$_{2}$}
  This corresponds to the introduction of the masses $\mu$ and $- \mu$ of the adjoint chirals 
  of $\SU(2)_{1}$ and $\SU(2)_{2}$ vector multiplets.
  Again the boundary conditions are
    \bea
    {\rm NS_{1}}: 
    w
    &\rightarrow& 0 ~~{\rm as}~~ t= - v^{2}/\Lambda_{1}^{2} ~~(v \rightarrow \infty),
    \nonumber \\
    {\rm NS_{2}}:
    w
    &\rightarrow& \mu v ~~{\rm as}~~ t= -1 ~~(v \rightarrow \infty),
    \nonumber \\
    {\rm NS_{3}}:
    w
    &\rightarrow& w_{3} ~~{\rm as}~~ t= - \Lambda_{2}^{2}/v^{2} ~~(v \rightarrow \infty),
    \eea
  The DV curve is not changed except that $w_{2}$ is now $w_{3}$
    \bea
    v
     =     \frac{P_{3}(w)}{\mu w (w - w_{3})}.
    \eea
  So the boundary conditions are $t = - (\frac{P_{3}(0)}{\mu \Lambda_{1}w_{3}})^{2} w^{-2}$ at $w = 0$
  and $t = - (\frac{\mu \Lambda_{2} w_{3}}{P_{3}(w_{R})})^{2} (w-w_{3})^{2}$ at $w = w_{3}$.
  Thus
    \bea
    t
     =   - \frac{(w - w_{3})^{2}}{w^{2}}.
    \eea
  However, the $w_{3}$ is determined to vanish by checking the $w^{1}$ order terms of the $\CN=2$ Seiberg-Witten curve
  (by substituting the above equations).
  Thus $t=-1$ which means that the NS$_{2}$-brane is detached from the rest.
  The rest part does not depend on the $w$-coordinate, namely on the fixed position of $w$, 
  as can be seen in figure \ref{fig:SU(2)x21}.
  This implies that the polynomial in the DV curve is $P_{3} = w^{3}$, and the DV curve is $v = w/\mu$.
  
  The detachment observed above means that the $\CN=2$ Seiberg-Witten curve is factorized to
    \bea
    0
     =     \Lambda_{1}^{2} t^{3} + P_{2,1}(v) t^{2} + P_{2,2}(v) t + \Lambda_{2}^{2}
     =     (t+1) (\Lambda_{1}^{2} t^{2} + (v^{2} + f) t + \Lambda_{2}^{2}),
    \eea
  where $f=u_{1} - \Lambda_{1}^{2} = u_{2} - \Lambda_{2}^{2}$, and $u_{1,2}$ are the Coulomb moduli of $\SU(2)_{1,2}$.
  Note that the moduli are related in these $\CN=1$ vacua, but still there is one (complex-)dimensional moduli space.
  Let us rewrite the latter part of the curve as
    \bea
    v^{2} + f + \Lambda_{1}^{2} t + \frac{\Lambda_{2}^{2}}{t}
     =     0.
           \label{N=1SWcurve}
    \eea
  
  As noticed in section \ref{subsec:SU(2)^{2}}, the theory with masses $\mu$ and $-\mu$ is 
  $\CN=1$ $\SU(2) \times \SU(2)$ gauge theory with two bifundamental chiral multiplets.
  This choice was exactly the two NS$_{1}$ and NS$_{3}$ branes are parallel.
  And after the detachment of the NS$_{2}$ brane, the two D4-branes freely move describing the moduli of this theory.

\subsection{$\SU(N)$ SQCD}
  Let us consider $\SU(N)$ SQCD with $N_{f}$ flavors.
  The $\CN=2$ curve is 
    \bea
    \Lambda^{N} t^{2} + P_{N}(v)t + \Lambda^{N-N_{f}}\prod_{f=1}^{N_{f}} (v - m_{f})
     =     0 ,
    \eea
  where $P_{N}$ is a polynomial of degree $N$.
  
  The boundary conditions are
    \bea
    {\rm NS_{1}}: 
    w
    &\rightarrow& 0 ~~{\rm as}~~ t= - v^{N}/\Lambda^{N} ~~(v \rightarrow \infty),
    \nonumber \\
    {\rm NS_{2}}:
    w
    &\rightarrow& \mu v ~~{\rm as}~~ t= -v^{N_{f}-N}/\Lambda^{N-N_{f}} ~~(v \rightarrow \infty).
    \eea
  The DV curve is again
    \bea
    v
     =     \frac{(w - w_{+})(w - w_{-})}{\mu w}.
    \eea
  The $\CN=1$ curve is obtained by observing that when $w \rightarrow \infty$, $t=-(\mu \Lambda/w)^{N-N_{f}}$,
  and when $w \rightarrow 0$, $t= - (w_{+} w_{-}/\mu \Lambda)^{N} w^{-N}$:
    \bea
    t
     =   - (\mu \Lambda)^{N-N_{f}} \frac{(w-w_{+})^{r}(w-w_{-})^{N_{f}-r}}{w^{N}},
    \eea
  where $r=0, 1, \ldots, [N_{f}/2]$.
  $w_{\pm}$ are determined by 
    \bea
    w_{+}^{N-r} w_{-}^{N-N_{f}+r}
    &=&    (-1)^{N_{f}} (\mu \Lambda)^{2N-N_{f}},
           \nonumber \\
    (N-N_{f} +r) w_{+} + (N-r)w_{-}
    &=&    0.
    \eea
  The $\CN=1$ curve in this vacuum can be written as
    \bea
    w^{N} + \sum_{k} V_{k}(t) w^{N-k}
     =     0,
           \label{N=1SU(N)SQCD}
    \eea
  where 
    \bea
    V_{1} 
    &=&    V_{2}
     =     \ldots
     =     V_{N-N_{f}-1}
     =     0, 
           \nonumber \\
    V_{N-N_{f}}
    &=&    \frac{(\mu \Lambda)^{N-N_{f}}}{t}, ~~~~~~
    V_{N-N_{f}+1}
     =     \frac{(\mu \Lambda)^{N-N_{f}}(-r w_{+} - (N_{f}-r)w_{-})}{t}, \ldots,
           \nonumber \\
    V_{N}
    &=&    \frac{(\mu \Lambda)^{N-N_{f}} (-1)^{N_{f}}w_{+}^{r} w_{-}^{N_{f}-r}}{t}.
    \eea
  
  In particular, for the pure SYM theory, we obtain the $\CN=1$ curve
  \eqref{N=1SU(N)SQCD} with
    \bea
    V_{1} 
     =     V_{2}
     =     \ldots
     =     V_{N-1}
     =     0, ~~~~~
    V_{N}
     =     \frac{(\mu \Lambda)^{N}}{t}.
    \eea
  Namely, $w^{N} = (\mu \Lambda)^{N}/t$.

\bibliographystyle{ytphys}

\bibliography{ref}

\end{document}